\documentclass[twocolumn]{aastex61}

\hypersetup{linkcolor=red,citecolor=blue,filecolor=cyan,urlcolor=magenta}

\shorttitle{Seyfert Lobes}
\shortauthors{Ghosh et al.}
\begin{document}
\title{Magnetic Field Structures In and Around Seyfert Galaxy Outflows}
\author[0009-0000-1447-5419]{Salmoli Ghosh}
\correspondingauthor{Salmoli Ghosh}
\email{salmoli@ncra.tifr.res.in}
\affiliation{National Centre for Radio Astrophysics (NCRA) - Tata Institute of Fundamental Research (TIFR), S. P. Pune University Campus, Pune 411007, Maharashtra, India}
\author[0000-0003-3203-1613]{Preeti Kharb}
\affiliation{National Centre for Radio Astrophysics (NCRA) - Tata Institute of Fundamental Research (TIFR), S. P. Pune University Campus, Pune 411007, Maharashtra, India}
\author[0000-0001-8428-6525]{Biny Sebastian}
\affiliation{Department of Physics and Astronomy, University of Manitoba, Winnipeg, Manitoba, Canada}
\affiliation{Space Telescope Science Institute, 3700 San Martin Drive, Baltimore, MD 21218, USA}
\author[0000-0002-6972-2760]{Jack Gallimore}
\affiliation{Department of Physics and Astronomy, Bucknell University, Lewisburg, PA 17837}
\author[0000-0003-1933-4636]{Alice Pasetto}
\affiliation{Instituto de Radioastronom\'ia y Astrof\'isica (IRyA-UNAM), 3-72 (Xangari), 8701, Morelia, 
Mexico}
\author[0000-0001-6421-054X]{Christopher P. O'Dea}
\affil{Department of Physics and Astronomy, University of Manitoba, Winnipeg, Manitoba, Canada}
\author{Timothy Heckman}
\affil{William H. Miller III Department of Physics \& Astronomy, Johns Hopkins University, Bloomberg Center for Physics and Astronomy, 3400 N. Charles Street, Baltimore, MD 21218, USA}
\author{Stefi A. Baum}
\affiliation{Department of Physics and Astronomy, University of Manitoba, Winnipeg, Manitoba, Canada}

\begin{abstract}
We present radio polarimetric images of 12 Seyfert and Low-Ionization Nuclear Emission-line Region (LINER) galaxies belonging to the Centre for Astrophysics (CfA)+12 micron sample exhibiting kiloparsec-scale radio outflows (KSRs). These observations have been carried out at 10 GHz with Karl G. Jansky Very Large Array (VLA) in D-array and at 1.4 GHz with the BnA$\rightarrow$A array configurations. We find signatures of organized magnetic (B-) field structures in the cores, jets and lobes of these galaxies. The linear polarization fraction varies from a few per cent in the cores to $47\pm18$ per cent in the lobes. The inferred B-fields are toroidal in the cores of several sources making them consistent with the presence of either a sheath-like or a wind-like component surrounding the jet. The in-band spectral index images typically show the presence of flat/inverted spectrum cores and steep spectrum lobes. Radio cores with flatter spectra are found to have lower Eddington ratios while the steeper ones have higher. A strong correlation is observed between the Seyfert/LINER radio outflow properties and the mass of the supermassive black holes (SMBHs); correlations with Eddington ratios are weaker. 
We find signatures of jet-medium interaction and both positive and negative AGN feedback in these sources. Overall, our study indicates that radio-quiet (RQ) AGN with KSRs possess radio outflows driven by magnetic fields anchored to their black holes - accretion disks, which significantly impact their environments.
\end{abstract}

\nopagebreak
\section{Introduction}
{Seyfert galaxies \citep{Seyfert1943} were the first identified class of galaxies known to host active galactic nuclei (AGN), a phenomenon now understood to result from mass accretion onto supermassive black holes (SMBH) \citep{Rees1984}. Seyeferts are known to host SMBH of lower masses, typically ranging from M $\sim10^6-10^8$M$_\sun$ \citep{Mclure2001}.} They are {traditionally} classified as radio-quiet (RQ) AGN {\citep{Padovani2017, Panessa2019}}, characterized by a ratio of radio flux density at 5 GHz to optical flux density in the B-band of $\le10$ \citep{Kellermann1989}. These galaxies exhibit high ionization emission lines in their optical and UV spectra and are typically found in spiral, lenticular or irregular host galaxies \citep{Peterson1997}. Low-Ionisation Nuclear Emission-line Region (LINER) galaxies show relatively lower ionization emission lines in their spectra \citep{Heckman1980, Ho1996}. They are typically distinguished by lower [OIII]$\lambda$5007/H$\beta$ values for similar [NII]$\lambda$6593/H$\alpha$ values in the {Baldwin Phillips Terlevich (BPT)} diagram \citep{Baldwin1981}. Seyfert and LINERs are categorized as low-luminosity AGN, indicating lower levels of accretion activity compared to more luminous AGN \citep{Peterson1997, Nagar2002b, Maoz2007}.

In roughly one-fourth of the Seyfert galaxies, the toroidal dusty region surrounding the black-hole-accretion-disk system obscures the faster-moving dense clouds called the broad line region (BLR; $n_e>10^9$~cm$^{-3}$, $v<10^4$~km~s$^{-1}$) from certain lines of sights giving rise to type 2 Seyferts \citep{Peterson1997,Ho1999}. When both the BLR and the slower-moving less-dense clouds, which are further away from the black hole - accretion disk system, also known as the narrow line region (NLR; $n_e\approx 10^3-10^6$~cm$^{-3}$, $v\sim100$~km~s$^{-1}$), remain visible from our line of sight, they are identified as type 1 Seyferts \citep{Antonucci1993,Urry1995,Peterson1997}. Type 2 Seyfert galaxies sometimes show broad lines in polarized light revealing the dust obscured BLR \citep[e.g.,][]{Tran1992,Nagao2004}, consistent with the picture of Unified Model \citep{Antonucci1993, UrryPadovani1995}.

\begin{table*}
\tiny{
\centering
\caption{Properties of the CfA+12 micron Seyfert \& LINER Sample}
\tabcolsep=0.11cm
\begin{tabular}{cclcclcccc}
\hline \hline
Source&Alternate&Type& R.A. & Dec. & Redshift&$\mathrm{log(M_{BH})}$&$\mathrm{log(L_{bol})}$&$\mathrm{log\big(\lambda_{Edd}\big)}$&log${R}$\\
name&name&&J2000&J2000& &$(M/M_\sun)$ & (erg~s$^{-1}$) & & \\
&&&hh mm ss.ss&dd mm ss.ss&&&&& \\\hline
NGC 262&Mrk 348&Seyfert 2, KSR& 00 48 47.14& +31 57 25.08&0.01503&7.21$^{[1]}$&44.27$^{[1]}$&-1.04&$-0.92^{[8]}$\\
Mrk 993&UGC 00987&Seyfert 2, point&01 25 31.46&+32 08 11.43&0.01553&\nodata&\nodata&\nodata&\nodata \\
IRAS 01475-0740&\nodata&Seyfert 2, KSR&01 50 02.69&-07 25 48.48&0.01767&$6.09^{[8]}$&$44.33$&$0.14^{[8]}$&$-0.79^{[8]}$ \\
NGC 1068&M 77&Seyfert 2, KSR&02 42 40.71&-00 00 47.81&0.00379& 7.23$^{[1]}$&44.98$^{[1]}$&-0.35&$-2.47^{[8]}$ \\
NGC 1194&UGC 02514&Seyfert 1, point&03 03 49.11&-01 06 13.48&0.01363&$7.32^{[8]}$&\nodata&$-1.02^{[8]}$&$-3.27^{[8]}$ \\
NGC 1241&\nodata&Seyfert 2, KSR&03 11 14.64&-08 55 19.7&0.01343&\nodata&\nodata&\nodata&\nodata \\
NGC 1320&Mrk 607&Seyfert 2, KSR&03 24 48.70&-03 02 32.20&0.00928&7.18$^{[1]}$&44.02$^{[1]}$&$-1.26$&$-3.58^{[8]}$ \\
IRAS 04385-0828&\nodata&Seyfert 2, point&04 40 54.96&-08 22 22.23&0.0151&$8.77^{[8]}$&\nodata&$-2.44^{[8]}$&$-2.27^{[8]}$ \\
Mrk 9&\nodata&Seyfert 1.5, point&07 36 56.97&+58 46 13.43&0.03914&$7.46^{[8]}$&\nodata&$-0.62^{[8]}$&$-2.92^{[8]}$ \\
NGC 2639&UGC 04544&LINER 1.9, KSR&08 43 38.07&50 12 20.00&0.011128&7.9$^{[2]}$&42.61$^{[3]}$&-3.39&$-0.48^{[8]}$ \\
NGC 2992&IRAS 09432-1405&Seyfert 2, KSR&09 45 42.05&-14 19 34.98&0.007710&7.72$^{[1]}$&43.92$^{[1]}$&-1.90&$-2.48^{[8]}$ \\
NGC 3079&UGC 05387&LINER, KSR&10 01 57.80&+55 40 47.24&0.003689&$7.97^{[8]}$&$43.68$&$-2.39^{[8]}$&$-1.66^{[8]}$ \\
NGC 3516&UGC 06153&Seyfert 1, KSR&11 06 47.49&+72 34 06.88&0.008836&7.36$^{[1]}$&44.29$^{[1]}$&-1.17&$-3.16^{[8]}$ \\
NGC 4051&UGC 07030&Seyfert 1.5, KSR&12 03 09.61&+44 31 52.80&0.002336&6.13$^{[1]}$&43.56$^{[1]}$&-0.67&$-3.80^{[8]}$ \\
NGC 4151&UGC 07166&Seyfert 1, KSR&12 10 32.58&+39 24 20.63&0.00884&7.13$^{[1]}$&43.73$^{[1]}$&-1.50&$-2.61^{[8]}$ \\
NGC 4235&UGC 07310&Seyfert 1, KSR&12 17 09.88&+07 11 29.67&0.007548&7.4$^{[2]}$&42.54$^{[4]}$&-2.96&$-1.74^{[9]}$ \\
NGC 4388&UGC 07520&Seyfert 1, KSR&12 25 46.74&+12 39 43.51&0.008419&6.92$^{[5]}$&44.40$^{[5]}$&-0.62&$-3.50^{[8]}$ \\
NGC 4593&Mrk 1330&Seyfert 1, KSR&12 39 39.42&-05 20 39.34&0.008312&6.91$^{[1]}$&44.09$^{[1]}$&-0.92&$-2.81^{[9]}$ \\
NGC 4594& M 104/Sombrero&LINER 1.9, KSR&12 39 59.43&-11 37 22.99&0.003639&8.9$^{[2]}$&42.15$^{[4]}$&-4.85& \nodata\\
NGC 5283&Mrk 270&Seyfert 2, point&13 41 05.76&+67 40 20.32&0.01014&7.6$^{[1]}$&43.37$^{[1]}$&-2.33& \nodata\\
NGC 5273&UGC 08675&Seyfert 1.9, point&13 42 08.34&+35 39 15.17&0.00361&6.51$^{[1]}$&43.03$^{[1]}$&-1.58&\nodata \\
NGC 5347&UGC 08805&Seyfert 2, KSR&13 53 17.83&+33 29 26.98&0.007899&6.79$^{[1]}$&43.81$^{[1]}$&-1.08&\nodata \\
NGC 5506&Mrk 1376&Seyfert 1.9, KSR&14 13 14.89&-03 12 27.28&0.006084&7.94$^{[6]}$&44.02$^{[7]}$&-2.02&$-2.35^{[8]}$ \\
NGC 5548&Mrk 1509/9027&Seyfert 1.5, KSR&14 17 59.53&+25 08 12.44&0.01717&8.03$^{[1]}$&44.83$^{[1]}$&-2.30&$-2.91^{[8]}$ \\
NGC 5695&Mrk 686& Seyfert 2, KSR&14 37 22.12&+36 34 04.11&0.01415&\nodata&\nodata&\nodata&\nodata \\
NGC 5929&I Zw 112&Seyfert 2, KSR&15 26 06.16&+41 40 14.40&0.00851&7.25$^{[1]}$&43.04$^{[1]}$&-2.31&$-1.70^{[8]}$ \\   \hline
\end{tabular}
\\

{Columns 1 \& 2: Source names. Column 3: Spectral type and radio morphology. Column 4, 5 \& 6: RA, Dec and redshifts of the sources. Column 7: Logarithm of the black hole masses. Column 8: Logarithm of the bolometric luminosities in erg~s$^{-1}$. Column 9: Logarithm of the Eddington ratios. Column 10: Logarithm of the radio loudness parameter.
{References:}
[1] \citet{Woo2002}
[2] \citet{Dong2006}
[3] using [OIII] line fluxes from [2] and using the relations from \citet{Heckman2004}
[4] \citet{Kharb2016}
[5] \citet{Ursini2019}
[6] \citet{Nikolajuk2009}
[7] \citet{Matt2015}
[8] \citet{Melendez2010}
[9] Obtained by dividing total radio luminosity (excluding galactic emission) extrapolated to 8.4~GHz by the [OIV]$\lambda$25.89 $\mu$m line luminosity from \citet{Weaver2010}.
}}
\label{tab:tab1}
\end{table*}

After radio interferometric imaging became routine in the 1960s and 1970s, it was clear that Seyfert galaxies hosted smaller radio structures that typically stayed confined to their spiral or lenticular host galaxies \citep{deBruyn1976, Wilson1980}. {Later, a population of Seyferts was identified with relativistic jets \citep[e.g.,][]{Abdo2009, Foschini2012,Lahteenmaki2018}, adding complexity to the conventional view of Seyferts as predominantly low-power, RQ AGN. Recently, $\sim100$ kpc-scale jets have been observed in a few Seyferts \citep{Vietri2022, Chen2024}, and superluminal jet motion has been reported in KISSR~872 \citep{Kharb2024}.} {A sensitive radio survey of Seyferts and LINERs with the VLA D-array at 5~GHz has detected kpc-scale radio outflows (KSRs) not associated with the optical emission from the host galaxies in at least $\approx40$\% of the RQ sources in a complete sample \citep{Gallimore2006}.} There has been extensive debate about the origin and nature of the radio emission observed in Seyfert/LINER galaxies: {AGN-driven jets or winds \citep[e.g.,][]{Colbert1996b,Gallimore2006,Kharb2006}, starburst-driven winds \citep[e.g.,][]{Baum1993}, coronal winds, gas powered by shocks \citep{Heckman1987, Filippenko1992, Panessa2019, Fischer2023}, or a combination of these \citep[e.g.,][]{Singh2015} are the suggested contenders.} Very long baseline interferometry (VLBI) observations that probe milli-arcsecond (mas) scales and hundred mas-scale with telescopes like the Very Long Baseline Array (VLBA), the European VLBI Network (EVN) and the Multi-Element Radio Linked Interferometer Network (MERLIN), have revealed tenuous connections between AGN jet emission on parsec-scales to the KSRs, in a few Seyfert galaxies \citep[e.g., Mrk~6, NGC~6764;][respectively]{Kharb2010b,Kharb2014}, though have been opposed by other studies \citep[e.g., NGC 1068;][]{Fischer2023}.

{In this paper, we have investigated the extended polarized radio structures of Seyfert and LINER galaxies on kpc-scales. More organized magnetic field structures are expected in the AGN outflows compared to the synchrotron emission associated with starburst activity \citep{Panessa2019, Sebastian2020}. However, disentangling the contributions of AGN jets and winds from the galactic synchrotron emission is challenging \citep[e.g.,][]{Silpa2021}. \citet{Mehdipour2019} suggest that in RQ AGN, the jets typically exhibit poloidal magnetic fields aligned with them, while the winds have toroidal magnetic fields perpendicular to the flow direction. We have analyzed polarized emission from these galaxies to elucidate the origin, propagation, and interaction of the radio outflows with the surrounding media. {This work follows the 5~GHz VLA B-array study on Seyfert galaxies by \citet{Sebastian2020}. Higher frequency observations were obtained to partially discount the effects of Faraday rotation induced depolarization. The D-array was chosen for improving the sensitivity to diffuse lobe emission in these galaxies (as well as galactic emission in NGC~4388) \footnote{Two sample sources presented here, viz., NGC~1068 and NGC~1320 were part of the previous $1.4$~GHz VLA study of Seyfert galaxies (Project 21B-291).}.} Additionally, we have investigated the radio spectral indices to understand the particle emission mechanisms. Emission-line data from a few nearby sources are examined to look for possible correlations with the polarized radio emission. 

{The paper consists of the following sections. Section~\ref{sec:sample} discusses the details of the selected sample for this study. Section~\ref{sec:dataanalysis} describes the observational details and data reduction procedures. Section~\ref{sec:results} reports the results obtained from the data analysis with sections \ref{individualnotes}, \ref{sec:jetpower}, \ref{energetics}, and \ref{sec:correlations} presenting the details of individual sources, AGN jet powers, energetics, and correlations found for the sample sources, respectively. Section~\ref{discussions} focuses on the interpretations and impact of our results. Finally, section \ref{summary} summarises the paper.}

In this paper, the spectral index $\alpha$ is defined such that flux density at a frequency $\nu$ is S$_\nu\propto\nu^\alpha$. The adopted cosmology is H$_0=$67.8~km~s$^{-1}$~Mpc$^{-1}$, $\Omega_{mat}=$0.308, $\Omega_{vac}=$0.692. {By way of a definition, we have considered a radio outflow to be a `jet' if its length exceeds its width by a factor of at least 2.5. This suggestion follows from a similar one made for RL sources by \citet{Bridle1994}. If the jet is not delineated, then a spectral index gradient or ordered polarization vectors are used for inferring its presence.}

\section{The Sample}{\label{sec:sample}}
The parent sample for this study comprises 26 Seyfert and LINER galaxies belonging to the Center for Astrophysics \citep[CfA;][]{Huchra1992} and the extended 12 micron \citep{Rush1993} sample. These sources were known to exhibit either a 'KSR' or a compact component identified as 'point sources' in the VLA study of \citet{Gallimore2006}. The sources have a redshift cutoff of $cz\leq5149$~km~s$^{-1}$ ($z\lesssim 0.02$) to resolve the KSRs of about 7 kpc from the radio nucleus, and declination $>20\degr$ for observing with the VLA. {Out of the 40 sources in \citet{Gallimore2006}, we excluded 14 galaxies whose radio continuum matched the optical disk closely, resulting in a sample of 26 galaxies hosting either a type 1 or a type 2 AGN (Seyferts and LINERs), thus decreasing the probability of the radio emission being powered by stellar activity in the selected sample.} The sample with some basic properties is presented in Table~\ref{tab:tab1}. 

In this paper, we present polarization-sensitive images along with the spectral index images of a sub-sample of 12 galaxies, from the sample of 26 (Table~\ref{tab:tab2}), identified as having KSRs, at 10~GHz with VLA D-array configuration {with a bandwidth of 4~GHz and a resolution of $\sim$7$\arcsec$} (Project ID: 19B-198; PI: Biny Sebastian), as well as 1.4~GHz with the VLA BnA$\rightarrow$A array configurations {with a bandwidth of 1~GHz and a resolution of $\sim$2$\arcsec$} (Project ID: 21B-291; PI: Biny Sebastian). {An additional dataset at 5.5~GHz from the VLA D-array from project 21B-291 has been analysed for one particular source (NGC 4388).} Three sources have been classified as LINER galaxies while nine are Seyfert galaxies of both type 1 and 2. The redshift distribution of the sources with respect to the total 1.4~GHz radio luminosity from the NRAO VLA Sky Survey \citep[NVSS;][]{Condon1998} is shown in Figure~\ref{fig:luminosityNVSSvsredshift}. The different coloured markers are representative of different galaxies and this convention is followed in the subsequent plots. {Polarization data of the remaining sample (14 of 26) sources have been recently acquired and will be the subject of another paper.}

We have defined the radio loudness parameter (${R}$) as the ratio of 8.4~GHz radio flux density to [OIV]$\lambda$25.89 $\mu$m line flux density as defined by \citet{Melendez2010}. This provides a more accurate measure of radio-loudness being unaffected by dust extinction as observed in other methods involving optical observations. We note that for our sample sources, the adopted ${R}$ parameter did not alter the designation of the sources into the RL and RQ categories, as obtained from the original definition of \citet{Kellermann1989} as used by \citet{Rush1996}. In the cases where the 8.4~GHz flux density was unavailable, the ${R}$ was calculated at 8.4~GHz by extrapolating its total radio luminosity at 10~GHz with its corresponding spectral index value, divided by its [OIV]$\lambda$25.89 $\mu$m line flux densities obtained from \citet{Weaver2010}.

\begin{figure*}
    \centering
    \includegraphics[height=7cm]{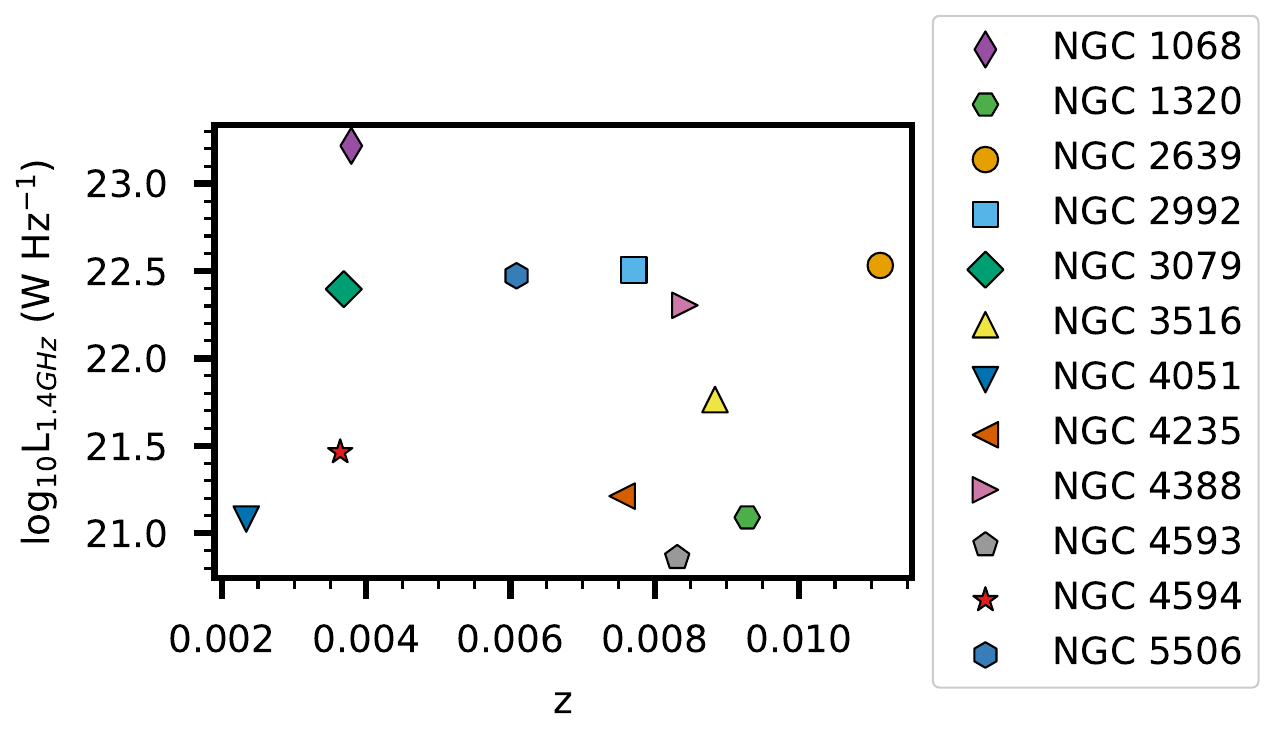}
    \caption{\small Redshift distribution of the sample presented in this paper with their corresponding VLA NVSS 1.4~GHz radio luminosities. Each Seyfert and LINER galaxy has been assigned a coloured marker that is preserved in the {subsequent} plots.}
    \label{fig:luminosityNVSSvsredshift}
\end{figure*}

\section{Radio Data Analysis}\label{sec:dataanalysis}
Polarization-sensitive data were acquired with VLA {\citep{Thompson1980}} D-array and BnA$\rightarrow$A array configurations at 10~GHz and 1.4~GHz, respectively. The observations were carried out in November 2019 with two sources observed in February - March 2022. Details of the observations and a list of calibrators are provided in Table~\ref{tab:tab2}. For the data analysis, the basic calibrations were carried out using the VLA recipes for {{Common Astronomical Software Application} \citep[CASA;][version 6.6]{CASA2022}.} This includes flagging, pre-determined calibrations like antenna position corrections and atmospheric opacity corrections, an iterative initial delay, bandpass, and gain calibrations. These are followed by additional flagging of the bad data, deriving the final delay and bandpass, obtaining the gain calibration solutions, and applying them to the data. 

Polarization calibration was carried out following the CASA-based polarization script\footnote{\url{https://github.com/astrouser-salm/radio-imaging/blob/main/VLA_polarization_pipeline.py}}. The polarization calibration primarily includes three steps: solving for the cross-hand delay, solving for the leakage terms, and solving for the polarization angle. For most cases, the polarized calibrator {i.e., 3C 286 or 3C 138}, observed over a large parallactic angle range ($\sim$60$^\circ$), has been used to derive the leakages. In some cases, the unpolarized calibrator {i.e., OQ 208 or 3C 84} has been used if the leakage solutions showed less scatter using the unpolarized ones. The leakages obtained for the VLA antennas were typically 5\% or less. For the delay and polarization angle calibration, polarized calibrator {(3C 138 or 3C 286)} with known polarization angles has been used.

After applying these calibration solutions to the data, {it was averaged over every $\sim$8 channels to reduce the data volume.} The target sources were then imaged using \textsc{multi-term multi-frequency synthesis} \citep[mtmfs][]{Rau2011} deconvolver, and \textsc{`briggs'} weighting with different robust parameters according to the requirement of weighting the longer or shorter baselines' data. \textsc{mtmfs} was run with terms = 2 to find out the in-band $\alpha$ at each pixel apart from the flux densities. A few rounds of phase-only and amplitude + phase self-calibration were carried out on the Stokes I images till convergence. The calibration results were applied to the data, and then the Stokes I, Q, U, and V images were obtained from the self-calibrated data. {The in-band spectral index image is obtained by blanking the spectral index image with its error image using an appropriate cut-off, which typically turns out to be $>30$\%.} The post-imaging analyses were carried out using {{Astronomical Image Processing System} \citep[\textsc{aips};][]{vanMoorsel1996}, \textsc{casa}} 
and \textsc{python}. The r.m.s. noise levels obtained for each image have been tabulated in Table \ref{tab:tab2}.

The linear polarization intensity (I$_p$), fractional polarization (f$_p$), polarization angle ($\chi$) values were obtained as $\sqrt{Q^2+U^2}$, $\sqrt{Q^2+U^2}/I$, and 0.5~tan$^{-1}$(U/Q). The I$_p$ images were generated in \textsc{aips} using task \textsc{comb} opcode \textsc{polc}, which accounts for the Ricean bias corrections, with a 3$\sigma$ cut-off. {The $\chi$ values} were obtained with opcode \textsc{pola} and a 10$^\circ$ angle error cut-off. The f$_p$ images were made by dividing the I$_p$ values with the I values, blanking with the 10\% error cut-off. The final electric vector position angle (EVPA) images were obtained by plotting the vectors with lengths proportional to f$_p$ and angle $\chi$. For optically thin {synchrotron emission}, the inferred magnetic field (B-field) is perpendicular to the observed EVPAs \citep{Pacholczyk1970, Ioannis2015}. Conversely, in optically thick regions (with optical depth $\tau \gtrsim 6$ and spectral index $\alpha \gtrsim 0.5$), inferred B-field vectors are parallel to the EVPAs \citep{Cawthorne2013, Wardle2018}. Therefore, for radio cores with an inverted spectrum with $\alpha \geq 0.5$, we interpret the B-fields to be aligned with the EVPAs. In all other situations, the B-fields are perpendicular to the EVPAs.

For estimating the core flux densities, we have obtained the peak intensity values (in mJy~beam$^{-1}$) by fitting each core with a Gaussian component. We considered the peak intensity values to be representative of the flux density for the compact core components. For the extended lobes/jets, the flux densities are obtained in mJy for the entire region. Mean values were obtained using CASA's polygon selector tool and running statistical analysis on the selected region. The errors in flux densities for the extended regions are derived by $\sqrt{\left(\sigma\sqrt{N_b}\right)^2+(\sigma_p S)^2}$ \citep{Kale2019}, where $S$ is the flux density, $\sigma_p$ is the percentage error in flux density (a conservative 5\% error is considered), $\sigma$ is the r.m.s. noise of the image, $N_b$ is the number of beams in the region. Estimates of the core spectral index $\alpha$, polarization fraction (f$_p$), and polarization angle ($\chi$) were obtained as average values in a box roughly equivalent to the FWHM of the synthesized beam centred on the core.

With the VLA D-array at 10~GHz, and BnA$\rightarrow$A array at 1.4~GHz, we observed 10 sources at a resolution of $\sim$7$\arcsec$ and 2 sources (NGC 1068, NGC 1320) at a resolution of $\sim$2$\arcsec$, respectively. {Considering the redshift range of this sample ($z=0.002-0.01$), the smallest spatial scales probed are between 300 pc and 1.5 kpc with D-array at 10~GHz and 85 pc to 500 pc with BnA array at 1.4~GHz, respectively.} {We note that while the `core' sizes are smaller in the 2$\arcsec$-scale versus the 7$\arcsec$-scale images, convolving the 2$\arcsec$-scale images with 7$\arcsec$ beams using \textsc{imsmooth} in \textsc{casa} resulted in similar core flux densities as obtained previously with the peak flux estimates at 2$\arcsec$ for the cores of the two sources. The peak flux is off by a factor of about 1.3 due to the inclusion of diffuse jet or lobe emission when convolved with the larger beam. We have therefore retained the original resolutions of the images for defining the `cores'.}

\section{Results}\label{sec:results}
The VLA $\sim$7$\arcsec$ images reveal extended features or lobe emission in all sample galaxies except NGC 2639 and NGC 4593; {extended lobe emission in these sources, if any, are on scales smaller than those probed in our observations.} In four galaxies, viz., NGC 4388, NGC 3079, NGC 4594, and NGC 1068, radio emission from the host galaxy is also detected (see Figure \ref{fig:opticalradiooverlay} in the APPENDIX). Linear polarization is detected in the cores, jets or lobes for all the 12 Seyfert and LINER galaxies. Below we discuss the results for individual sources.

\subsection{Notes on Individual Sources}\label{individualnotes}
\begin{table*}
\centering
\small{
\caption{Observational Details.}
\tabcolsep=0.1cm
\begin{tabular}{ccccccc}
\hline \hline
Source&Observation&Frequency&VLA Array& r.m.s. noise&Polarization Angle&Leakage\\
&Date&GHz&Configuration &($\mu$Jy/beam)&Calibrator& Calibrator \\ \hline
NGC 1068&26-Feb-2022&1.4&BnA$\rightarrow$A&250&3C 138&3C 138\\
NGC 1320&03-Mar-2022&1.4&BnA$\rightarrow$A&40&3C 138&3C 138\\ \hline
NGC 2639&11-Nov-2019&10&D&15&3C 138&3C 84\\
NGC 2992& 10-Nov-2019&10&D&12&3C 138&3C 84\\
NGC 3079&08-Nov-2019&10&D&30&3C 286& 3C 286 \& OQ 208\\
NGC 3516&09-Nov-2019&10&D&16&3C 286&3C 286 \& OQ 208\\
NGC 4051&11-Nov-2019&10&D&15&3C 286&3C 286 \& OQ 208\\
NGC 4235&08-Nov-2019&10&D&15&3C 286&3C 286 \& OQ 208\\
NGC 4388&20-Nov-2019&10&D&20&3C 286&3C 286 \& OQ 208\\\hline
NGC 4388&11-Nov-2019&5.5&D&17&3C 286&3C 286 \& OQ 208\\\hline
NGC 4593&12-Nov-2019&10&D&45&3C 286&3C 286 \& OQ 208\\
NGC 4594&10-Nov-2019&10&D&22&3C 286&3C 286 \& OQ 208\\
NGC 5506&08-Nov-2019&10&D&9&3C 286&3C 286 \& OQ 208\\\hline

\end{tabular}
}
\label{tab:tab2}
\end{table*}

\subsubsection{NGC 1068}
NGC 1068 is a well-known spiral galaxy identified early-on to host a Seyfert nucleus. It later classified as a type 2 Seyfert due to the presence of an obscured BLR \citep{Khachikian1974,Antonucci1985}.
The AGN in this source has been suggested to drive a biconical molecular outflow almost perpendicular to the nuclear disk causing a deficit of molecular gas at the central 130 pc, leaving out only a 400 parsec sized, ring-shaped circumnuclear disk \citep{Gallimore2016,Garciaburillo2024b}.
The high-resolution images of this source show a $\sim13\arcsec$ {($\sim$0.8 kpc)} linearly oriented jet-like structure from the core ending at a bow-shock-like edge \citep{Wilson1982,Ulvestad1987}. 

Our observations reveal both the galactic disk emission and the Seyfert outflow at the center. The galactic emission is most likely due to synchrotron radiation from supernova remnants associated with starburst activity \citep{Wilson1982}. The Seyfert emission extends approximately {$26\arcsec$ ($\sim$1.7~kpc)}. We observe an offset in the polarization angle at the northern bowshock or hotspot in the Seyfert outflow at 1.4 GHz compared to that found by \citet{Wilson1982} at 15 GHz (see Figure \ref{fig:NGC1068fpol}, top panel). This offset can be attributed to the RM, which is more significant at lower frequencies. 

The in-band spectral index image centered around 1~GHz (see Figure \ref{fig:NGC1068fpol}, bottom panel) presents a jet-core-jet structure with a steep spectrum core ($\alpha \approx -0.67$) and an even steeper spectrum jet/outflow ($\alpha \approx -0.88$). A new feature towards the east of the core is highlighted in polarized emission, exhibiting a high polarization fraction of $f_p = 3.2 \pm 0.4$\%. Interestingly, the spectral index image reveals that this highly polarized region has a flat spectrum ($\alpha \approx 0$) {with a 24$\sigma$ detection in total intensity and an error of 0.001 in the spectral index.} This region may result from the sweeping away of relic radio plasma by ram-pressure bending along the galaxy's spiral motion {\citep[see][]{Kapferer2009}. Ram-pressure on relic plasma may have re-energized the charged particles \citep[see ][]{Lebedev2014} flattening the spectrum \citep[e.g.,][for galaxy cluster environments]{Kang2011}}.

\begin{figure*}
\centering
\includegraphics[width=8.6cm]{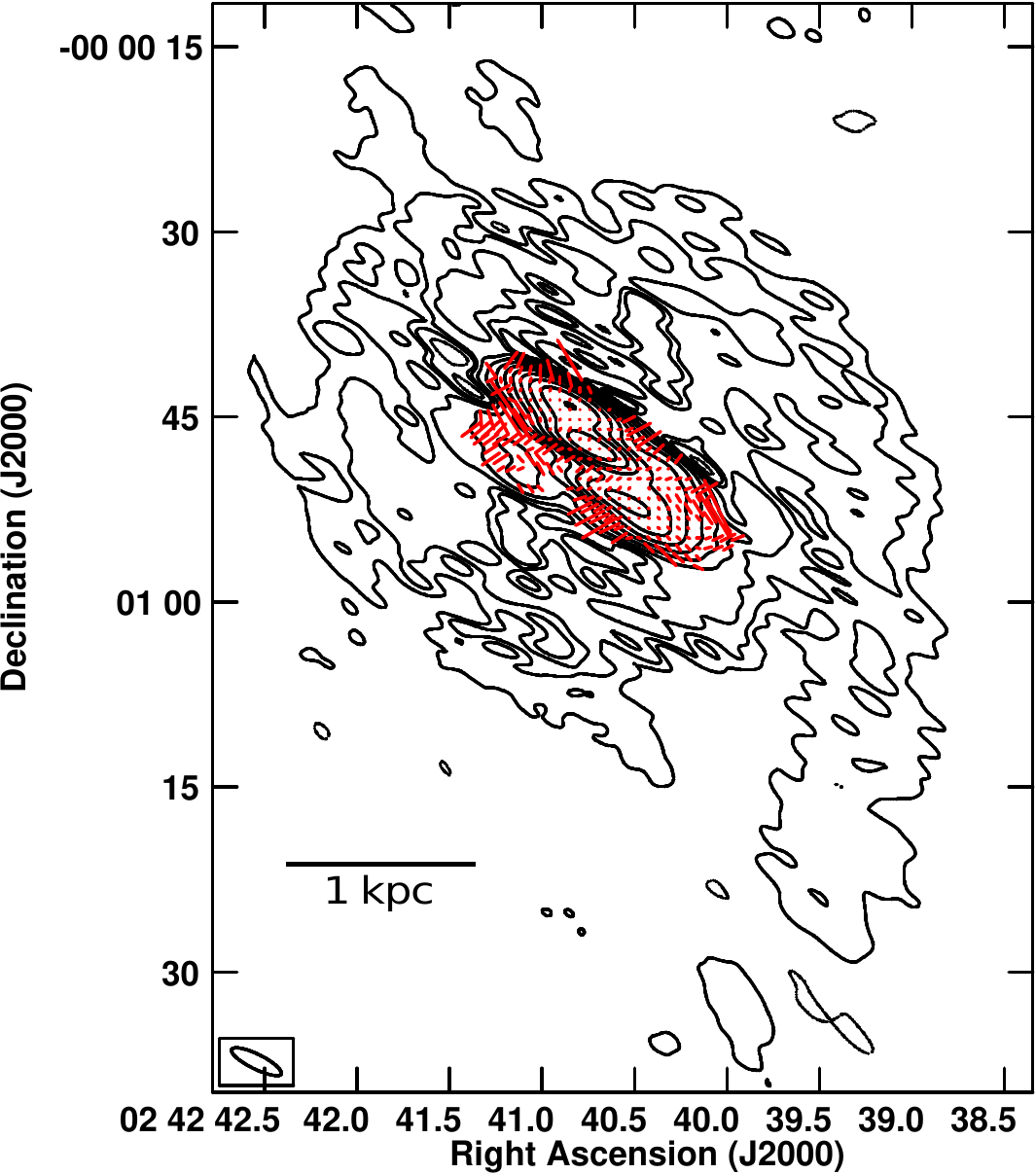}
\includegraphics[width=9.2cm]{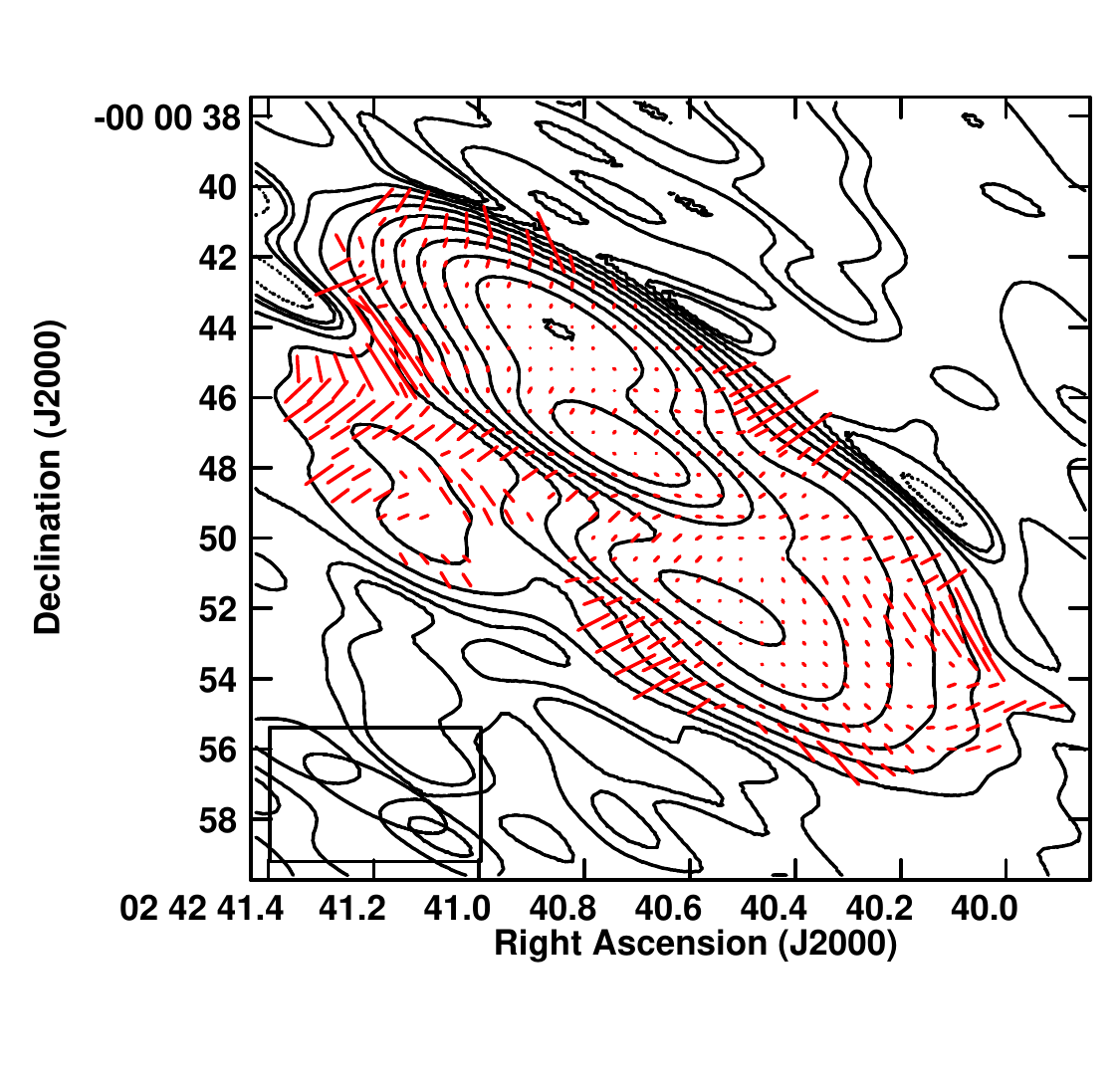}
\includegraphics[width=8.6cm]{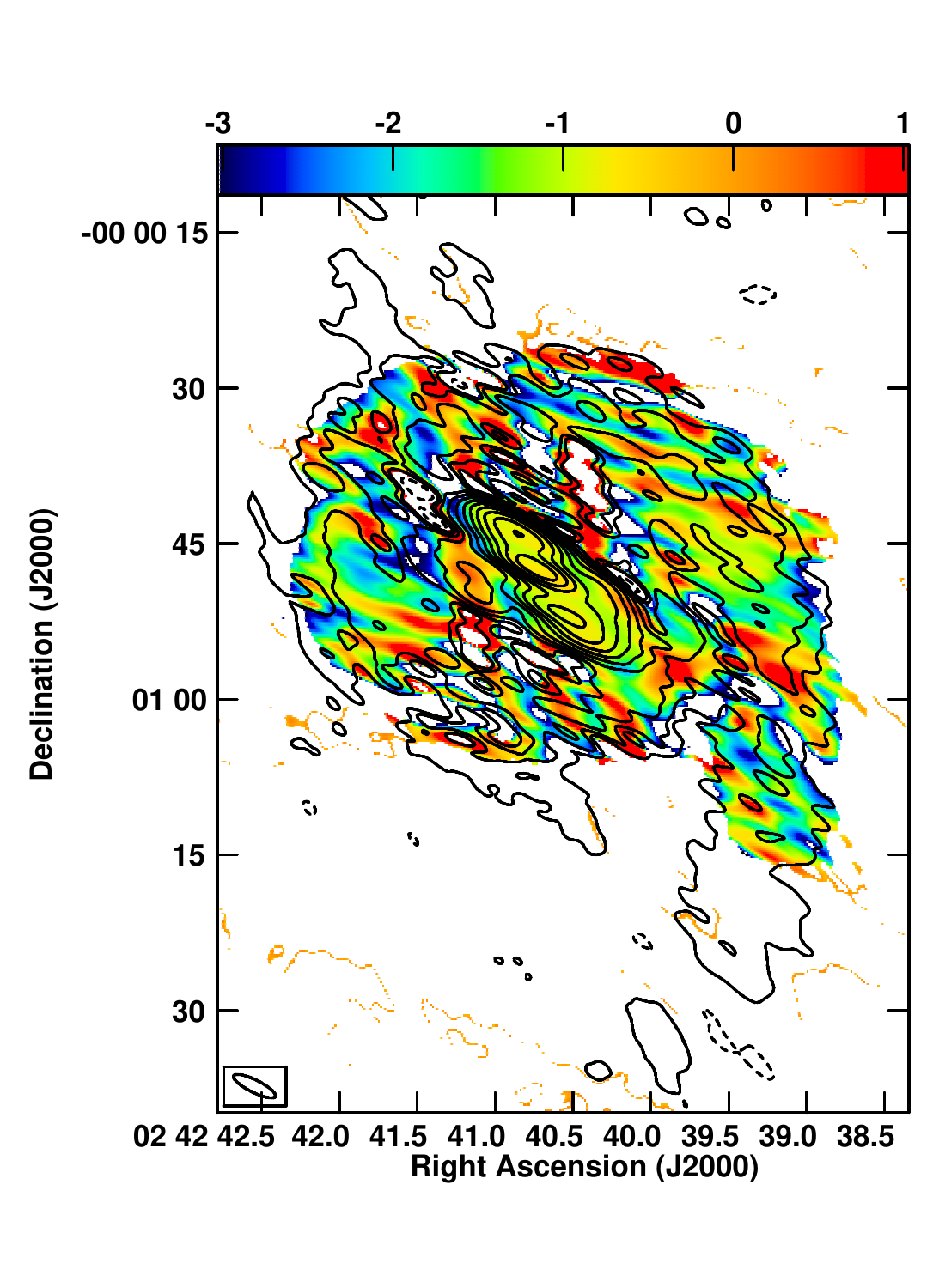}
\includegraphics[width=9.2cm]{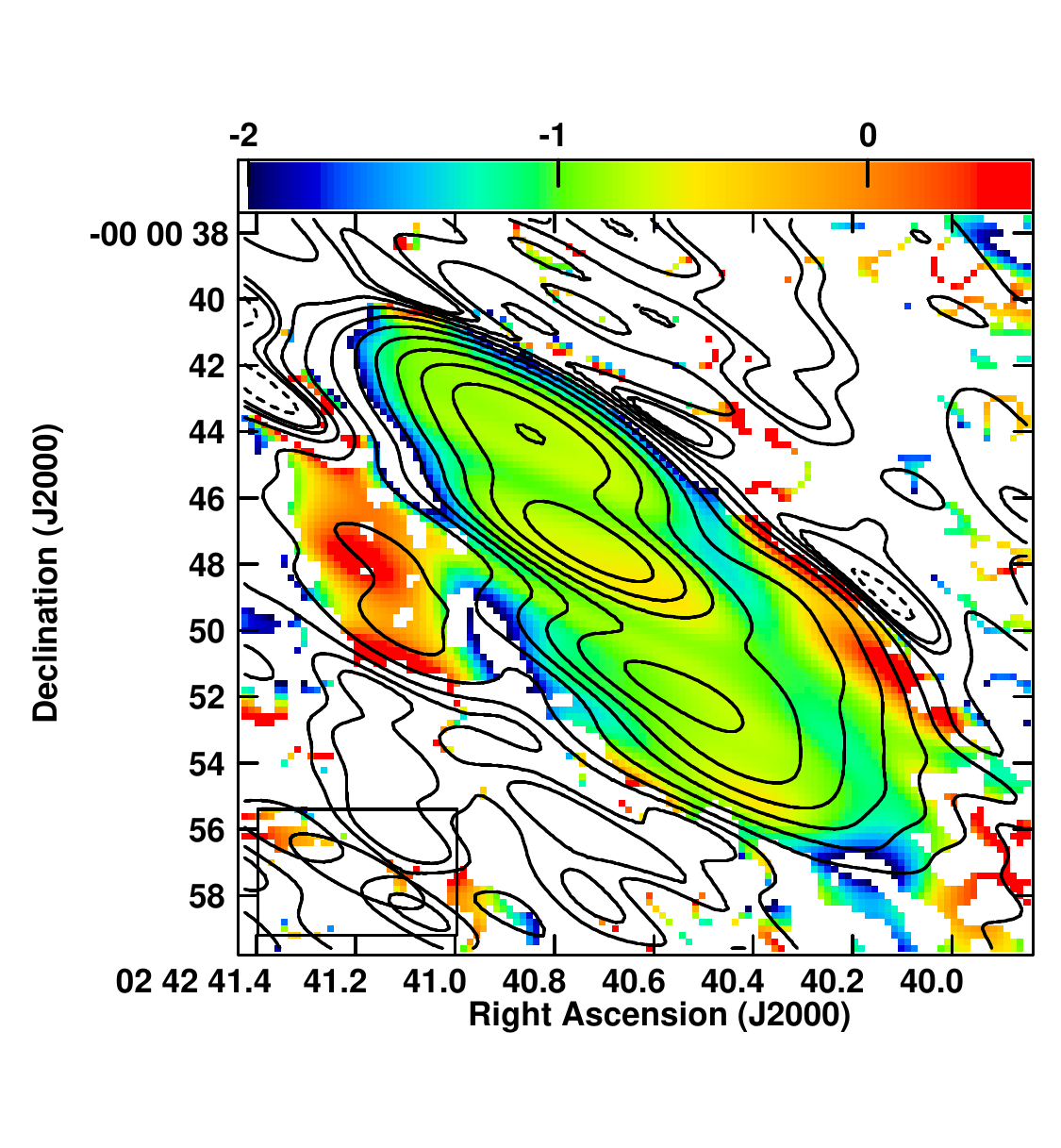}
\caption{\small (Top left) NGC 1068 image at 1.52~GHz with the VLA BnA$\rightarrow$A array at a resolution of $\sim2\arcsec$. The contour levels {in all the panels} are at 3$\sigma \times(-1, 1, 2, 4, 8, 16, 32, 64, 128, 256, 512, 1024)$ with $\sigma = 250~\mu$Jy beam$^{-1}$. {Similarly, the} synthesized beam size is 4.57$\arcsec \times 1.23\arcsec$ at a position angle (PA) of $63.8\degr$. The tick lengths are proportional to fractional polarization with 5 arcsec = 15.6\%. {\it{Radio emission is detected from the host galaxy along with the KSR.}} (Top right) Zoomed-in image showing the KSR. 
The tick lengths are proportional to fractional polarization with 4 arcsec = 25\%. (Bottom left) In-band spectral index image of NGC 1068. The spectral index values in colour range from $-3.0$ to $+1.0$. (Bottom right) Zoomed-in image showing the KSR. The spectral index values in colour range from $-2.0$ to $+0.5$.}
\label{fig:NGC1068fpol}
\end{figure*}


\subsubsection{NGC 1320}
NGC 1320 is an edge-on spiral galaxy with a Seyfert type 2 nucleus having narrow emission lines \citep{DeRobertis1986}. Faint radio emission ($\sim3.5$~mJy at 5~GHz) of extent 0.9~kpc and 2.6~kpc towards the south from the core was observed by \citet{Colbert1996b} and \citet{Gallimore2006}, respectively. We detect a clearer core-lobe structure with emission clearly extended to the north-south as well as the north-east. A total flux density of 5.6 mJy at 1.5~GHz and a total radio extent of $\sim$9.7$\arcsec$ ($\sim$2 kpc) is detected in this $\sim$2$\arcsec$ {($\sim$370 pc)} resolution image. High fractional polarization (47$\pm$18\%) is observed at the end of the southern lobe (Figure~\ref{fig:NGC1320fpol&spix}, left panel). Fractional polarization of $20\pm6$\% is observed to the north of the core. The [OIII]$\lambda$5007 emission as observed by \citet{Schmitt2003} is concentrated at the core and has an extent of $\sim$660~pc in the north-west direction, which also coincides with the host galaxy major axis \citep[see][]{Ferruit2000}. 

The spectral index image (Figure \ref{fig:NGC1320fpol&spix}, right panel) suggests that the true core {(with a flatter spectral index)} is to the west of the peak emission and that the jet emerges to the south-east before bending towards the south. Overall, an S-shape is observed in the north-south jet and the inferred magnetic fields are aligned with the radio outflow both in the north and south. The origin of the extended emission to the north-east is unclear. 

\subsubsection{NGC 2639} \label{sec:NGC2639}
NGC 2639 is a spiral galaxy with an extension of $\sim$60 kpc possessing both inner and outer rings, hosting a Seyfert 1.9 or a LINER nucleus \citep{Veron2006}.
\citet{Xanthopoulos2010} identified an S-shaped bent morphology in the radio emission from this source at 5 GHz, suggesting jet precession along with strong jet-medium interaction impeded jet propagation. However, \citet{Sebastian2019b} later ruled out jet precession for this source and discovered episodic activity, with earlier episodes exhibiting greater extents than previously reported. Recently, the source has been observed to undergo multiple episodes of activity, exhibiting the highest frequency of such episodes among Seyferts, as evidenced by observations at various frequencies and resolutions \citep{rao2023}.

At a resolution of $\sim7\arcsec$ {($\sim$2~kpc)} at 10~GHz, the Seyfert outflow of extent of 32$\arcsec$($\sim$8~kpc) is not resolved into individual features. The structure resembles the Giant Metrewave Radio Telescope (GMRT) 325 MHz image of this source \citep{rao2023}, which is considered to originate from the first episode of activity. The inferred B-field, perpendicular to the EVPAs shown in Figure \ref{fig:NGC2639} (left panel), is oriented along the northeast direction in the central region of the source, suggesting a jet-like structure in that direction. This is confirmed by the GMRT 735 MHz image \citep[resolution $\sim$5$\arcsec$;][]{rao2023}, which reveals a jet along the same direction. 

The spectral index image (Figure \ref{fig:NGC2639}; right panel) shows a gradient perpendicular to the jet direction with flat {($\alpha\sim$0)} to inverted {($\alpha\sim$0.5)} spectral index values. This may be because the source hosts multiple radio lobes in different directions, sometimes perpendicular to each other, {with the newest episode being oriented along the northwest-southeast direction. The positive value of $\alpha$ may arise due to the combined effect of poor resolution and free-free absorption by the plasma ejected during different episodes}. We estimated the integrated spectral index to be $-0.48\pm0.04$ by comparing the flux densities at 325 MHz \citep{rao2023} and 10 GHz. Additionally, we find that the kinetic jet power of the NGC 2639 is comparable to that of other Seyferts/LINERs and does not stand out as particularly special even though it produces multiple jet episodes.

\subsubsection{NGC 2992} \label{sec:NGC2992}
NGC 2992 is changing-look (CL) Seyfert galaxy with extreme variability \citep{Gilli2000, Marinucci2018, Almeida2009}, hosting a nuclear ultra-fast outflow (UFO) with a velocity up to $\sim$0.4c \citep{Marinucci2018,Luminari2023}. An 8-shaped AGN outflow almost perpendicular to the galactic disc is detected at 5 GHz \citep{Ulvestad1984b}. This galaxy is also part of a tidally interactive system along with three other galaxies, namely, NGC 2993, Arp 245, and dwarf A245 \citep{Duc2000}. 


\begin{figure*}
\centering
\includegraphics[width=7cm]{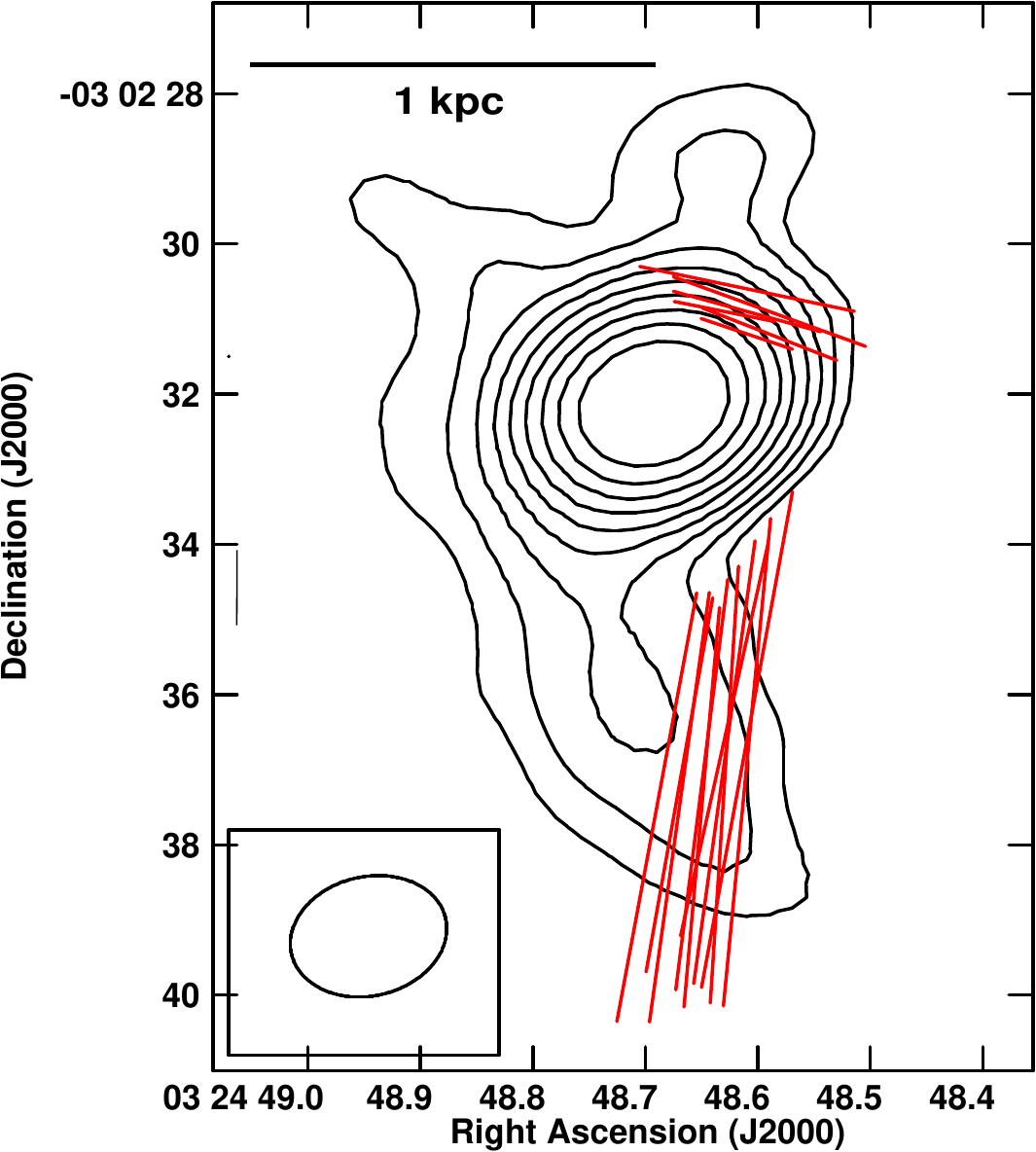}
\includegraphics[width=6.5cm]{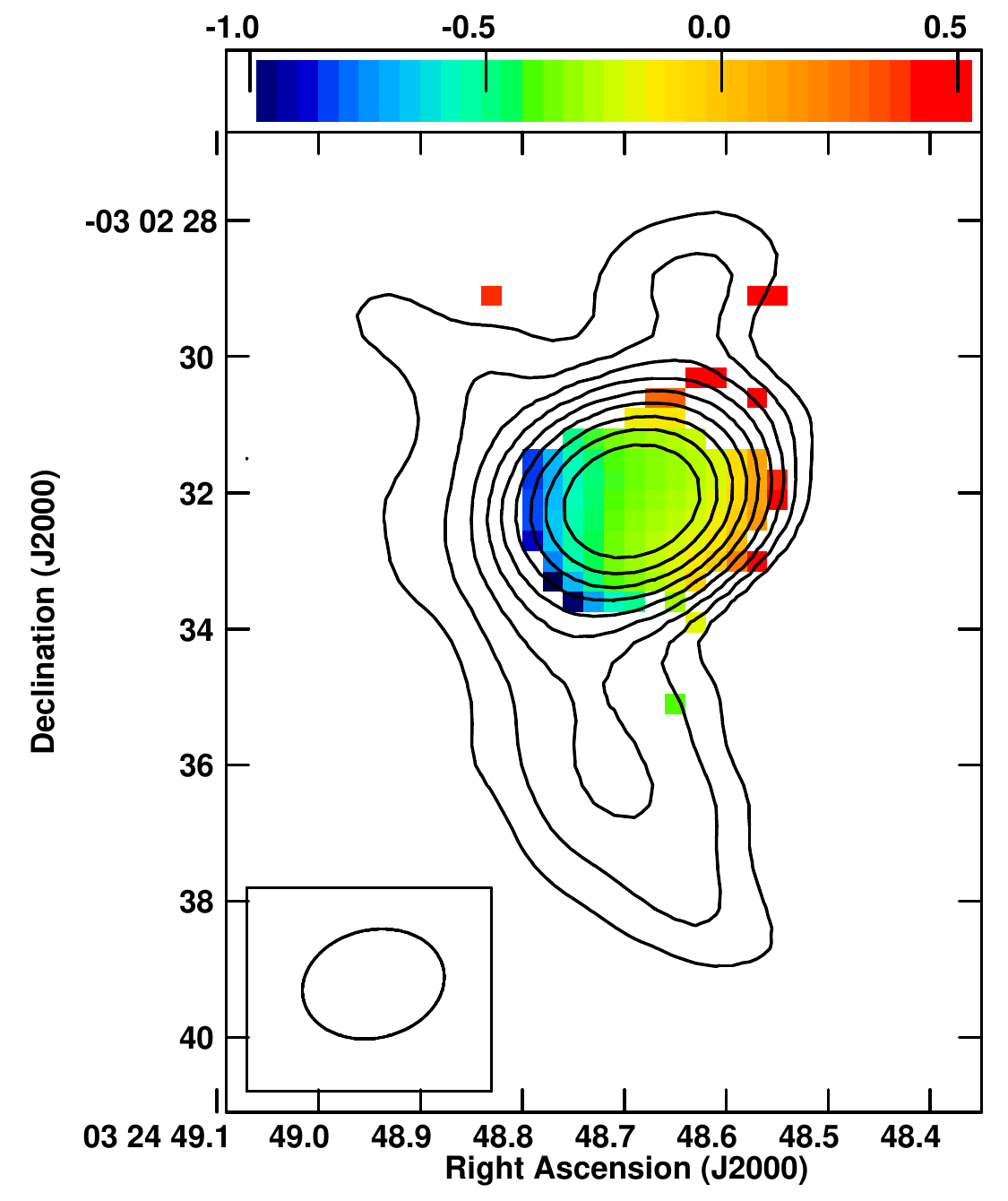}    
\caption{\small (Left) NGC 1320 image at 1.46~GHz with the VLA BnA$\rightarrow$A array. For all the panels the contour levels are at 3$\sigma \times(-1, 1, 1.414, 2, 2.828, 4, 5.657, 8, 11.31, 16)$ with $\sigma = 40~\mu$Jy beam$^{-1}$. Similarly, the synthesized beam size is 2.11$\arcsec$ $\times$ 1.59$\arcsec$ at a PA of $-77.0\degr$. The tick lengths are proportional to fractional polarization; $3\arcsec = 1$\%. Fractional polarization of errors $>35\%$ have been blanked. (Right) The VLA in-band spectral index image of NGC 1320. The spectral index values are in colour ranging from -1.0 to 0.5.}
\label{fig:NGC1320fpol&spix}
\end{figure*}


We observe a core-halo-like radio structure at 10 GHz with the VLA D-array. The H$\alpha$ image from the Continuum Halos in Nearby Galaxies - an EVLA Survey \citep[CHANG-ES;][]{Vargas2019} shows two components that may be due to the two rotating arms of the spiral host galaxy. \citet{Wehrle1988} suggest that the prominent division in the H$\alpha$ map into two bulges might be due to obscuration by the dust lane along the disk of the galaxy. Alternately, the fainter component which coincides with the higher fractional polarization region ($10.9\pm0.3\%$) with a steeper spectral index (see Figure \ref{fig:2992fpolspixHalpha}, top panels {\& bottom left panel}) might represent the relic lobe discovered in polarized light by \citet{Irwin2017}. The H$\alpha$ emission is brightest in the nuclear region as seen in the VLA 0.5$\arcsec$ resolution image in Figure \ref{fig:2992fpolspixHalpha} (bottom {right} panel) and follows the spiral arms of the galaxy \citep{Durret1987,Wehrle1988}. 

\begin{figure*}
    \centering
\includegraphics[width=8.6cm]{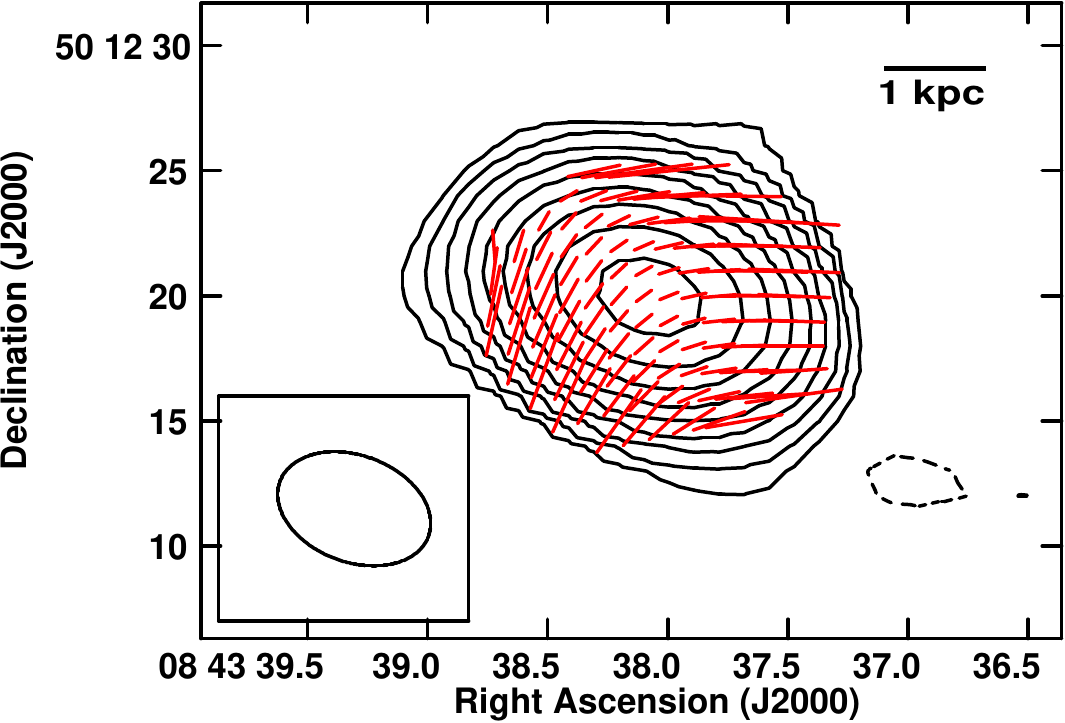}
\includegraphics[width=7.8cm]{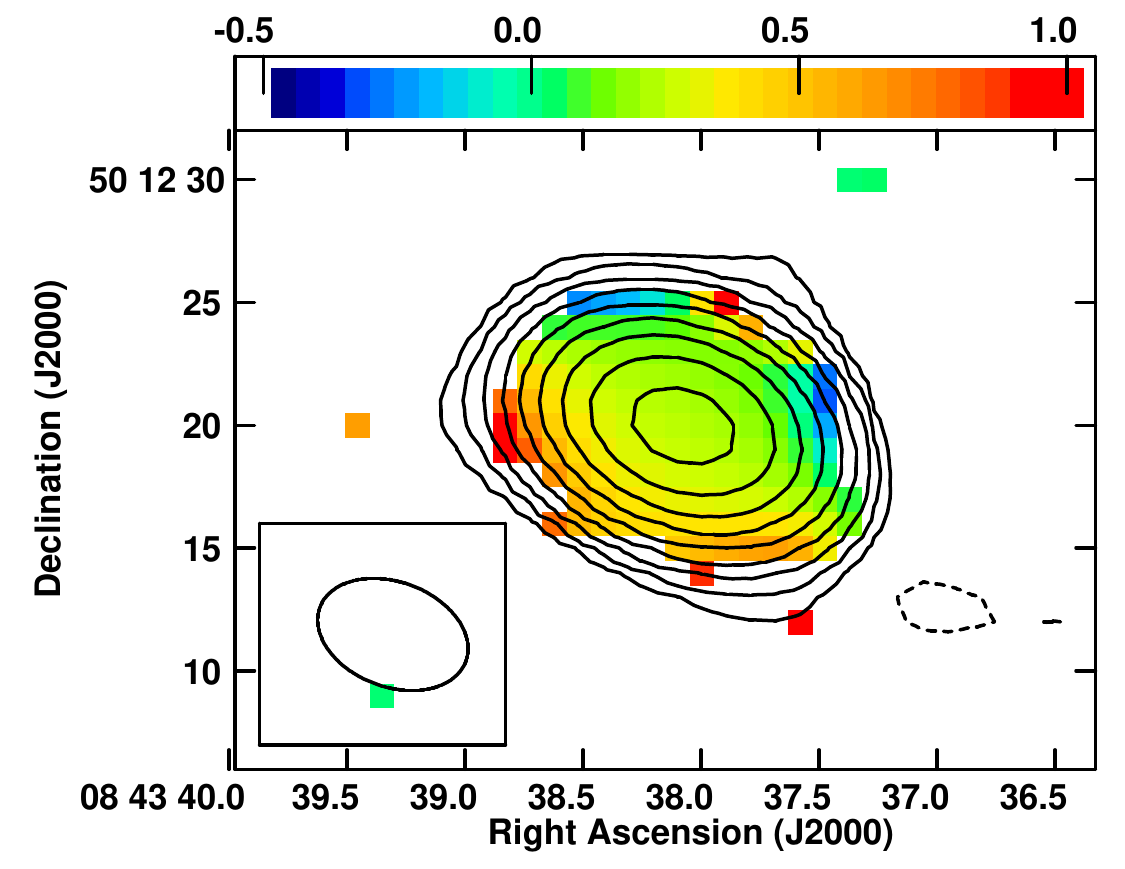}
\caption{\small (Left) NGC 2639 image at 10~GHz with the VLA D-array. For all the panels the contour levels are at 3$\sigma \times(-0.4, 1, 2, 4, 8, 16, 32, 64, 128, 256)$ with $\sigma = 15~\mu$Jy beam$^{-1}$. Similarly, the synthesized beam size is 6.33$\arcsec$ $\times$ 4.26$\arcsec$ at a PA of $69.7\degr$.The tick lengths are proportional to fractional polarization; $2\arcsec = 2$\%. (Right) In-band spectral index image of NGC 2639 centred at 10 GHz. The spectral index values are in colour ranging from -0.5 to 1.0. }
\label{fig:NGC2639}
\end{figure*}
The 8-shaped lobed structure observed in the high-resolution VLA image (see Figure \ref{fig:2992fpolspixHalpha}, bottom right panel) may result from bi-directional spherical shells of shocked gas or from magnetic loops/arches \citep{Wehrle1988}. \citet{Xu2024} found an enhanced young stellar population at the boundary of the nuclear bubble, as well as an anti-correlation between AGN fraction and young stellar population up to 1 kpc from the nucleus, indicating low SFR in AGN outflow regions. These findings suggest that both positive and negative AGN feedback coexist in NGC 2992. 

Our observations reveal more diffuse radio emission of extent 39$\arcsec$ ($\sim$7 kpc) compared to the $\sim1.5$ kpc long 8-shaped radio lobes. The polarization vectors show that the inferred B-fields are perpendicular to the direction of the 8-shaped Seyfert outflow. Given the outflow direction and a steep spectrum core (mean $\alpha\approx -0.8\pm0.1$), the inferred B-fields are toroidal at the jet base. Toroidal B-fields are consistent with AGN winds along with the jets/lobes \citep{Mehdipour2019}. AGN-driven winds also have been suggested in this source by \citet{Zanchettin2023}.

\begin{figure*}
\centering
\includegraphics[width=8.2cm]{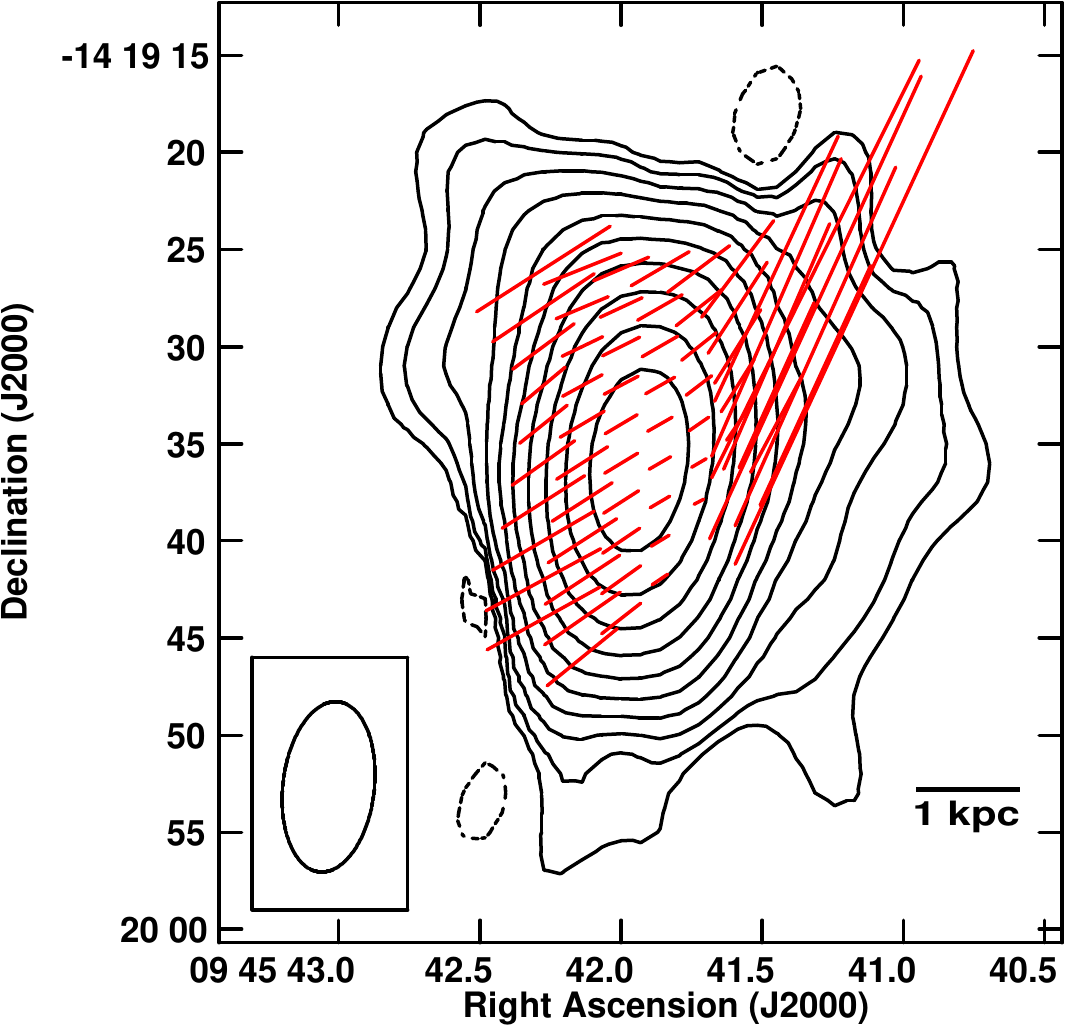}
\includegraphics[width=7.8cm]{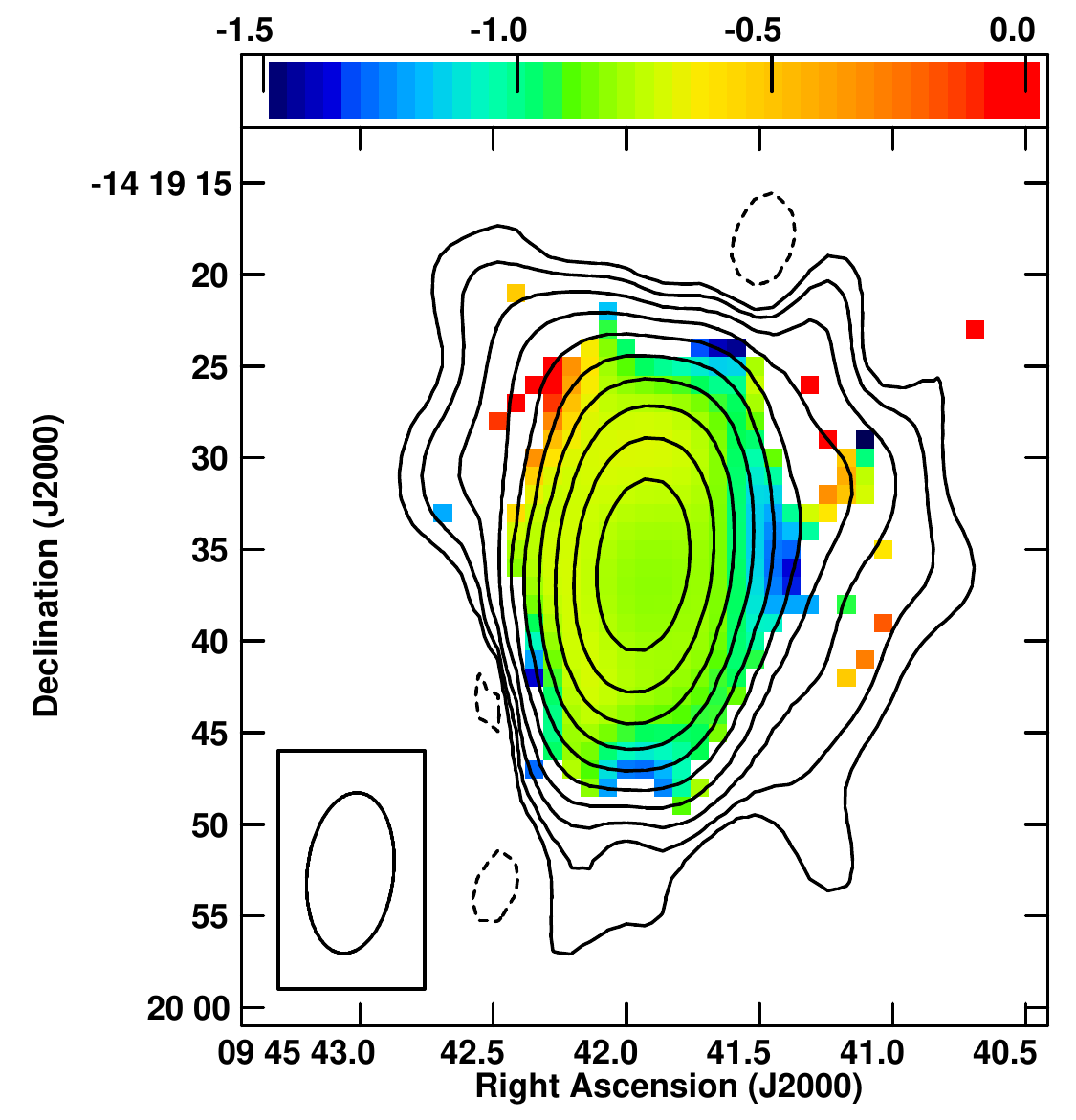}
\includegraphics[width=7.8cm]{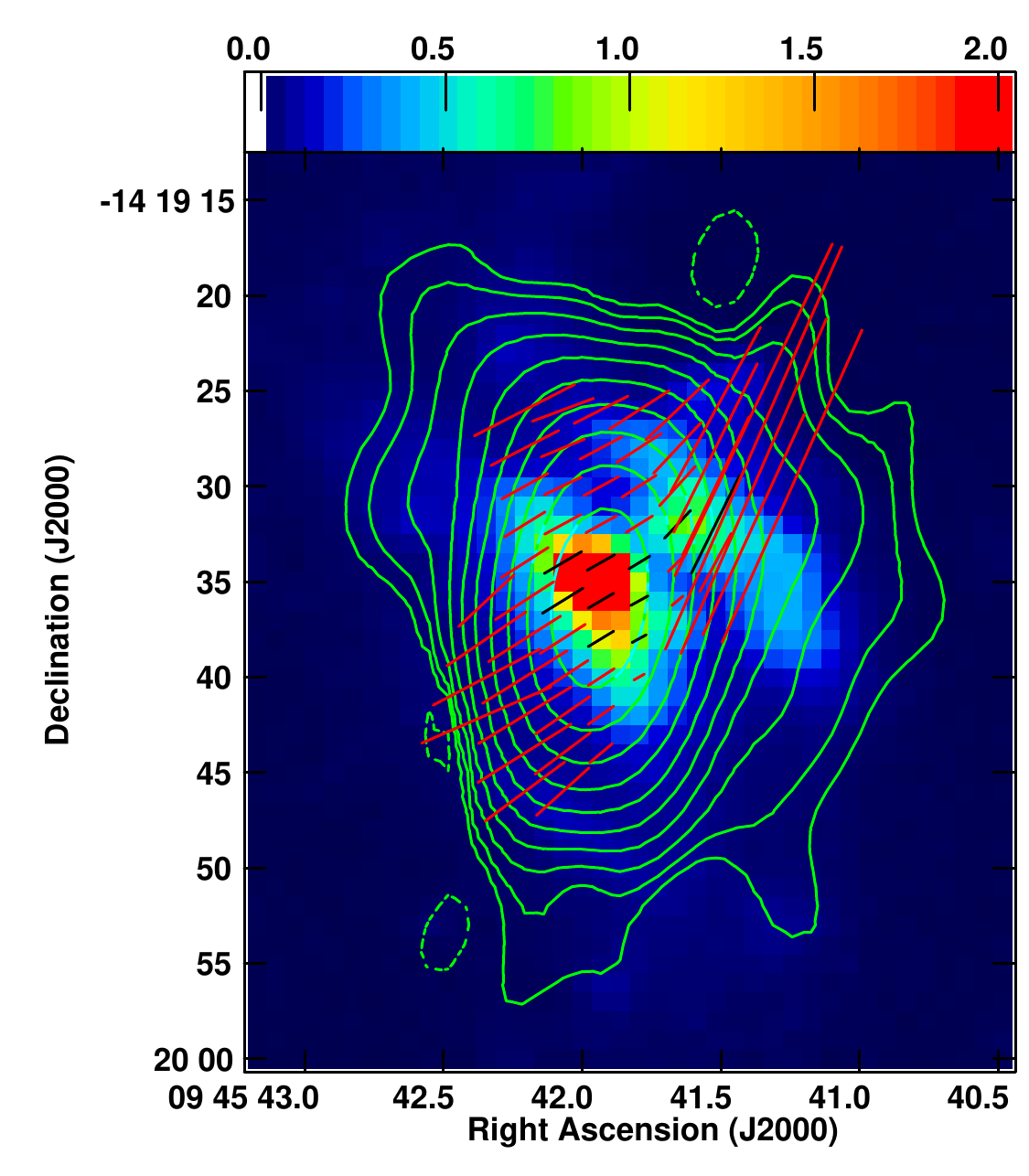}
\includegraphics[width=9cm]{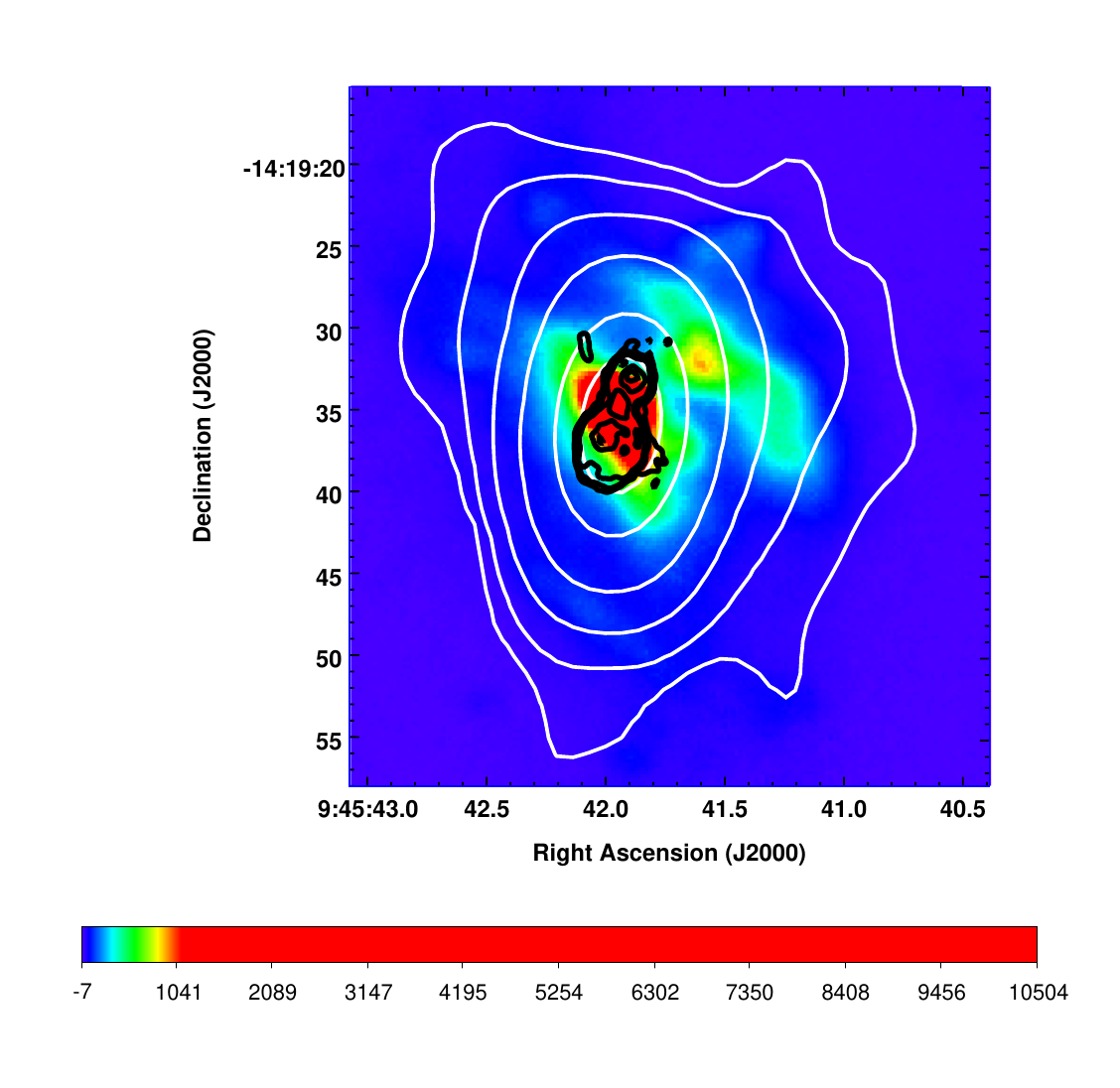}
\caption{\small (Top left) Seyfert galaxy NGC 2992 image at 10~GHz with the VLA D array. For all the panels the contour levels are at 3$\sigma \times(-1, 1, 2, 4, 8, 16, 32, 64, 128, 256, 512)$ with $\sigma= 12~\mu$Jy beam$^{-1}$. Similarly, the synthesized beam size is $8.81\arcsec\times4.71\arcsec$ at a PA of $-6.5\degr$. The tick lengths are proportional to fractional polarization with $5\arcsec = 2.5$\%. (Top right) In-band spectral index image of NGC 2992 centred at 10 GHz. The spectral index values are in colour ranging from $-1.5$ to 0. (Bottom left) Radio contours and polarization vectors from our work over-plotted on the H$\alpha$ image from the CHANG-ES survey for NGC 2992. The H$\alpha$ gas is in colour ranging from (0.0 to 2000.0)$\times$ 2.217$\times10^{-19}$ erg s$^{-1}$cm$^{-2}$. The tick lengths are the same as the top left panel. (Bottom right) Radio contours over-plotted on the H$\alpha$ image from CHANG-ES. White contours are from our work. Black contours show an archival VLA 5 GHz image at a resolution of $\sim$0.5$\arcsec$ with contour levels at 5$\sigma \times(-1, 1, 2, 4, 8)$ with $\sigma= 2~\mu$Jy beam$^{-1}$. }
\label{fig:2992fpolspixHalpha}
\end{figure*}

\subsubsection{NGC 3079} \label{sec:NGC3079}
NGC 3079 is an edge-on barred spiral LINER + starburst composite galaxy \citep{Heckman1980,Irwin2003}. The parsec-scale jet is found to be misaligned to the kpc-scale radio bubble by about $65\degr$ \citep{Irwin1988}. Given the negligible projection effects \citep[e.g.,][]{Readhead1983} for such an edge-on galaxy, such large misalignments are rarely observed except for new findings like NGC~2639 (for details, see Section \ref{sec:NGC2639}). Although, the images at all scales put together construct an image of steady jet bending till the outflow manages to channel out along the minor axis \citep{Irwin1988}. This idea is also supported by simulations, which find that highly inclined jets interact with the disc causing a wide outflow by drawing materials (gas clouds and filaments) from the disc and thereby affecting the ISM on kpc-scales {then channelising its way out along the minor axis of the host galaxy}\citep{Mukherjee2018}. 

Alternately, based on a VLBI study, \citet{Middelberg2007} suggested that the jet of NGC 3079 undergoes a strong interaction with the surrounding clumpy gas within a few parsec radius. The simulations by \citet{Saxton2005} show that this interaction causes the jet to form a large expanding radio bubble reaching kpc-scales, which explains the {observed H$\alpha$ and X-ray superbubbles of size 13$\arcsec$ ($\sim$1 kpc) \citep{Veilleux1994, Cecil2002}}. \citet{Strickland2004} argued that the H$\alpha$ and X-ray bubbles are unrelated to the 1.5 kpc radio jet and lobe, attributing their partial coincidence to projection effects. {However, \citet{Li2024} has recently demonstrated that the energy injection required for maintaining pressure balance in the bubble cannot be explained solely by starburst activity but rather AGN activity is required}. Unlike the H$\alpha$ gas that forms a superbubble in the nuclear region, the [OIII] gas is concentrated in compact regions along the galactic disk, peaking near the nucleus. These compact [OIII] emissions are thought to arise from photoionization by disk winds, with the brightest emission resulting from dust extinction near the nucleus \citep{Robitaille2007}.

From our work, radio emission is detected from the galactic disc extending about $152\arcsec$ {($\sim14$~kpc)} and the Seyfert outflow in a bipolar lobe, perpendicular to the galactic emission, extending about $53\arcsec$ {($\sim5$~kpc)}. The radio emission is coincident with the H$\alpha$ image \citep[CHANG-ES;][]{Vargas2019}, highlighting the synchrotron emission from the high star-forming regions (see Figure \ref{fig:NGC3079fpol&spix}, {bottom left panel}). The core (f$_p$ $\approx$ 2\%) and the lobe (f$_p\approx$16\%) stand out in polarization intensity compared to the galactic disk. The core is steep (mean $\alpha=-0.9\pm0.2$), the lobe and the galactic disk are steeper and have similar mean spectral indices ($\alpha=-1.5\pm0.2$) as in Figure \ref{fig:NGC3079fpol&spix} {(top right panel)}. The inferred B-fields in the galactic plane (see Figure \ref{fig:NGC3079fpol&spix}, {top left panel}, for the polarization vectors), are along the rotating disk. 

{The B-fields in the western lobe bend along the lobe's curvature, indicating either confinement by the surrounding medium \citep[e.g.,][]{Laing1980a,Ghosh2023} or the presence of initially sheared B-field tangential to the surface without any outward component \citep{Blandford1978, Laing1980b, Laing1981}.} In the Eastern lobe, the B-fields follow the bends initially but become perpendicular to the boundary at the edge of the radio bubble, differing from typical shocked regions \citep[see also][]{Cecil2001, Sebastian2019a}. The inferred B-field at the core is aligned with the {VLBI} jet direction. The atomic neutral hydrogen (HI) map \citep[CHANG-ES;][]{Zheng2022} reveals a lack of neutral hydrogen along the Seyfert outflow (Figure \ref{fig:NGC3079fpol&spix}; {bottom right panel}). This may suggest that AGN activity is depleting or ionizing the neutral gas. The presence of H$\alpha$ (Figure \ref{fig:NGC3079fpol&spix}; {bottom left panel}) in the same region as the HI deficit supports this ionization hypothesis. The reduced amount of neutral gas, being the fuel reservoir for star formation, in the AGN outflow region, acts as a negative AGN feedback.

\subsubsection{NGC 3516}
NGC 3516 is a barred lenticular galaxy hosting a CL AGN \citep{Mehdipour2022}. This galaxy is also identified as an ambiguous type 1 Seyfert \citep{Gallimore2006} with an S-shaped NLR which is more common to type 2s \citep[see][]{Fischer2013}. The gradual weakening of the broad lines and their complete disappearance in 2014 has been noted by \citet{Shapovalova2019}. \citet{Veilleux1993} have proposed two models for the S- or Z-shaped NLR: a precessing twin-jet model, where the precessing jet entrains NLR clouds, and a model where the gas-entraining jet is deflected by the ram pressure of the rotating spiral galaxy. In both scenarios, the radio jet interacts with the emission line gas, with the jet entraining the gas clouds rather than the gas ionizing the medium to produce synchrotron emission. 


We observe the projected Seyfert outflow to be nearly aligned with the major axis of the host galaxy and extending up to $\sim54\arcsec$ ($\sim10$~kpc; Figure \ref{fig:NGC3516_fpol&spix}). We observe a flat-spectrum, depolarized core with a double-lobed structure in the north-east and south-west direction. After subtracting the core's contribution, the approaching jet/lobe \citep[see][]{Miyaji1992, Mundell2009} is found to be three times brighter than the receding lobe, with flux densities of $\sim2.9$~mJy and $\sim0.96$~mJy, respectively. The inferred B-field structure observed in the north-east lobe is similar to that observed in a Fanaroff-Riley (FR) type II source; that is, the B-fields are aligned along the jet/outflow and become perpendicular in a hotspot-like region which also has a flatter spectral index. This is suggestive of a terminal shock that compresses B-fields and causes particle reacceleration. 

\begin{figure*}
\centering
\includegraphics[width=6.9cm]{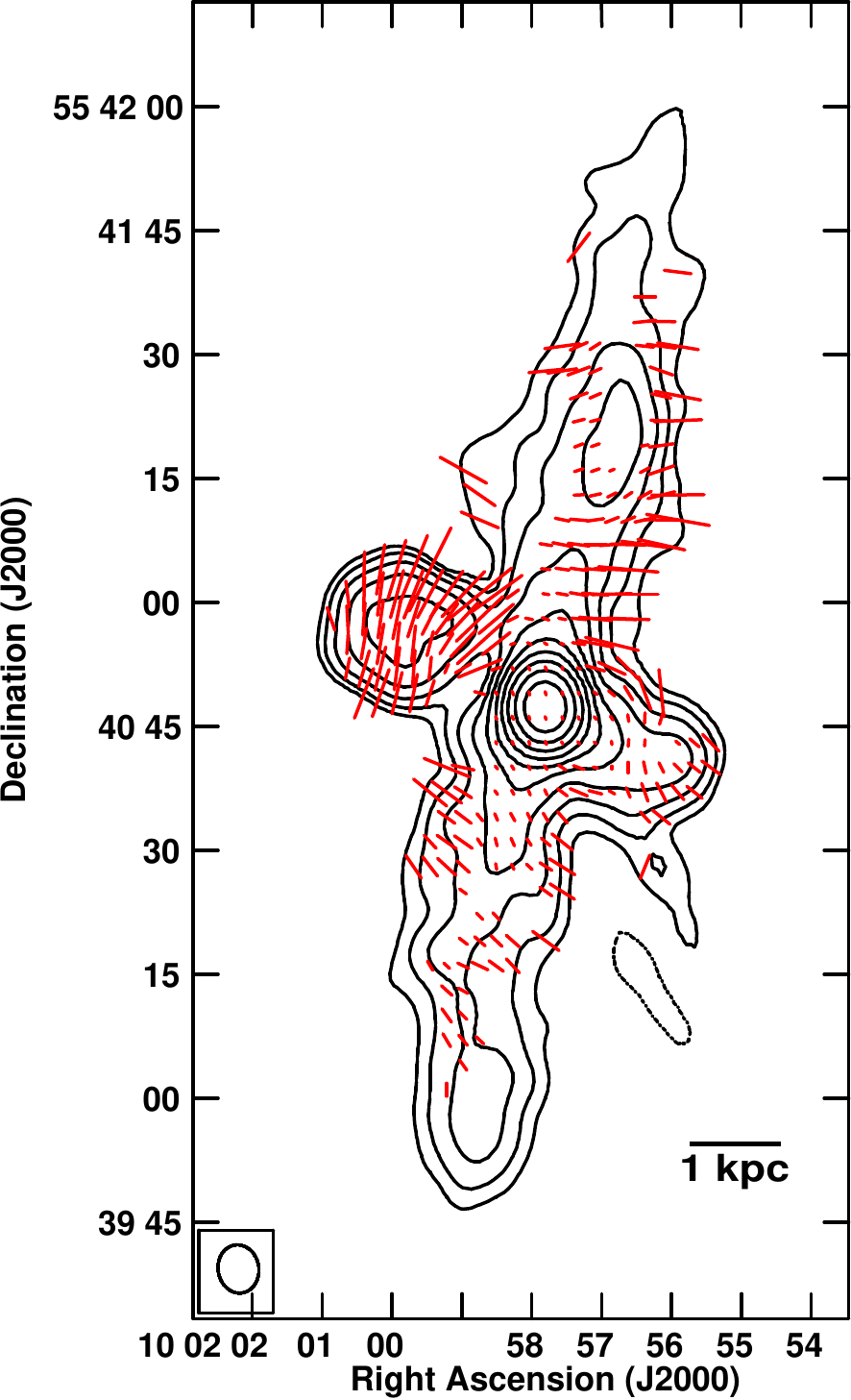}
\includegraphics[width=6.7cm]{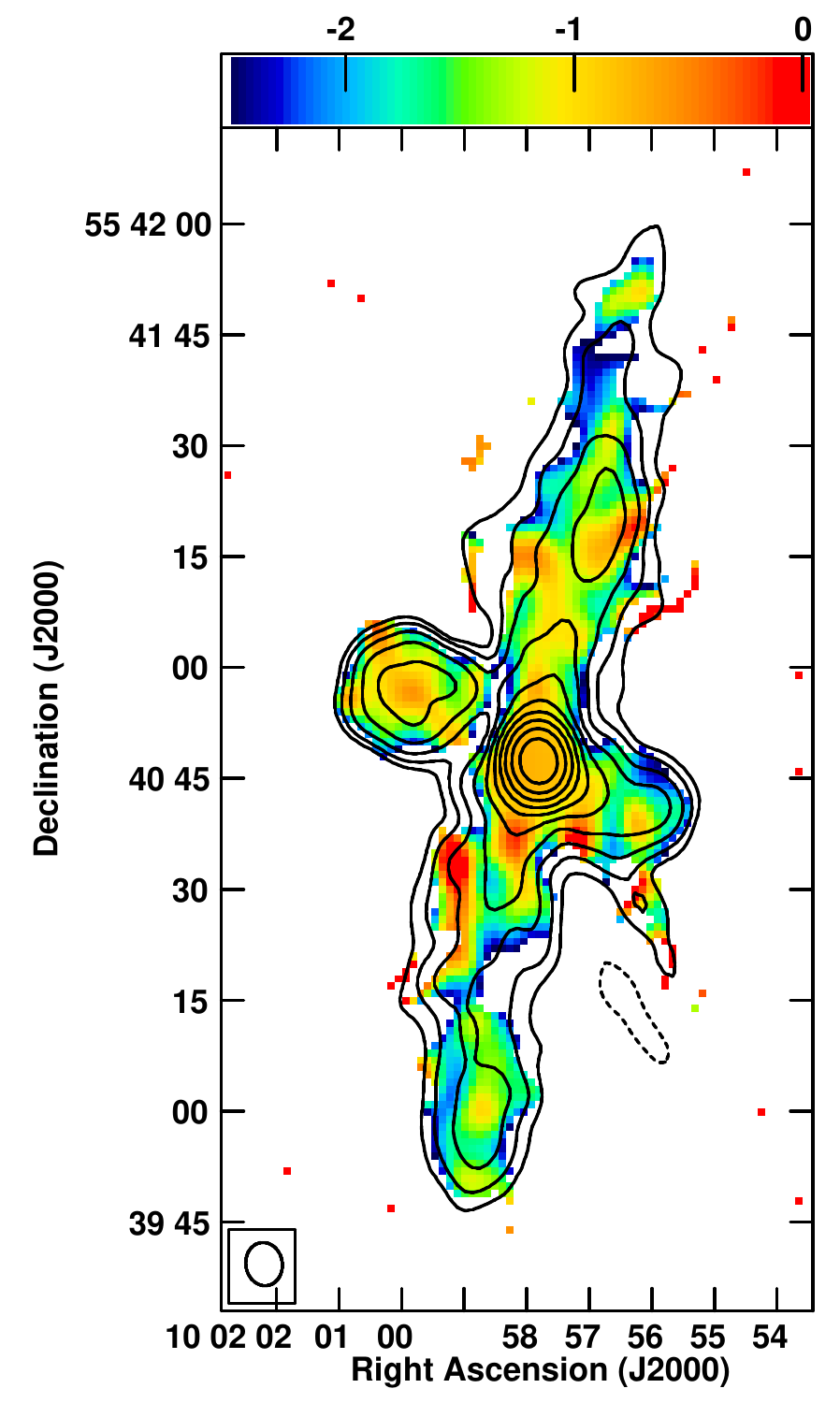}
\includegraphics[width=5.5cm]{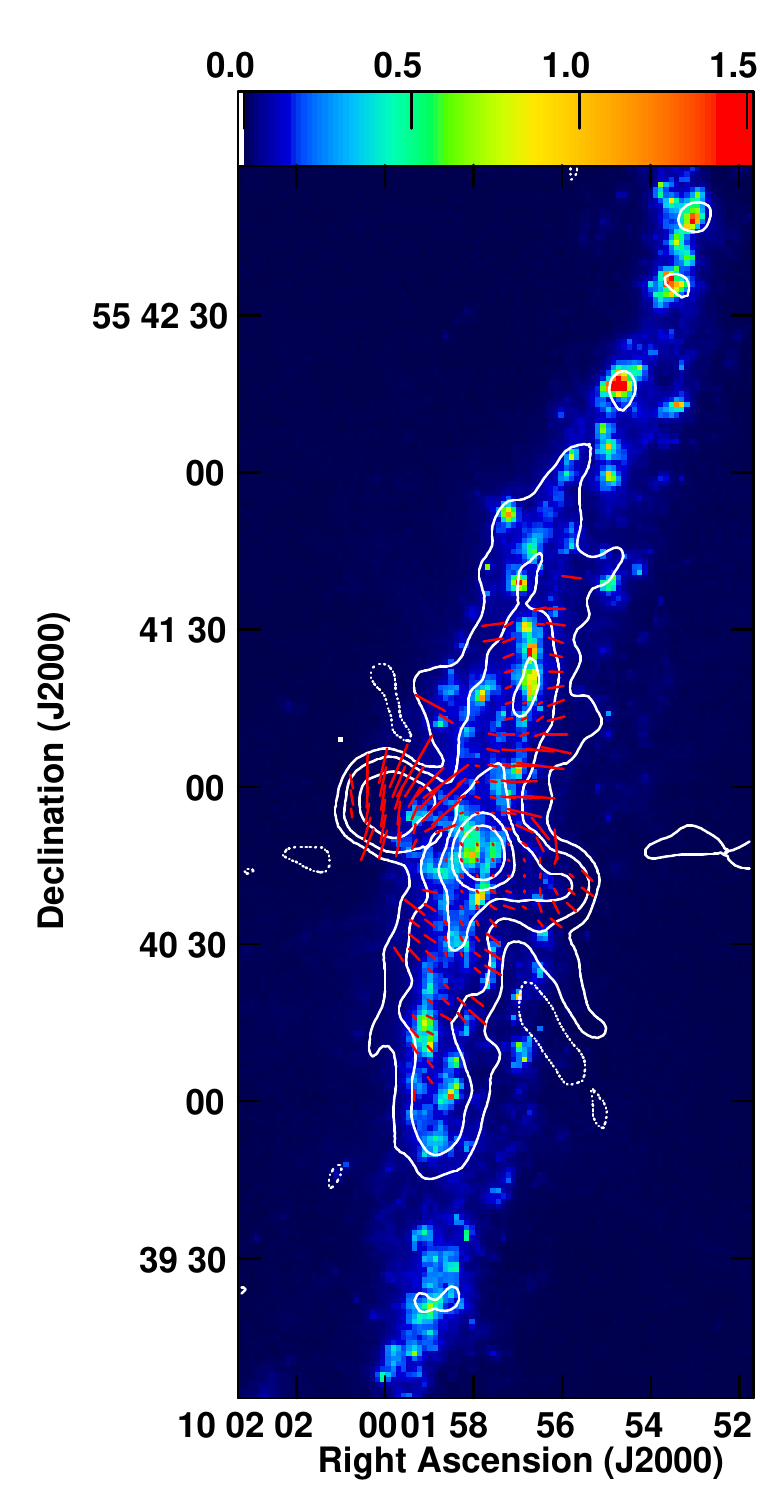}
\includegraphics[width=4.5cm,trim=0 28 0 0]{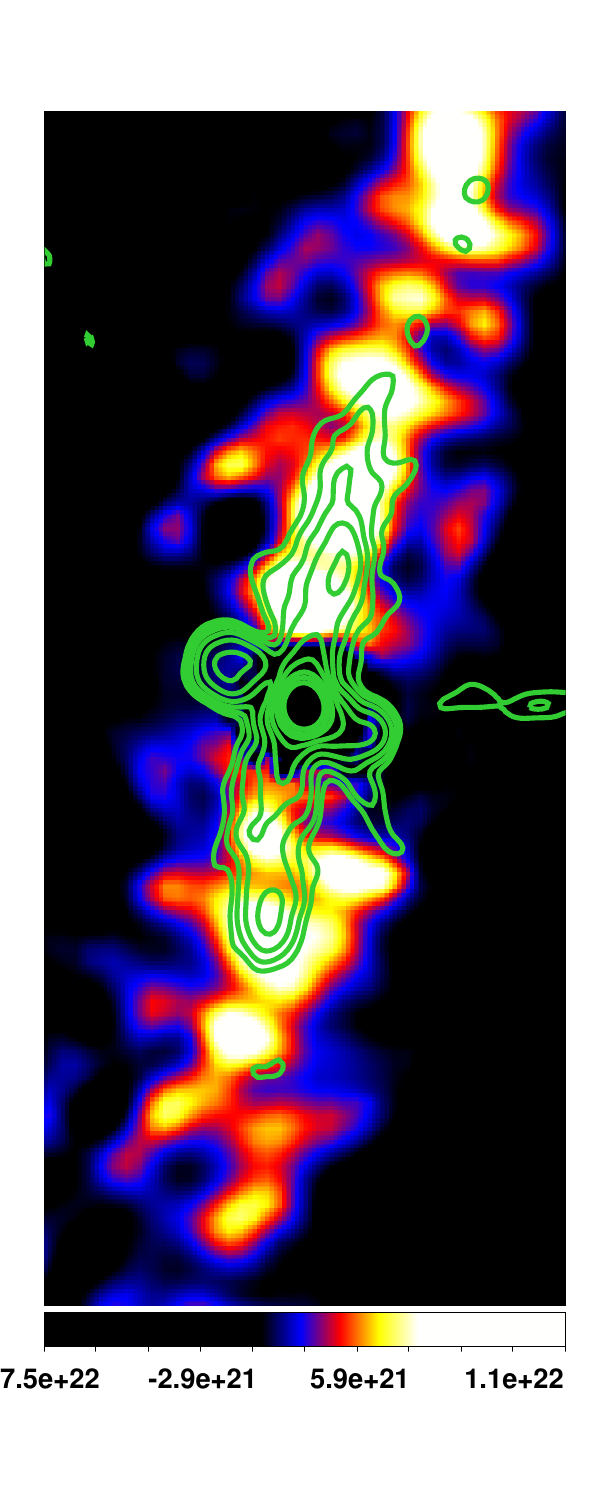}
\caption{\small {(Top left) NGC 3079 image at 10~GHz with the VLA D-array. {\it Galactic radio emission is detected along with the KSR in this galaxy.} For all the panels the contour levels are at 3$\sigma \times(-1, 1, 2, 4, 8, 16, 32, 64, 128, 256, 512)$ with $\sigma = 50~\mu$Jy beam$^{-1}$. Similarly, the synthesized beam size is $5.86\arcsec\times 4.93\arcsec$ at a PA of $9.31\deg$. The tick lengths are proportional to fractional polarization with $5\arcsec = 25$\%. (Top right) In-band spectral index image of NGC 3079 centred at 10 GHz. The spectral index values are in colour ranging from -2.5 to 0.0. (Bottom left) Radio contours with polarization vectors over-plotted on the H$\alpha$ image from CHANG-ES for the Seyfert galaxy NGC 3079. The H$\alpha$ is in colour ranging from (0.0 to 1500.0)$\times2.086\times10^{-19}$~erg~s$^{-1}$~cm$^{-2}$. The tick lengths are the same as the top left panel. (Bottom right) Radio contours are overplotted on HI moment 0 map from CHANG-ES \citep{Zheng2022}.}}
\label{fig:NGC3079fpol&spix}
\end{figure*}

\subsubsection{NGC 4051}
NGC 4051 is a narrow-line Seyfert 1 (NLS1) inclined at an angle of 12$^\circ$ to our line of sight \citep{Fischer2013, Khachikian1974}. It has also been identified as a Seyfert type 1.5 \citep{Gallimore2006} hosted by a weakly barred spiral galaxy \citep{Dumas2007}. {Similar to NGC 2992, this galaxy is known to harbour an UFO with velocity $\sim$1.2c \citep{Pounds2013}.} The X-ray power spectrum shows similarity with the soft state of Cygnus X-1 and its low mass suggests a high accretion rate of about a few tens of per cent of the Eddington rate \citep{Taylor2003}. This source is found to be an analogue of a soft-state galactic black hole binary where jets are typically not found \citep{Jones2011}. However, VLBI observations reveal three compact regions in a line with an extension of $\sim1\arcsec$ ($\sim$0.07~kpc) at a PA of $73\degr$ at 1.5 GHz, confirming the existence of small-scale jets in this source \citep{Christopoulou1997, Giroletti2009,Jones2011}.

At a resolution of $\sim$7$\arcsec$ {($\sim0.5$ kpc)}, we observe extended radio emission from this Seyfert galaxy with extent $\sim$20$\arcsec$ ($\sim$1.3~kpc) without resolving the core-jet structure. The spectral index image (Figure \ref{fig:NGC4051_fpol&spix}, right panel) shows that the `core' is to the north of the peak radio emission, and a jet-core-jet structure is revealed by steep spectrum emission on either side of the flat spectrum core. \citet{Sebastian2020} reported a non-detection of polarization at 5~GHz with an upper limit of f$_p$ = 14.3\%. We detect polarised emission from the northern {lobe} (f$_p$ = $7\pm2$\%, Figure~\ref{fig:NGC4051_fpol&spix}, left panel), with B-fields aligned with the {possible jet direction}. No polarization is detected towards the south which could be due to the presence of a greater Faraday rotating medium to the south. 

\begin{figure*}
\centering
\includegraphics[width=8.2cm]{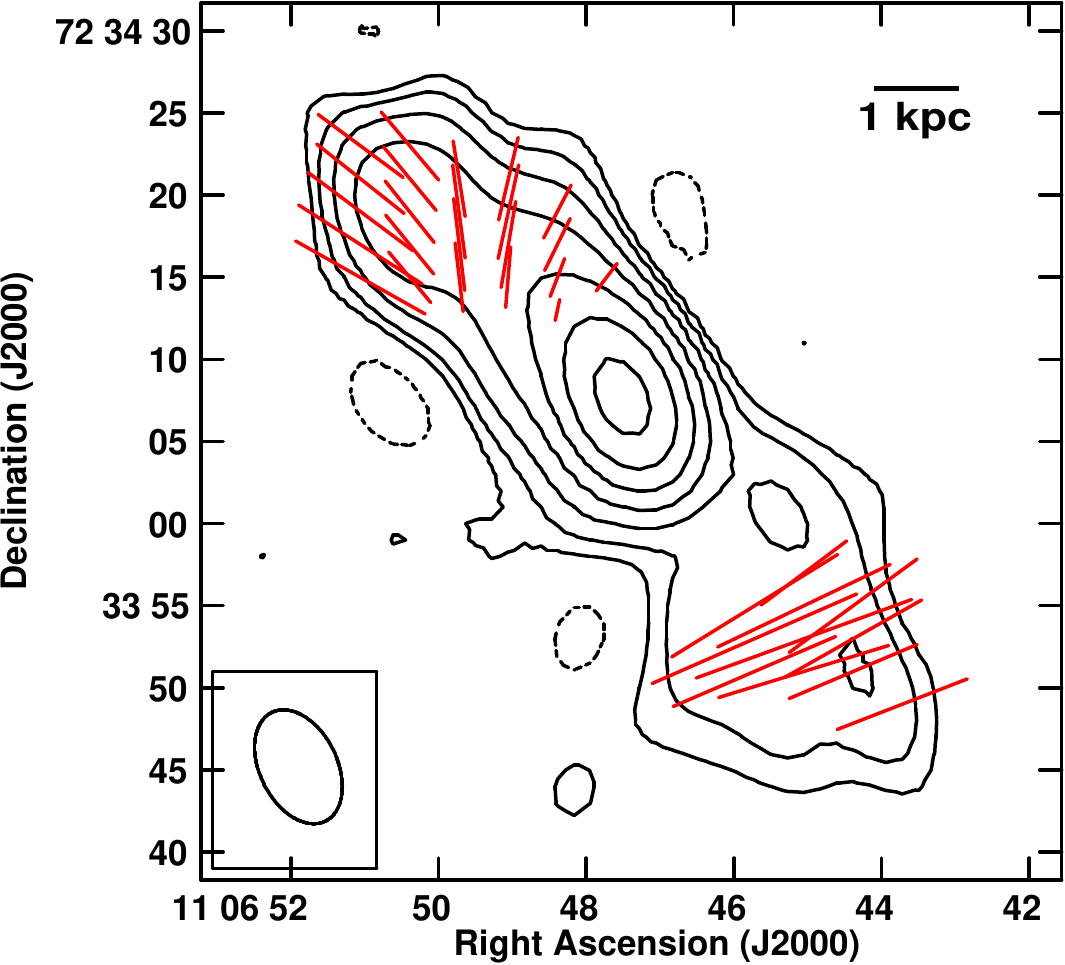}
\includegraphics[width=7.5cm]{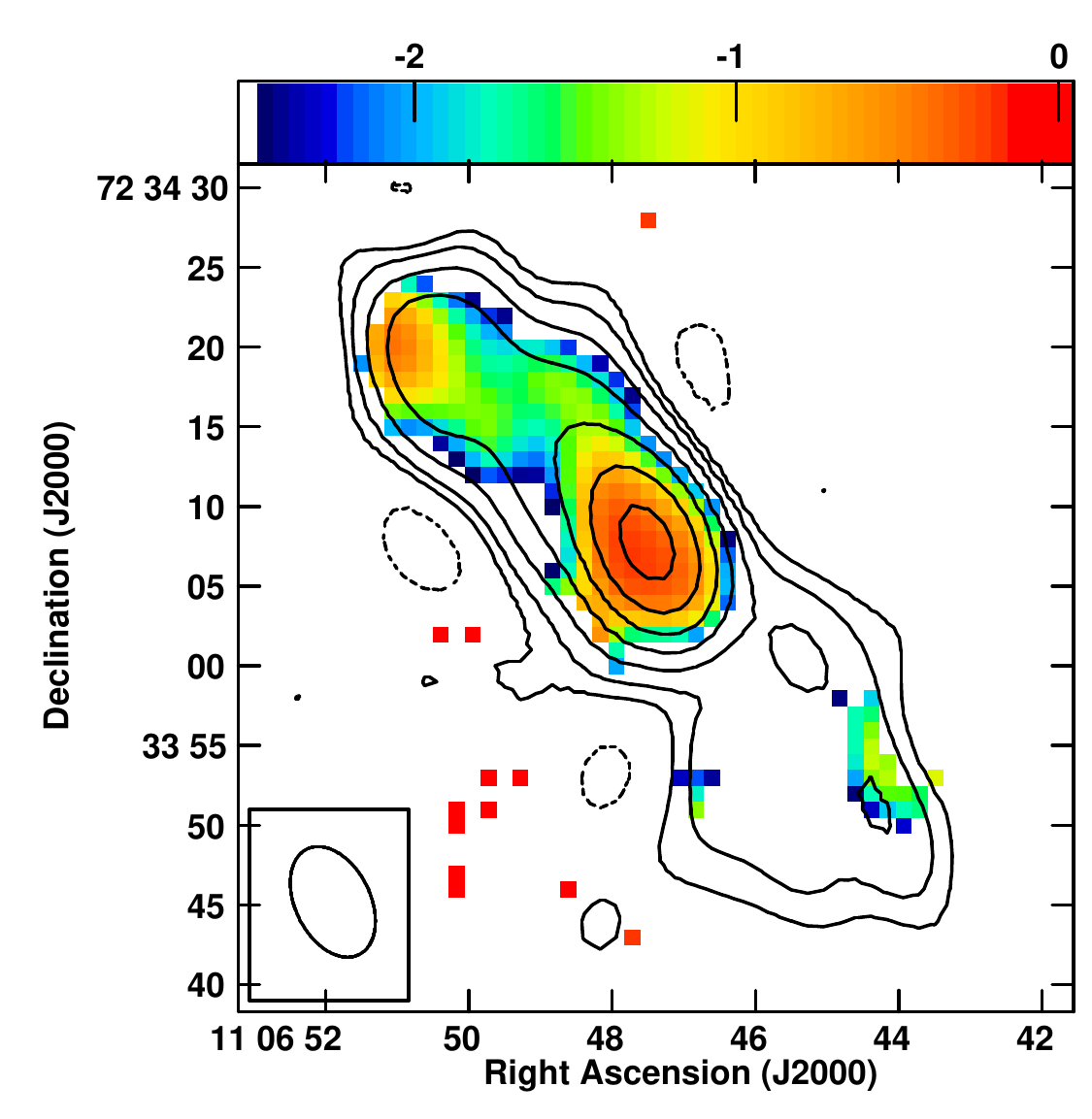}   
    \caption{\small (Left) NGC 3516 at 10~GHz with the VLA D-array. For all the panels the contour levels are at 3$\sigma \times(-1, 1, 2, 4, 8, 16, 32, 64)$ with $\sigma = 16~\mu$Jy beam$^{-1}$. The synthesized beam size is 7.38$\arcsec \times 4.71\arcsec$ at a PA of $26.29\degr$. The tick lengths are proportional to fractional polarization with $5\arcsec = 12.5$\%. (Right) In-band spectral index image of NGC 3516 centred at 10 GHz. The spectral index values are in colour ranging from -2.5 to 0.0. }
    \label{fig:NGC3516_fpol&spix}
\end{figure*}

\subsubsection{NGC 4235}
NGC 4235 is a nearly edge-on spiral galaxy without a bar hosting a Seyfert 1 nucleus \citep{Abell1978,Jimenez-Benito2000,Gallimore2006}. 
The H$\alpha+$ [NII], [OIII]$\lambda$5007\r{A} emission lines are observed to have an extension along the major axis of the galaxy. 
No HII has been detected in the disk and no ionised gas has been detected above the galactic plane \citep{Pogge1989b}.

Our images show arc-like regions that are part of bubble-like lobes in the east-west direction, roughly perpendicular to the galactic disk \citep[see also][]{Colbert1996b}. The maximum observed extent of the source is $\sim$118$\arcsec$ ($\sim$22~kpc). The western lobe (flux density $\sim$0.9~mJy) is brighter than the eastern lobe (flux density $\sim$0.3~mJy). \citet{Kharb2016} have found a bent jet-like feature connecting the core to the western lobe, which is consistent with the sub-arcsecond-scale image of \citet{Kukula1995} showing a north-west jet. 

The eastern lobe emission is seen from behind the host galaxy disk \citep{Colbert1996a, Malkan1998} which can explain the lack of significant polarization being due to the presence of greater Faraday rotating medium. Polarisation was not detected in this source at 5~GHz with an upper limit of 4.7\% \citep{Sebastian2020}. At 10~GHz we detect polarization both in the core and the eastern and western lobes at the level of $2.5\pm0.5\%$ and $5\pm1\%$, respectively. The inferred B-fields are aligned with the edges of the lobes (see Figure \ref{fig:NGC4235fpol&spix}, top panel). Given the optically thick core ($\alpha=+0.11\pm0.02$; Figure \ref{fig:NGC4235fpol&spix}, bottom panel), the inferred B-fields are perpendicular to the sub-arcsec-scale jet direction. 

\subsubsection{NGC 4388}
NGC 4388 is an edge-on unbarred spiral galaxy near the Virgo cluster centre with a highly inclined disk to our line of sight ($\sim78^\circ$), with the Northern part being closer to us \citep{Phillips1982,Chung2009}. The galaxy is found to host a Seyfert type 2 nucleus \citep{Phillips1982}. The radio emission from this source is found to be extra-planar, with high excitation gas like H$\alpha$+[NII]$\lambda$6548 and [OIII]$\lambda$4959, $\lambda$5007 showing large extensions to the north-east, not exactly coinciding with the jet 
\citep{Corbin1988, Veilleux2003}. A tail of neutral hydrogen gas is observed away from the galaxy, leading to an HI deficiency in NGC 4388, supported by persistent ram-pressure stripping in the ICM \citep{Vollmer2003, Oosterloo2005}. Studies indicate the presence of a precessing jet either due to a dual AGN or a precessing disk in this source 
\citep[see][]{Mueller2022,Sargent2024}.

The galactic emission with the Seyfert outflow along the minor axis of the galaxy is detected in our observations at 6~GHz and 10~GHz. The galactic disk exhibits a slightly steeper spectral index ($\alpha\sim-1.2$) compared to the Seyfert outflow ($\alpha\sim-0.8$; see Figures \ref{fig:NGC4388_Xband_fpol&spix&Halpha} and \ref{fig:NGC4388_Cband_fpol&spix&Halpha}, middle panel). The Seyfert outflow also stands out in having higher fractional polarization, with f$_p$ = $13\pm1\%$ at 10~GHz ($5.6\pm0.8\%$ at 6~GHz), compared to the galactic disk's f$_p$ = $9\pm2\%$ at 10~GHz ($4.1\pm0.7\%$ at 6 GHz; see Figures \ref{fig:NGC4388_Xband_fpol&spix&Halpha} and \ref{fig:NGC4388_Cband_fpol&spix&Halpha}, top panel). The B-fields, assumed to be perpendicular to the EVPAs, are aligned along the galactic disk, with RM tracing the rotating disk. In contrast, the B-fields along the Seyfert lobe is observed to change directions over short scales ($\sim$0.5 kpc). 

This rapid change in B-field directions observed may indicate jet wiggling caused by precession or strong jet-medium interactions. \citet{Stone1988} reported continuous bending of the radio emission along the length of the extra-planar ejecta from their multi-frequency radio observations. They attributed this to various processes such as gravitational pull, ram pressure bending by galactic gas or ICM, buoyancy maintenance due to the density gradient between the radio plasma and the surroundings, the presence of curved B-fields, and jet nozzle precession. 
From our observations, we conclude that either the curved B-field structures or the precessing jet nozzle model best explains the data. 

The 10 GHz image detects no polarization at the core whereas the 6 GHz image detects a fractional polarization of $\sim$1\% at the core. Figures~\ref{fig:NGC4388_Xband_fpol&spix&Halpha} and  \ref{fig:NGC4388_Cband_fpol&spix&Halpha} (bottom panel) show that the H$\alpha$ gas \citep{Vargas2019} is extended perpendicular to the disk but parallel to the B-field lines in the jet. The H$\alpha$ emission is bright along the galaxy's spiral arms. 
The 6~GHz image shows some radio-association with the H$\alpha$ brightening and the jet.
The H$\alpha$ emission could be linked to young star formation activity driven by the radio jet. The lack of atomic hydrogen near the radio jet could be explained by excess star formation in that region, suggesting positive AGN feedback.

The RM image from CHANG-ES \citep{Krause2020} shows an increase in RM at the core, where the linear polarization intensity drops and becomes depolarized at 10 GHz. Our preliminary analysis also detects low circular polarization (circularly polarized flux density of $\sim$0.12 mJy) at the core of this source at 6 GHz. We obtain a maximum circular fractional polarization of 0.4\% at the core at 6 GHz \citep[$>0.3\%$ is considered a strong signal;][]{Homan2006}. The region of circular polarization is coincident with the region lacking linear polarization. This might indicate Faraday conversion happening at the core of this source. \citet{Irwin2018} have recently found a circularly polarized structure in this source at 1.4 GHz. The detection of circular polarization in this source makes it interesting as it might be intrinsic to the source (i.e., not due to Faraday conversion) and could help in the determination of B-field strengths unambiguously and obtain limits on cosmic ray electron energies.  

\begin{figure*}
\centering
\includegraphics[width=8.4cm]{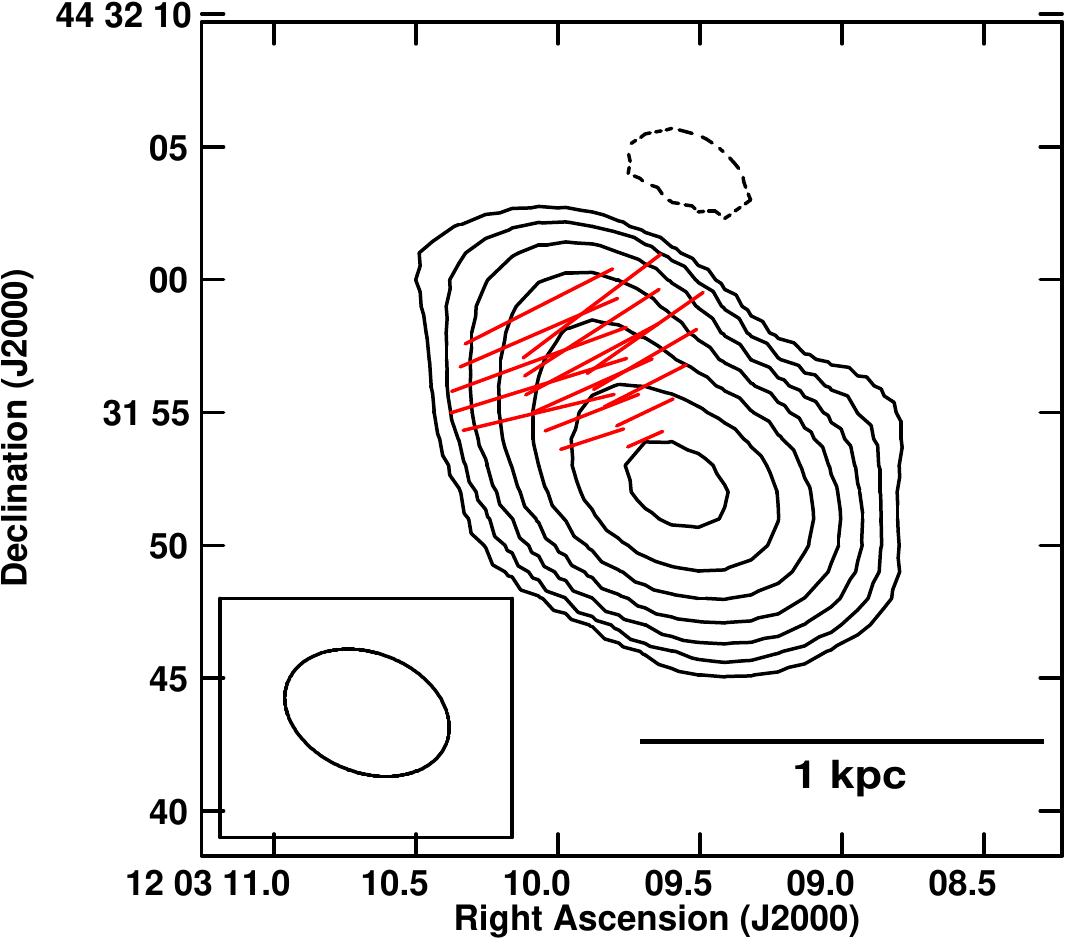}
\includegraphics[width=7.8cm]{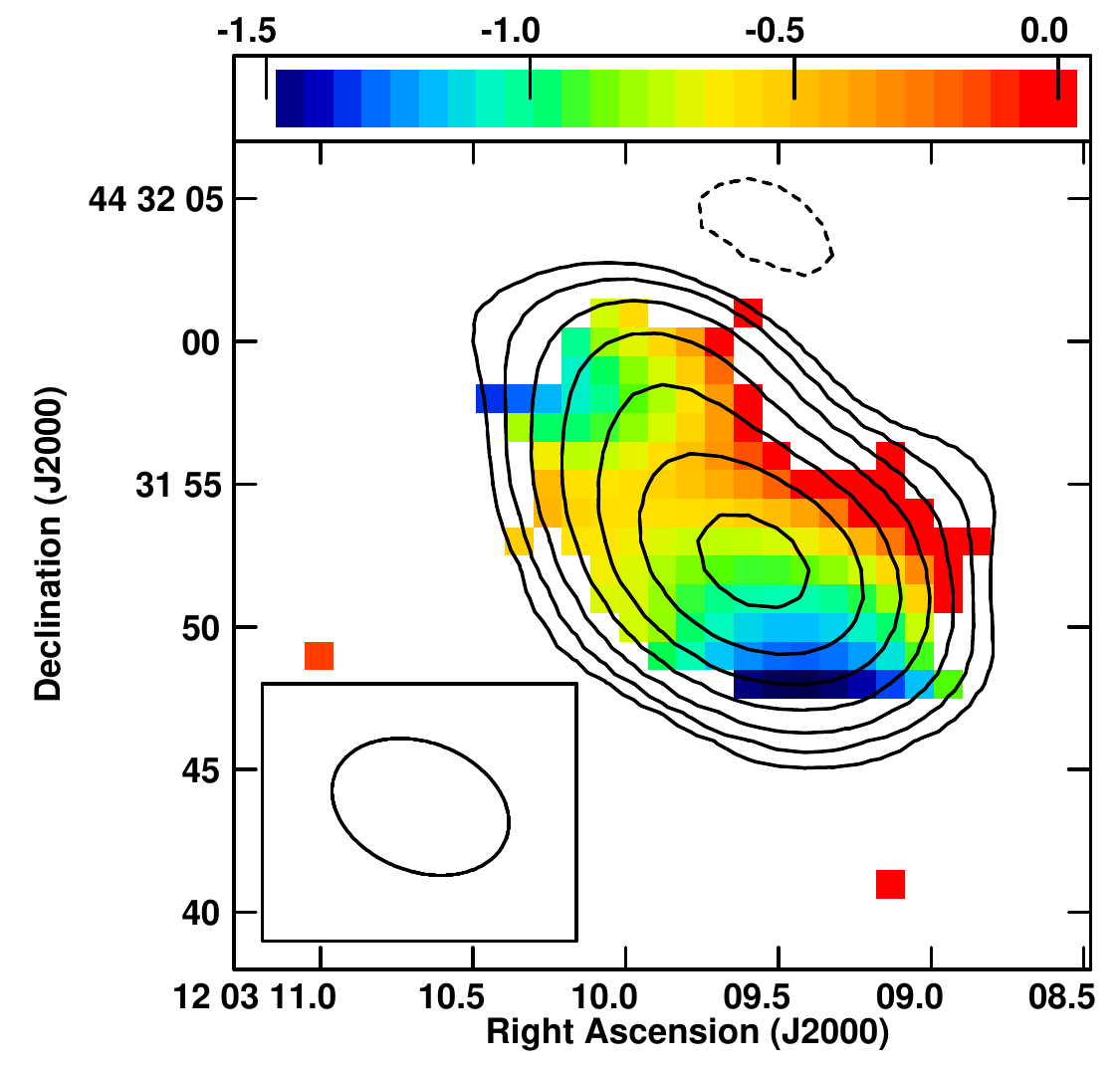}
\caption{\small (Left) NGC 4051 image at 10~GHz with the VLA D array. For all the panels the contour levels are at 3$\sigma \times(-1, 1, 2, 4, 8, 16, 32, 64)$ with $\sigma = 15~\mu$Jy beam$^{-1}$. Similarly, the synthesized beam size is 6.40$\arcsec$ $\times$ 4.51$\arcsec$ at a PA of $69\degr$. The tick lengths are proportional to fractional polarization with $2\arcsec = 2.5$\%. (Right) In-band spectral index image of NGC 4051 centered at 10~GHz. The spectral index values are in colour ranging from $-1.5$ to 0.0. }
\label{fig:NGC4051_fpol&spix}
\end{figure*}

\begin{figure*}
\centering
\includegraphics[width=10cm]{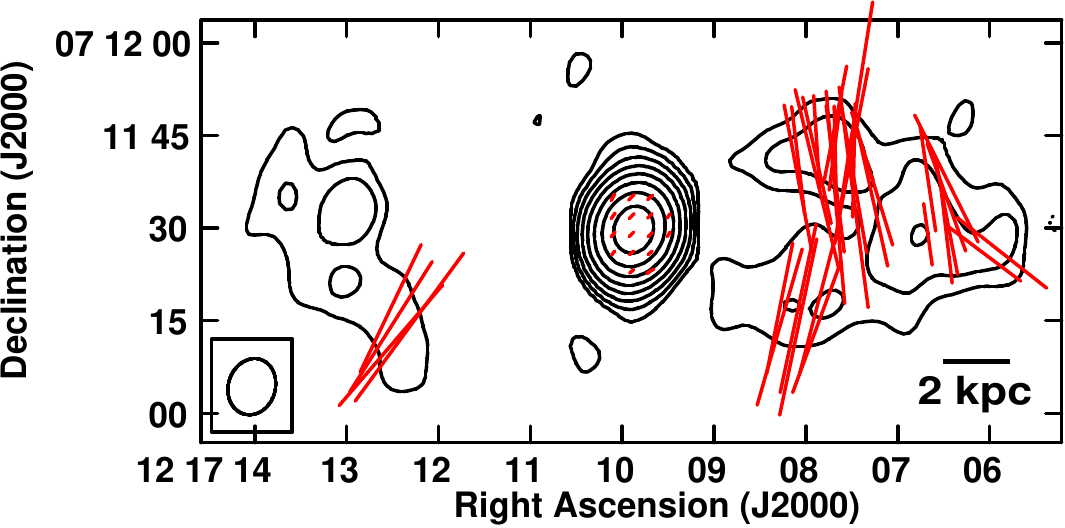}
\includegraphics[width=10cm]{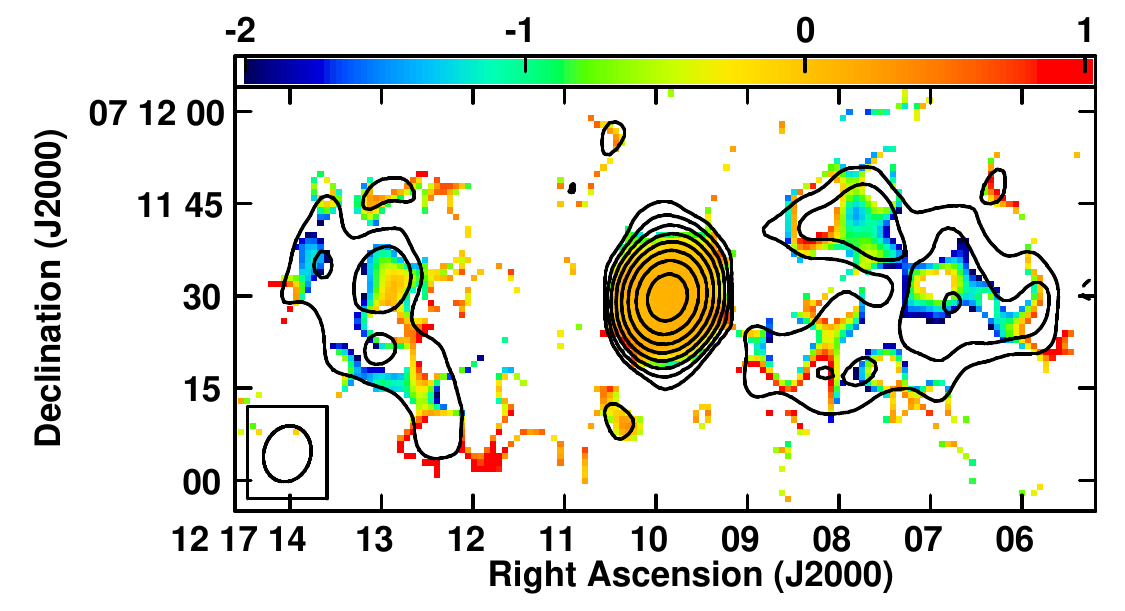}
\caption{\small (Top)  NGC 4235 image at 10~GHz with the VLA D array. For all the panels the contour levels are at 3$\sigma \times(-1, 1, 2, 4, 8, 16, 32, 64, 128)$ with $\sigma = 10~\mu$Jy beam$^{-1}$. Similarly, the synthesized beam size is 9.24$\arcsec$ $\times$ 7.60$\arcsec$ at a PA of $-17.18\degr$. The tick lengths are proportional to fractional polarization; $10\arcsec = 25$\%. Fractional polarization of errors $>30\%$ have been blanked. (Bottom) In-band spectral index image of NGC 4235. The spectral index values are in colour ranging from -2.0 to 1.0. }
\label{fig:NGC4235fpol&spix}
\end{figure*}

\subsubsection{NGC 4593}
NGC 4593 is a barred spiral galaxy with outer and inner rings and a nuclear dust ring hosting a Seyfert 1 nucleus \citep{Balzano1983, Kollatschny1985} 
exhibiting extreme variability, especially in UV and X-ray spectra \citep{Clavel1983, Santos1995}.  
NGC~4593 is thought to possess one of the smallest BLR known \citep{Dietrich1994}. Compared to the high-resolution images of this source which showed a compact feature in radio \citep{Schmitt2001}, \citet{Gallimore2006} found a lobed structure of extent of $\sim$5 kpc in this source. 

\begin{figure*}
\centering
\includegraphics[width=10cm]{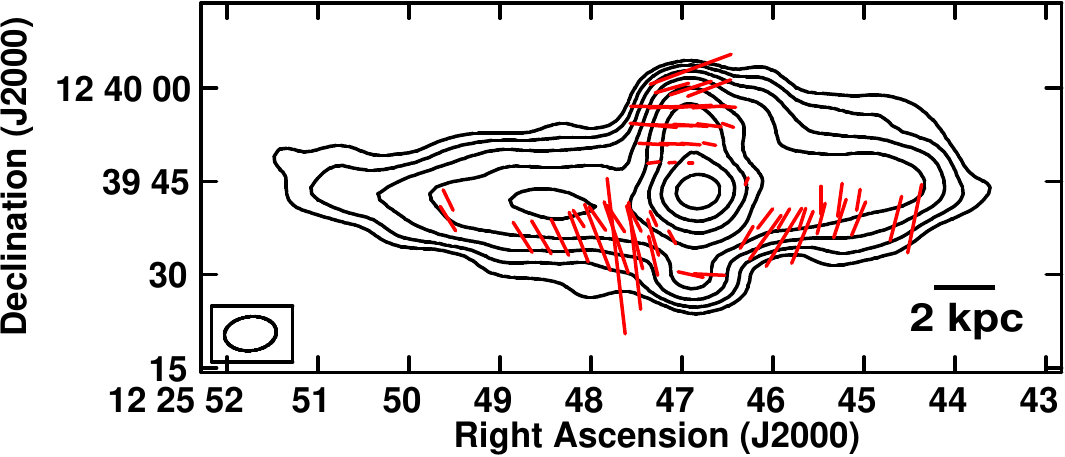}
\includegraphics[width=10cm]{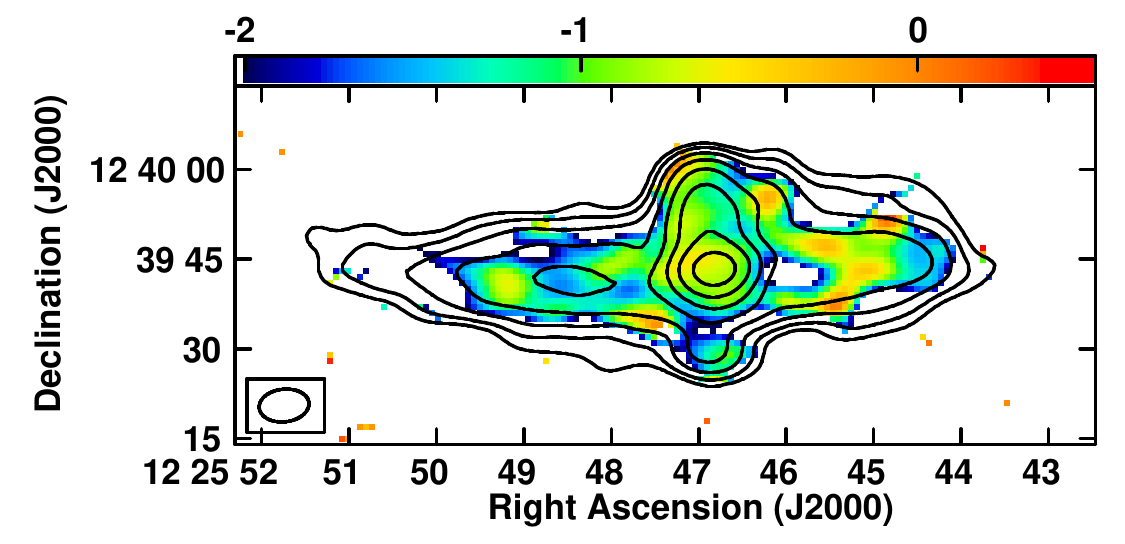}
\includegraphics[width=10cm]{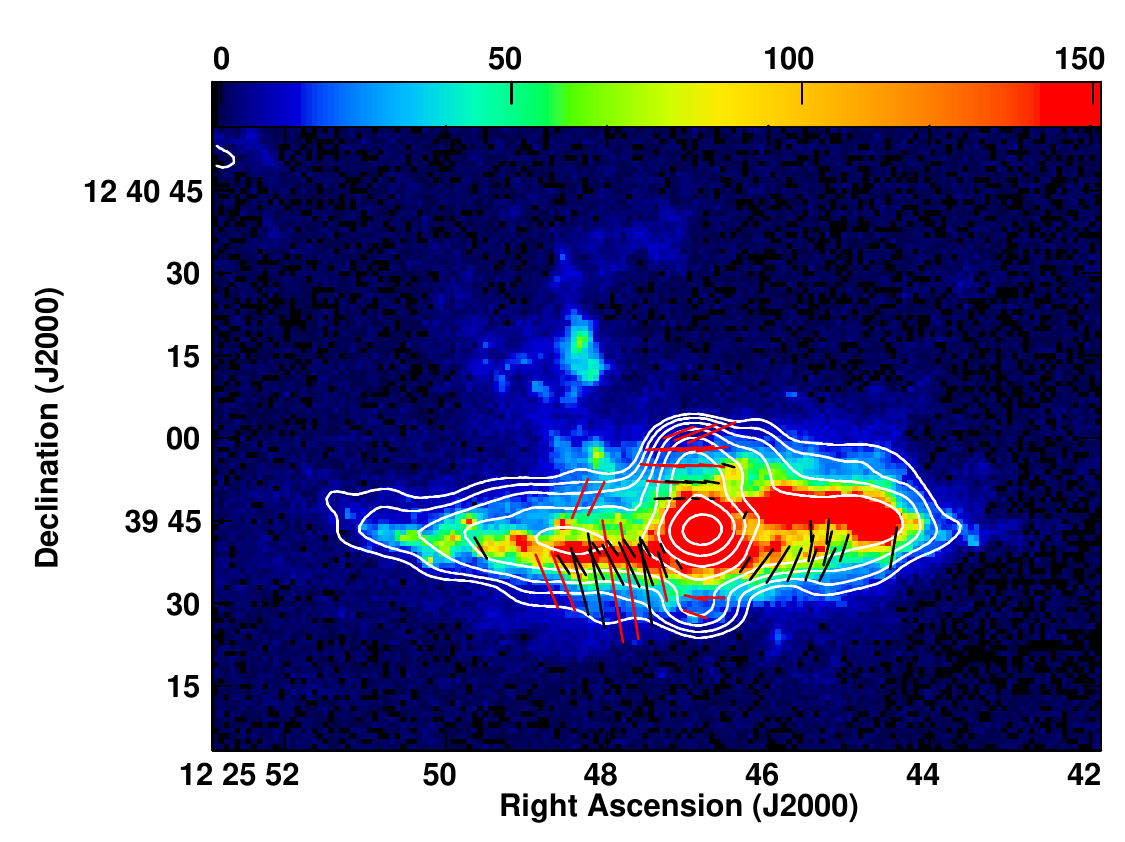}
\caption{\small (Top) NGC 4388 image at 10 GHz with the VLA D array. {\it Galactic radio emission is  detected along with the KSR in this galaxy.} For all the panels the contour levels are at 3$\sigma \times(-1, 1, 2, 4, 8, 16, 32, 64, 128)$ with $\sigma = 20~\mu$Jy~beam$^{-1}$. Similarly, the synthesized beam size is 8.36$\arcsec \times 5.47\arcsec$ at a PA of $-83.7\degr$. The tick lengths are proportional to fractional polarization with $10\arcsec = 25$\%. (Middle) In-band spectral index image of NGC 4388 centred at 10 GHz. The spectral index values in colour range from $-2.0$ to $+0.5$. (Bottom) Radio contours and polarization vectors at 10 GHz overplotted on H$\alpha$ image from CHANG-ES. The H$\alpha$ is in colour ranging from (0.0 to 150.0) $\times 2.145\times10^{19}$ erg s$^{-1}$ cm$^{-1}$.The tick lengths are the same as the top panel. }
    \label{fig:NGC4388_Xband_fpol&spix&Halpha}
\end{figure*}

\begin{figure*}
\centering
\includegraphics[width=9cm]{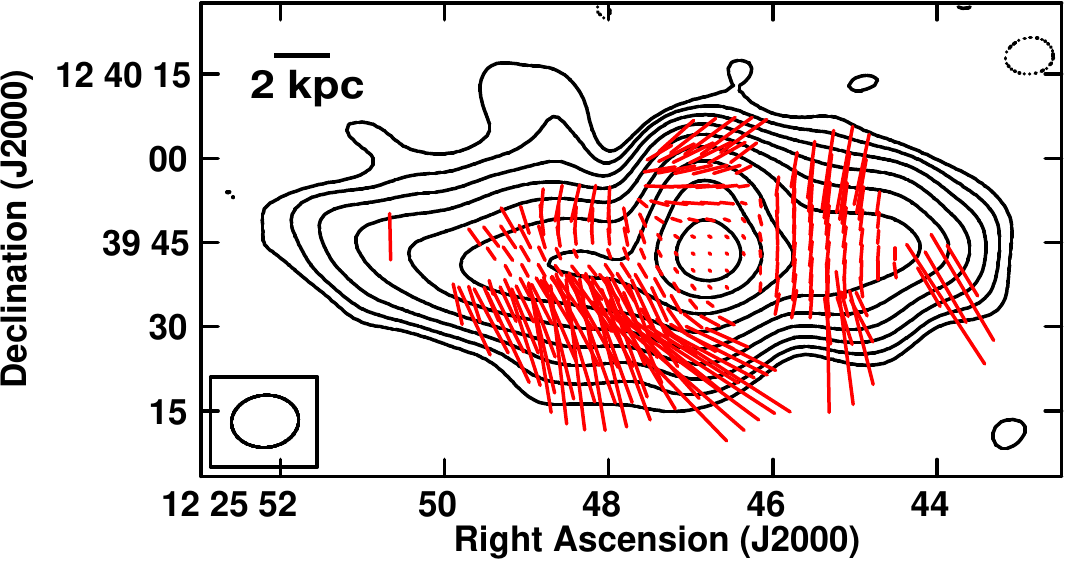}
\includegraphics[width=9cm]{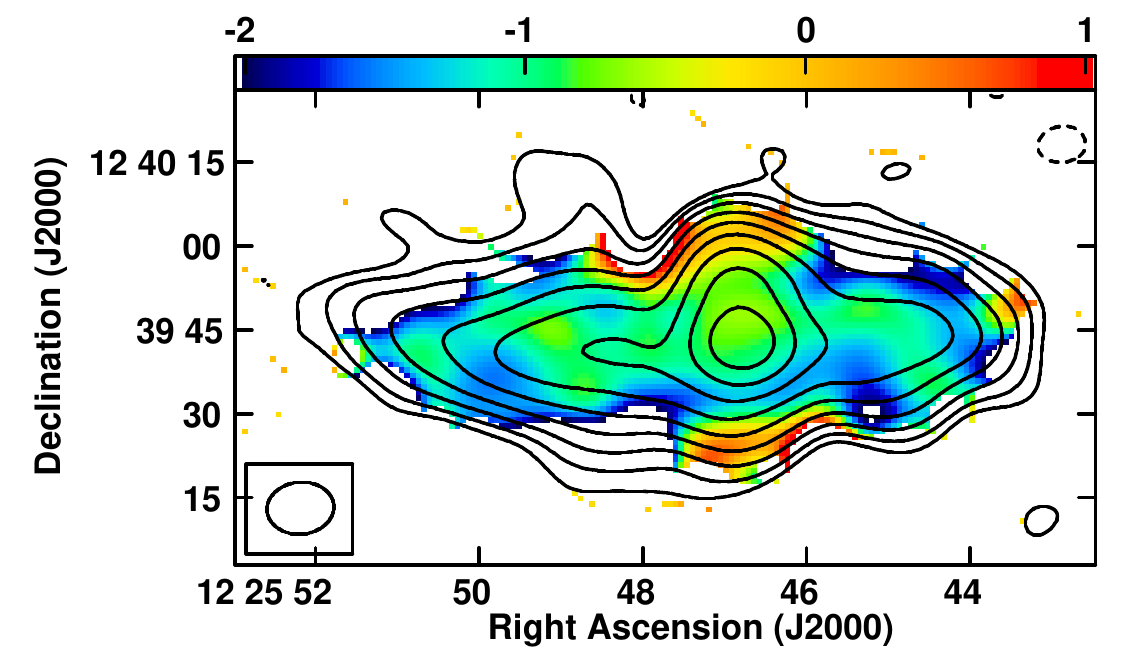}
\includegraphics[width=9cm]{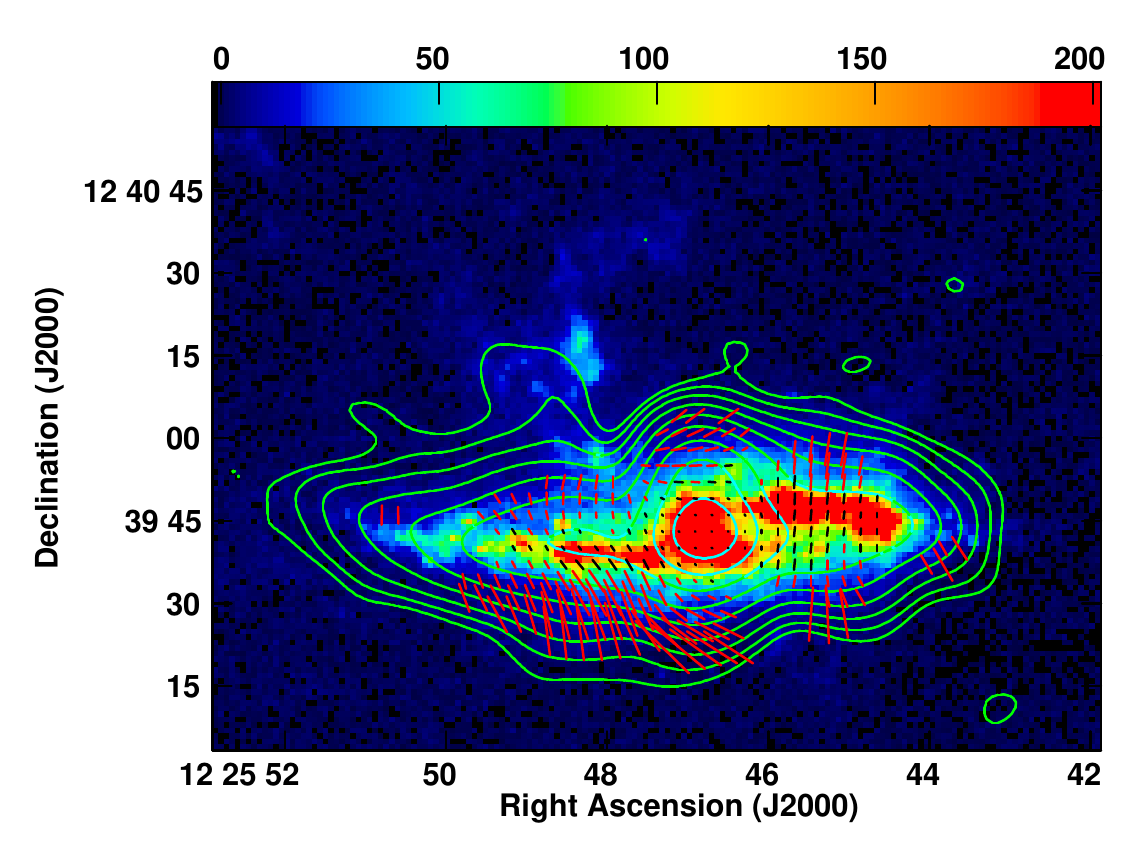}
\caption{\small (Top) NGC 4388 image at 5.3 GHz with the VLA D array. {\it Galactic radio emission is  detected along with the KSR in this galaxy.} For all the panels the contour levels are at 3$\sigma \times(-1, 1, 2, 4, 8, 16, 32, 64, 128, 256)$ with $\sigma = 17~\mu$Jy beam$^{-1}$. Similarly, the synthesized beam size is 11.99$\arcsec \times 9.31\arcsec$ at a PA of $-85.0\degr$. The tick lengths are proportional to fractional polarization with $10\arcsec = 10$\%. (Middle) In-band spectral index image of NGC 4388 at C-band. The spectral index values are in colour ranging from -2.0 to 1.0. (Bottom) Radio contours and polarization vectors at C-band over-plotted on the H$\alpha$ image from CHANG-ES. The H$\alpha$ is in colour ranging from (0.0 to 200.0) $\times 2.145\times10^{19}$ erg s$^{-1}$ cm$^{-2}$. The tick lengths are proportional to fractional polarization with $10\arcsec = 25$\%.}
    \label{fig:NGC4388_Cband_fpol&spix&Halpha}
\end{figure*}

\begin{table*}

\small{
\caption{Summary of the Polarization Properties.}
\tabcolsep=0.11cm
\center{
\begin{tabular}{cccccc}
\hline \hline
         Source&Frequency&Region&I$_P$&FP&$\chi$\\
          & & &(mJy)&(\%) &(degrees) \\
         \hline
         NGC 1068&{1.4 GHz}& Core&$3.0\pm0.7$&$0.51\pm0.01$&\nodata\\
          & &Northern lobe&$5.7\pm0.2$&$3.9\pm0.4$&$7\pm2$\\
          & &Southern lobe&$1.9\pm0.1$&$0.9\pm0.04$&$-28\pm2$\\
          & &{East polarized region}&$0.8\pm0.1$&$3.2\pm0.4$&\nodata\\
          & &{Total}&$24\pm1$&$1.7\pm0.3$&\nodata\\\hline
         {NGC 1320}&{1.4 GHz}& Core&5$\times$10$^{-3}$(5)&0.73(u)&\nodata\\
          & &North-west lobe&$0.019\pm0.006$&$20\pm6$&$72\pm9$\\
          & &Southern lobe&$0.030\pm0.009$&$47\pm18$&$-8\pm9$\\
          & &{Total}&$0.05\pm0.01$&$36\pm13$&\nodata\\\hline
         {NGC 2639}&{10 GHz}& Core&{$0.37\pm0.01$}&{$1.2\pm0.4$}&$-35\pm3$\\
         & &{Total}&$0.72\pm0.04$&$2.2\pm0.2$&\nodata
         \\\hline
         {NGC 2992}&{10 GHz}& Core&{$0.20\pm0.05$}&{$1.0\pm0.4$}&$-53\pm4$\\
         & &{Total}&$0.51\pm0.05$&$2.9\pm0.6$&\nodata\\\hline
         {NGC 3079}&{10 GHz}& Core&{$2.0\pm0.1$}&{$2.2\pm0.1$}&$28\pm2$\\
          & &East-West lobe&$7.3\pm0.2$&$16\pm2$&\nodata\\
          & &Galactic disk&$0.9\pm0.2$&$12\pm2$&$7\pm6$\\
          & &{Total}&$7.3\pm0.3$&$14.4\pm1.2$&\nodata\\\hline
         {NGC 3516}&{10 GHz}& Core&27$\times$10$^{-4}$(u)&0.06(u)&\nodata\\
          & &Extended lobe&$0.28\pm0.05$&$17\pm4$&\nodata\\
          & &{Total}&$0.27\pm0.05$&$17\pm0.4$\\\hline
         {NGC 4051}&{10 GHz}& Core&3.9$\times$10$^{-3}$(u)&0.32(u)&\nodata\\
         & &Extended lobe&$0.06\pm0.01$&$7\pm2$&$-66\pm7$\\ 
         & &{Total}&$0.06\pm0.01$&$6.3\pm1.8$&\nodata\\\hline
         {NGC 4235}&{10 GHz}& Core &$0.13\pm0.02$&$2.5\pm0.5$&$-42\pm5$\\
          & &Western lobe &$0.39\pm0.08$&$5\pm1$&\nodata\\
          & &Eastern lobe &$0.23\pm0.01$&$2.5\pm0.5$&\nodata\\ 
          & &{Total}&$0.37\pm0.08$&$5\pm1$&\nodata\\\hline
         {NGC 4388}&{10 GHz}& Core &6$\times$10$^{-3}$(u)&0.04(u)&\nodata \\
          & &Northern lobe&$0.81\pm0.09$&$13\pm1$&\nodata\\
          & &Galactic disk&$0.25\pm0.05$&$9\pm2$&$\pm18\pm7$\\ 
          & &{Total}&$0.72\pm0.07$&$12.3\pm1.7$&\nodata\\\hline
         {NGC 4593}&{10 GHz}& Core&12$\times$10$^{-3}$(u)&0.5(u)&\nodata\\
          & &Extended lobe&$0.07\pm0.02$&$16.9\pm0.5$&$-26\pm8$\\
          & &{Total}&$0.08\pm0.02$&$16.9\pm0.5$&\nodata\\\hline
         {NGC 4594}&{10 GHz}& Core&{$0.50\pm0.06$}&{$0.6\pm0.3$}&$54\pm3$\\ & &{Total}&$0.71\pm0.02$&$0.9\pm0.1$&\nodata\\\hline
         {NGC 5506}&{10 GHz}& Core&{$0.05\pm0.01$}&{$0.10\pm0.02$}&$-28\pm7$\\ 
         & &{Total}&$0.12\pm0.03$&$5\pm1$&\nodata\\\hline
\end{tabular}}
\\

{\small Column 1: Name of the sources. Column 2: Observing frequency. Column 3: Source region selected for analysis. Column 4: Mean polarized intensity in that region in mJy. Column 5: Mean polarization fraction in that region in \%. Column 6: Mean angle of the polarized vectors in degrees in that region. Sources with polarized intensity and polarization fraction values but not with polarization angle values presented here imply that the vectors are oriented at multiple angles within the selected region. (u) denotes an upper limit.
}}
\label{tab:tab3}
\end{table*}

\begin{figure*}
\centering
\includegraphics[width=8.5cm]{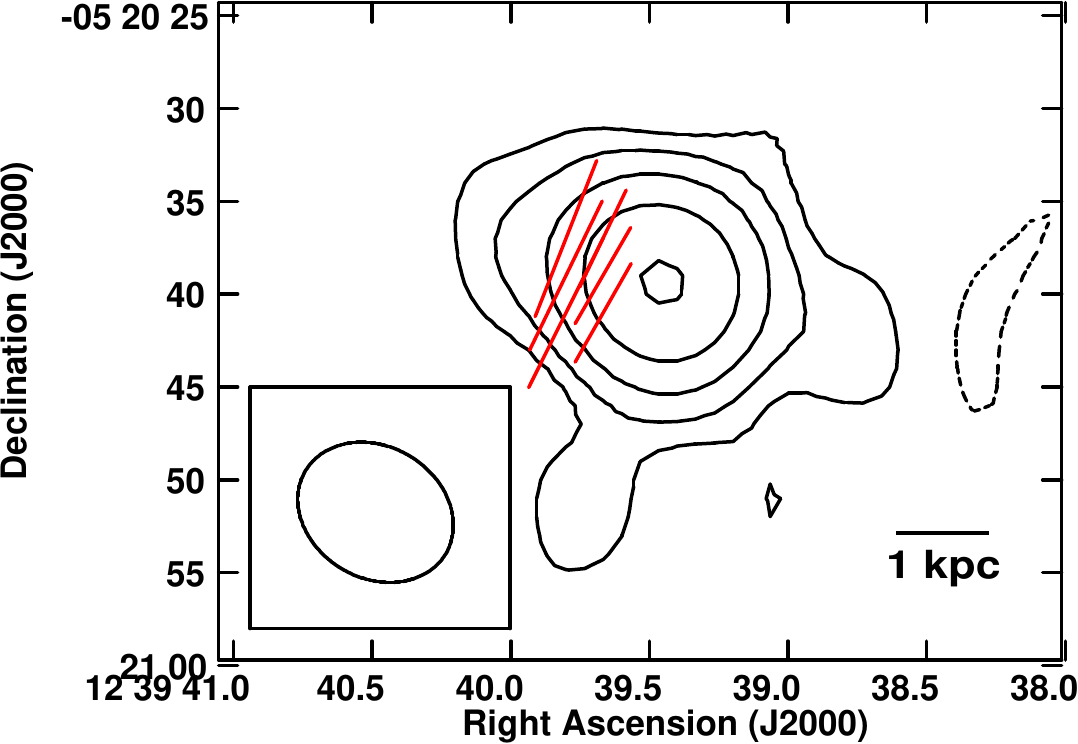}
\includegraphics[width=7.7cm]{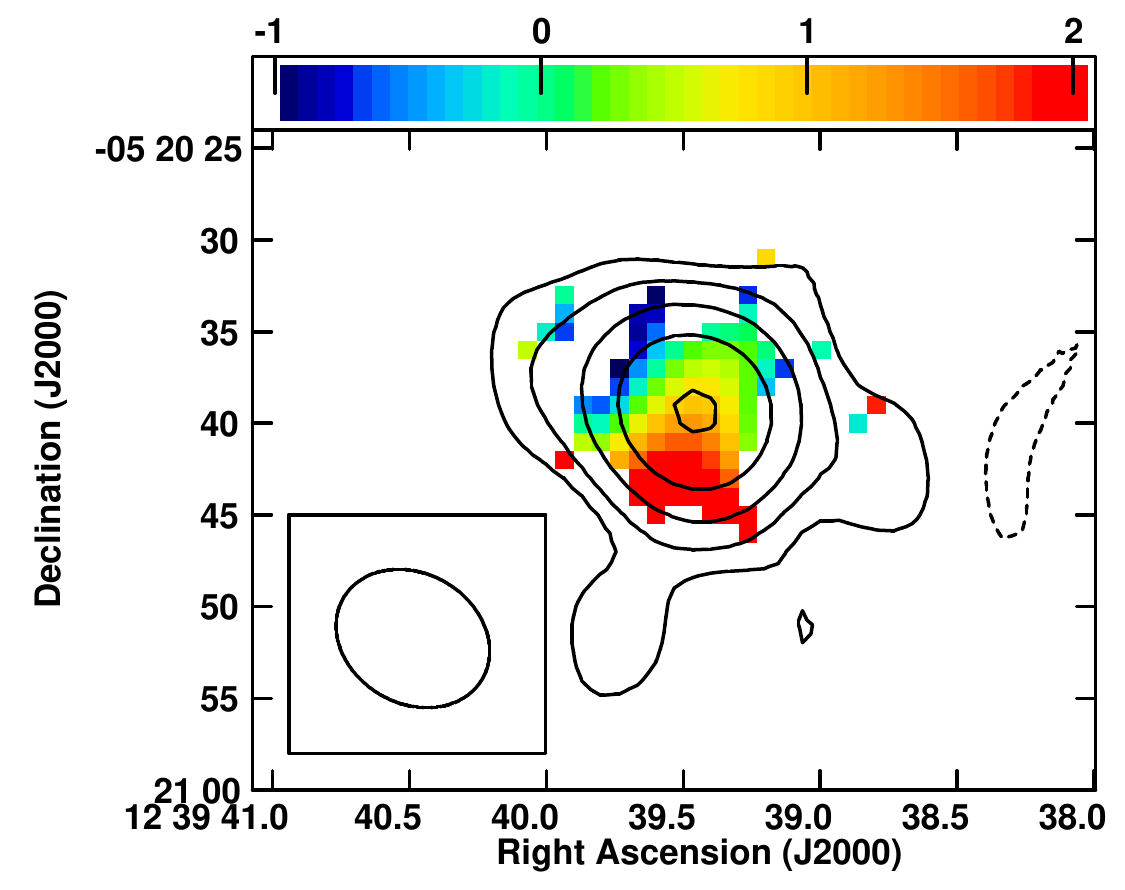}
\caption{\small (Top) NGC 4593 image at 10 GHz with the VLA D array. For all the panels the contour levels are at 3$\sigma \times(-1, 1, 2, 4, 8, 16)$ with $\sigma = 45~\mu$Jy beam$^{-1}$. Similarly, the synthesized beam size is 8.77$\arcsec$ $\times$ 7.09$\arcsec$ at a PA of $59.74\degr$. The ticks are proportional to fractional polarization; $5\arcsec = 12.5$\%. (Bottom) In-band spectral index image of NGC 4593 centred at 10 GHz. The spectral index values are in colour ranging from $-1.0$ to $+2.0$.}
    \label{fig:NGC4593fpol&spix}
\end{figure*}

Our observations show a weak extension in the radio emission of $\sim26\arcsec$ ($\sim$5 kpc) in the east-west direction. Polarization with an inferred B-field aligned with the radio extension is observed in a region to the east of the core (see Figure \ref{fig:NGC4593fpol&spix}; {left panel}). A spectral index steepening is observed in the same direction (Figure \ref{fig:NGC4593fpol&spix}, {right} panel), more prominently towards the north-east and less prominently towards the north-west. These results are consistent with a small nearly east-west oriented double-lobed structure present in NGC 4593 in the 5~GHz image of \citet{Gallimore2006}. This structure is not resolved in our image but is reflected in the spectral index image suggesting the presence of a jet or lobe in the east-west direction. However, an inversion in the spectrum observed in the southern part of the source remains unexplained. The [OIII]$\lambda$5007 emission \citep{Schmitt2003} coincides with the extended structure, with stronger emission on the west than the east of the core. The lack of polarization detection in the eastern lobe could be a consequence of a greater Faraday rotating medium (cold gas) in that direction. 

\subsubsection{NGC 4594}
NGC 4594 a.k.a. the Sombrero galaxy is an unbarred, edge-on, spiral galaxy with a large stellar bulge and a containing LINER type 2 nucleus \citep{Heckman1980,Giovanelli1986,Irwin2012}. This galaxy is found to be a poorly star-forming galaxy in comparison to other galaxies of similar type, although dominated by Type 1a supernovae activity in the ISM \citep{Li2011, Li2016, Jiang2024}. \citet{Jiang2024} suggested that this galaxy might have gone through a starburst stage at an earlier time, and has now become a quiescent one. The accretion flow in this source 
partially resolvable by the Event Horizon Telescope \citep[EHT;][]{Bandyopadhyay2019}.

\citet{Gallimore2006} observed faint radio emission \citep[highly polarised;][]{Bajaja1988} obliquely oriented along the galactic minor axis which is suggested to be arising from the weak KSR. Our sensitive observations also show a stretch of radio emission along the galactic plane similar to that observed by \citet{Kharb2016,Yang2024}, which is supposedly coming due to synchrotron emission from the previous supernova activity. The Seyfert emission lies perpendicular to the galactic disk. We detect a flat spectrum ($-0.13\pm0.06$) core of $112\pm6$ Jy consistent with the findings by \citet{Hummel1984}. In Figure \ref{fig:NGC4594fpol&spix} ({bottom panel}), the steepening of the spectrum towards the north probably indicates the presence of an outflow in that direction. The core is polarized and the B-field directions are oblique to the extended outflow (see Figure \ref{fig:NGC4594fpol&spix}; top panel). A sub-parsec-scale jet (brighter towards the north) is observed \citep{Hada2013, Mezcua2014} in an oblique direction, perpendicular to our inferred B-field direction, suggesting that the B-fields are toroidal in the jet.

\subsubsection{NGC 5506}
NGC 5506 is an edge-on spiral galaxy hosting a Seyfert nucleus. The nucleus was initially identified as a Seyfert type 2 by \citet{Rubin1978} but later classified as an NLS1 by \citet{Nagar2002a} and as a Seyfert 1.9 by \citet{Gallimore2006, Fischer2013}. 

Our observations of NGC 5506 show a diffuse radio halo surrounding the core. It shows very faint polarized structures (f$_p$ = $0.5\pm0.1\%$) away from the central region. The directions of the EVPAs as shown in Figure \ref{fig:NGC5506fpol&spix} (left panel) appear to vary across different regions and appear to rotate around the core's circumference, indicating a complex polarized structure, as also observed by \citet{Sebastian2020}. This complexity may result from multiple individual components within the unresolved core. The inferred B-field orientation in an ordered radial pattern could suggest jet precession, varying jet orientations at different scales, or a toroidal B-field at the jet base. 

\begin{figure*}
\centering
\includegraphics[width=10cm]{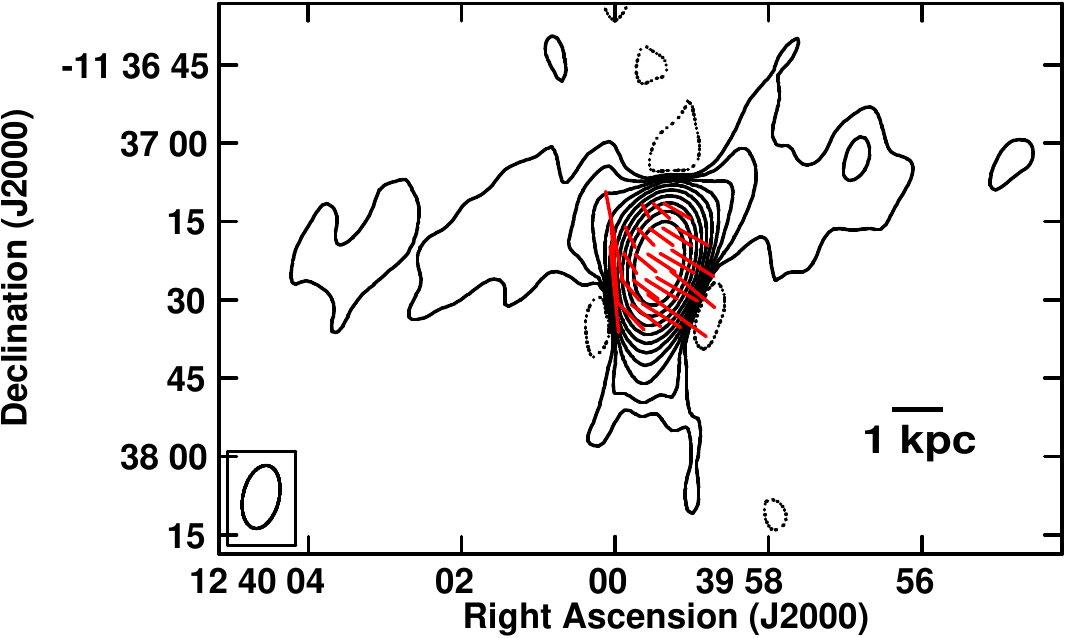}
\includegraphics[width=10cm]{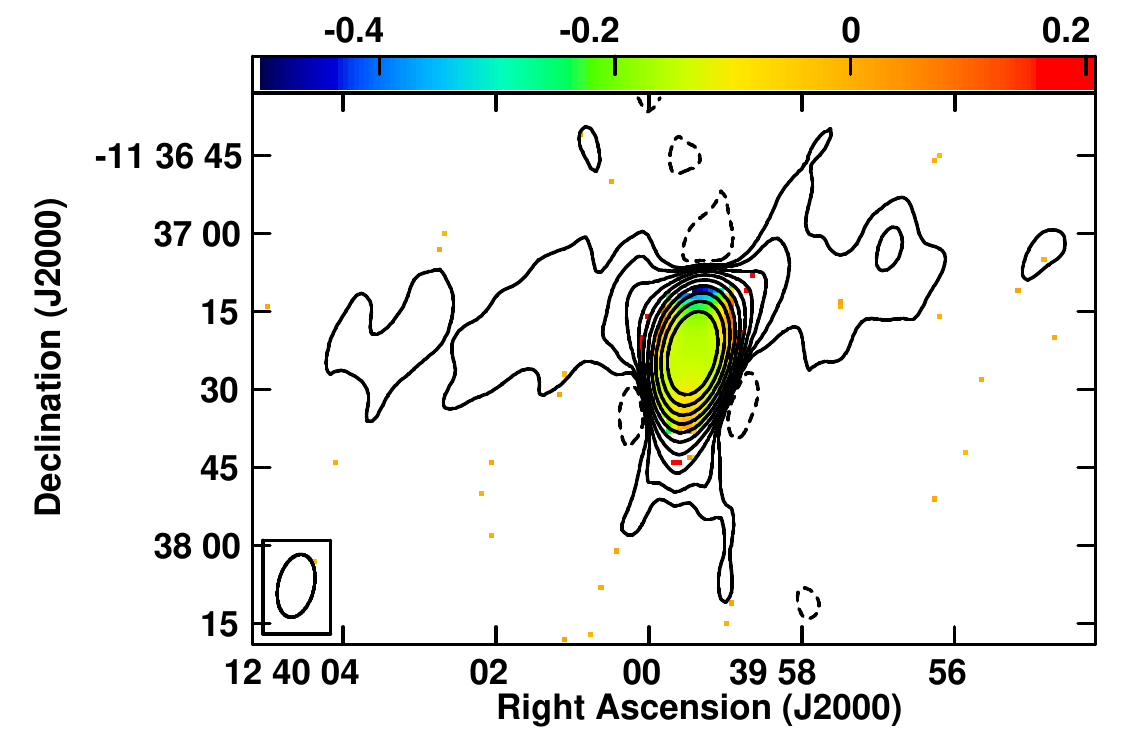}
\caption{\small (Top) NGC 4594 image at 10 GHz with the VLA D array. {\it Galactic radio emission is detected along with the KSR in this galaxy.} For all the panels the contour levels are at 3$\sigma \times(-1, 1, 2, 4, 8, 16, 32, 64, 128, 256, 512)$ with $\sigma = 22~\mu$Jy beam$^{-1}$. Similarly, the synthesized beam size is $12.26\arcsec\times6.90\arcsec$ at a PA of $-12.96\degr$. The ticks are proportional to fractional polarization; $10\arcsec = 1$\%. (Bottom) The VLA in-band spectral index image of NGC 4594. The spectral index values are in colour ranging from $-1.5$ to $+0.2$.}
    \label{fig:NGC4594fpol&spix}
\end{figure*}
The spectral index image (see Figure \ref{fig:NGC5506fpol&spix}; {right panel}) does not reveal any particular directionality of the outflow, rather the absence of a flat core signifies the presence of unresolved outflow at this resolution. 
Our observations reveal extended emissions towards the south, also marked by a higher degree of polarization (f$_p=19\pm5\%$) and a spectral flattening {(disregarding the edge pixels $\alpha\sim-0.2$)} at its base. It is interesting to note that this southern extension is also observed in the [OIII]$\lambda$5007 image presented by \citet{Esposito2024} 
which is suggested to be arising due to the biconical outflow as observed in NLRs.

\begin{figure*}
\centering
\includegraphics[width=8.2cm]{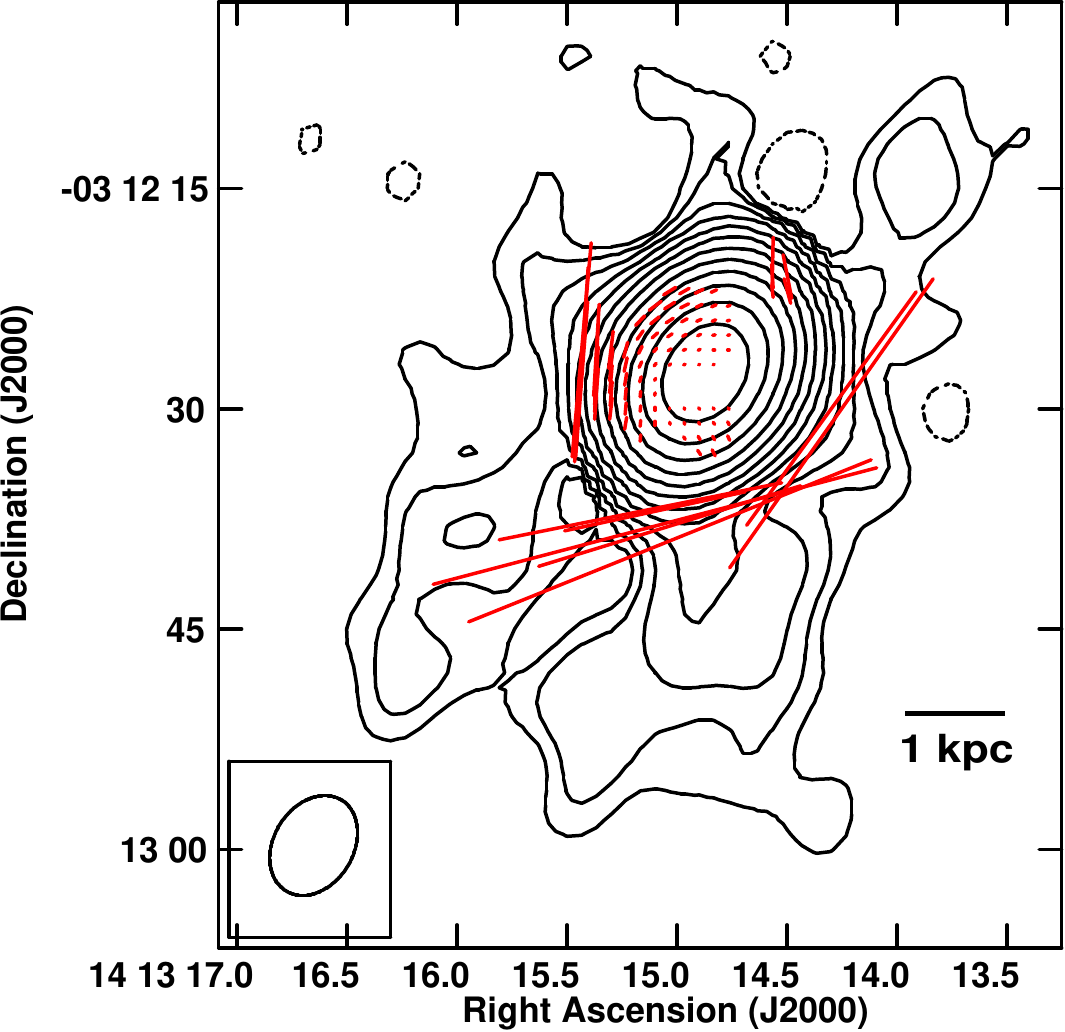}
\includegraphics[width=8cm]{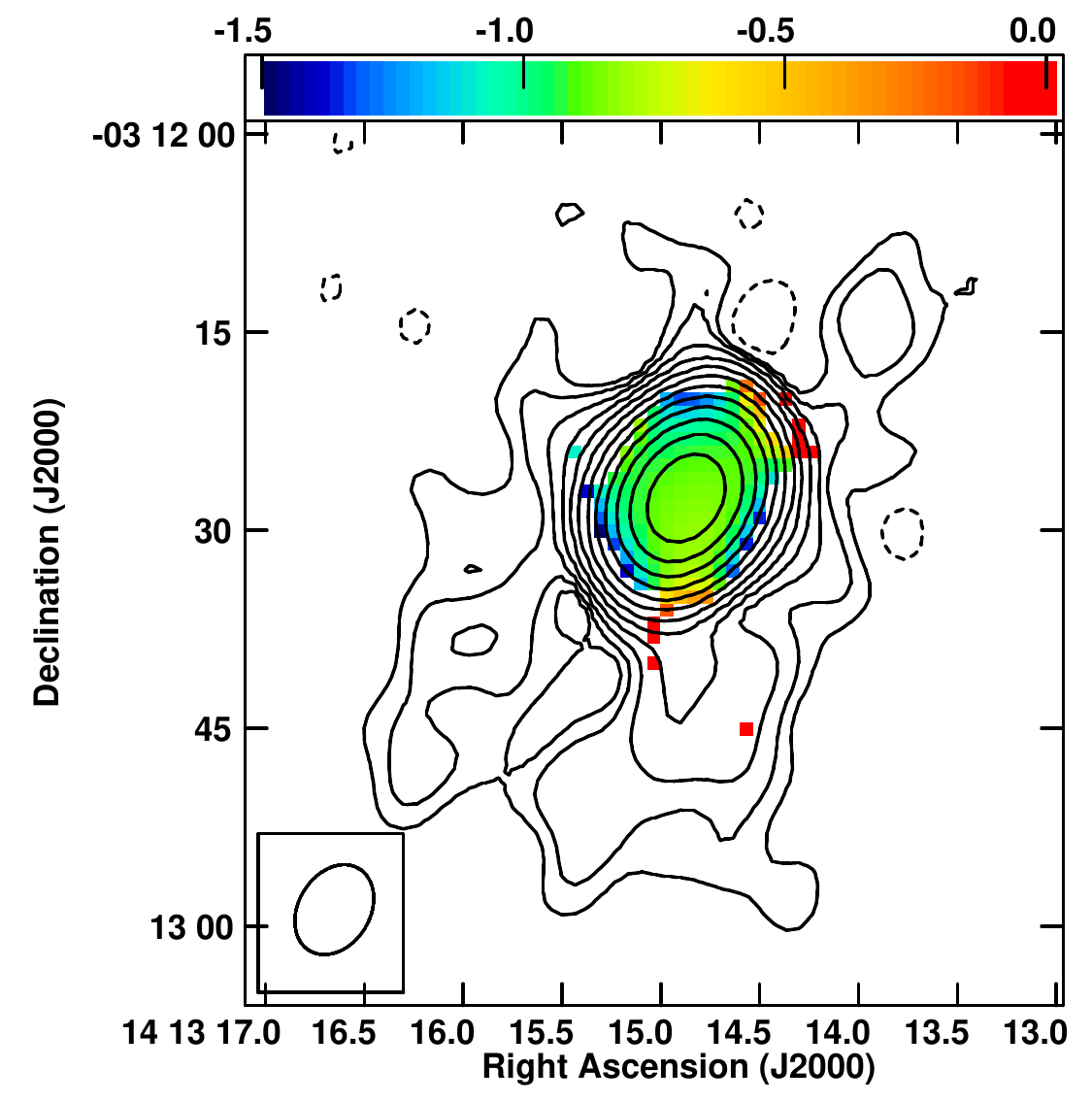}
\caption{\small (Left) NGC 5506 image at 10 GHz with the VLA D array. For all the panels the contour levels are at 3$\sigma \times(-1, 1, 2, 4, 8, 16, 32, 64, 128, 256, 512, 1024, 2048)$ with $\sigma = 9~\mu$Jy beam$^{-1}$. Similarly, the synthesized beam size is $7.25\arcsec\times5.42\arcsec$ at a PA of $-31.47\degr$. The tick lengths are proportional to fractional polarization; $5\arcsec = 2.5$\%. (Right) In-band spectral index image of NGC 5506 centred at 10~GHz. The spectral index values are in colour ranging from $-1.5$ to 0. }
    \label{fig:NGC5506fpol&spix}
\end{figure*}

\begin{table*}
\small{
\caption{Jet power estimates}
\tabcolsep=0.1cm
\begin{center}
\begin{tabular}{ccccc}
\hline \hline
Source&$L_{\rm{core,10GHz}}$&$\alpha_C$&$\log P_{\rm{jet,rad}}$ (erg s$^{-1}$)&$\log P_{\rm{jet,kin}}$ (erg s$^{-1}$)\\
name &(10$^{28}$ W Hz$^{-1}$ sr$^{-1}$)&&\citep{Foschini2014}& \citep{Merloni2002}\\ \hline
NGC 1068&0.35&$-0.371\pm0.04$&$40\pm3$&$43\pm6$\\
NGC 1320&9.46&$-0.808\pm0.001$&$41\pm3$&$42\pm6$\\
NGC 2639&15.51&$0.247\pm0.003$&$42\pm3$&$43\pm6$\\
NGC 2992&4.81&$-0.773\pm0.006$&$41\pm3$&$43\pm6$\\
NGC 3079&4.88&$-0.821\pm0.002$&$41\pm3$&$43\pm6$\\
NGC 3516&0.74&$-0.44\pm0.03$&$41\pm3$&$42\pm6$\\
NGC 4051&0.05&$-0.80\pm0.04$&$40\pm3$&$41\pm6$\\
NGC 4235&0.82&$0.11\pm0.02$&$41\pm3$&$42\pm6$\\
NGC 4388&1.96&$-0.78\pm0.01$&$41\pm3$&$43\pm6$\\
NGC 4593&0.372&$0.70\pm0.05$&$40\pm3$&$42\pm6$\\
NGC 4594&3.48&$-0.128\pm0.005$&$41\pm3$&$43\pm6$\\
NGC 5506&9.06&$-0.9\pm0.2$&$41\pm3$&$43\pm6$\\
\hline
\end{tabular}
\end{center}

{\small Column 1: Names of the sources. Column 2: Core luminosity in the units of 10$^{28}$ erg s$^{-1}$ Hz$^{-1}$ Sr$^{-1}$ at 10 GHz using  $4\pi D_L^2 S_{\rm{peak,10GHz}}/(1+z)^{1+\alpha_C}$, where $S_{\rm{peak,10GHz}}$ is the peak brightness in units of mJy~beam$^{-1}$ with a beam area of 1 unit. Column 3: Average in-band spectral index evaluated at the core (highest one or two contour levels).
Column 4: Logarithm of jet radiative power in erg~s$^{-1}$ using relations from \citet{Foschini2014}. Column 5: Logarithm of jet kinetic power in erg~s$^{-1}$ using relations from \citet{Merloni2007}.}
}
\label{tab:tab5}
\end{table*}

\begin{table*}

\small{
\caption{Summary of the Energetics}
\tabcolsep=0.11cm
\begin{center}
\begin{tabular}{lcccccc}
\hline\hline
{Source}&$\alpha$&$S_C$&$L_\mathrm{rad}$ &$B_\mathrm{min}$ &$E_\mathrm{total}$&$\tau$ \\  
name & (in-band) &(mJy)&(10$^{38}$~erg~s$^{-1}$)&(10$^{-6}$ G)&(10$^{54}$ ergs)&(10$^7$ yr)\\\hline
NGC 1068&$-0.85\pm0.01$&$365\pm18$&$40\pm2$&$63\pm7$&$1.3\pm0.2$&$0.052\pm0.009$\\
NGC 1320&$-0.37\pm0.04$&$1.9\pm0.1$&$0.76\pm0.06$&$6.2\pm0.6$&$0.38\pm0.05$&$1.2\pm0.1$\\
NGC 2639&$0.28\pm0.09$ &$54\pm3$&$21\pm1$&$5.4\pm0.3$&$4.3\pm0.4$&$1.31\pm0.05$\\
NGC 2992&$-0.78\pm0.01$ &$42.0\pm2.1$&$19\pm1$&$8.5\pm0.4$&$12.4\pm0.8$&$0.86\pm0.04$\\
NGC 3079&$-0.9\pm0.2$&$172\pm9$&$23\pm1$&$33\pm3$&$2.4\pm0.3$&$0.14\pm0.02$\\
NGC 3516&$-0.66\pm0.04$&$3.5\pm0.2$&$1.8\pm0.1$&$7.1\pm0.6$&$1.2\pm0.1$&$1.02\pm0.08$\\
NGC 4051&$-0.73\pm0.07$ &$5.2\pm0.2$&$0.23\pm0.02$&$13\pm1$&$0.08\pm0.01$&$0.52\pm0.07$\\
NGC 4235&$-0.03\pm0.1$ &$6.3\pm0.3$&$1.3\pm0.1$&$2.7\pm0.1$&$1.5\pm0.1$&$1.793\pm0.007$\\
NGC 4388&$-0.88\pm0.02$ &$15.5\pm0.8$&$9.4\pm0.6$&$11.6\pm0.9$&$4.5\pm0.5$&$6.2\pm0.6$\\
NGC 4593&$0.5\pm0.2$ &$2.5\pm0.2$&$0.51\pm0.04$&$1.62\pm0.08$&$0.32\pm0.03$&$1.75\pm0.02$\\
NGC 4594&$-0.13\pm0.06$ &$112\pm6$&$5.6\pm0.4$&$6.7\pm0.3$&$1.9\pm0.1$&$1.10\pm0.04$\\
NGC 5506&$-0.9\pm0.2$ &$108\pm5$&$38\pm15$&$15\pm2$&$14\pm3$&$0.4\pm0.07$\\

\hline
\end{tabular}
\end{center}

{\small Column 1: Names of the sources. For the following columns estimates have been obtained for the core and surrounding diffuse emission excluding the extended jets/lobes. Column 2: In-band spectral index from either VLA $8-12$ GHz or $1-2$ GHz VLA data. Column 3: Flux density in mJy at 10 GHz. Column 4: Radio luminosity in unit of 10$^{38}$~erg~s$^{-1}$ using $S_C$ and $\alpha$. Column 5: Magnetic field value in the $\mu$G at minimum pressure under equipartition condition. Column 6: Total energy in units of 10$^{54}$ erg, including particle energy and field energy, at minimum energy. Column 7: Electron lifetimes in the units of 10 Myr due to synchrotron and IC CMB losses.}}
\label{tab:tab4}
\end{table*}

\subsection{Jet Powers}\label{sec:jetpower}
The power transported by the radio jets in Seyferts and LINERs, either kinetically or radiatively, can provide us a measure of the power ejected by the central engine. It is expected that radio core emission and the kpc-scale jet kinetic power must be correlated since both arise from the jets \citep{Punsly2005, Merloni2007, Cavagnolo2010}. Therefore, the radio luminosity of the core becomes an important tool to probe the energy transported from the AGN to the jet as it propagates in the surrounding medium. Theoretical models for AGN with conical jets predict that the radio luminosity of the flat-spectrum optically thick radio core emission depends on the jet power as $L_{\rm{core}}\propto P_{\rm{jet}}^{17/12}$ \citep{Blandford1979, Heinz2003}. Later, observational measurements gave best-suited empirical relations between the core radio luminosity and jet power. \citet{Foschini2014} found a correlation between the observed core radio power and the jet power obtained from spectral energy distribution (SED) modeling by \citet{Ghisellini2009}:
\begin{equation}
    \log P_{\rm{jet,rad}}=(0.75\pm0.04)\log (\nu L_{\rm{core,15GHz}})+(12\pm2)
    \label{eq:PJrad_Foschini}
\end{equation}
\begin{equation}
    \log P_{\rm{jet,kin}}=(0.90\pm0.04)\log (\nu L_{\rm{core,15GHz})}+(6\pm2)
    \label{eq:PJkin_Foschini}
\end{equation}
where, $P_{\rm{jet,rad}}$ is the radiative jet power, $P_{\rm{jet,kin}}$ is the kinetic (electrons, protons, and B-field) jet power, and $L_{\rm{core,15GHz}}$ the radio core luminosity at 15 GHz. Though the coefficient does not entirely match the theoretical value, {the values are within 2$\sigma$ of each other.} The $L_{\rm{core,15GHz}}$ is calculated from 10~GHz core (peak) flux ($S_{\rm{peak}}$) as $L_{\rm{core,\nu}}=4\pi D_L^2 S_{\rm{peak,\nu}}/(1+z)^{1+\alpha}$, and $S_{\rm{peak,\nu}}=(\nu/10)^{\alpha_C}\times S_{\rm{peak}}$, considering K-correction, $\nu$ as 15~GHz, and $\alpha$ as the spectral index at the core. {These relations have been derived for high-resolution (mostly VLBI) data but have also been applied for single dish observations \citep[e.g.,][]{Angelakis2015}. Typically only a fraction of the flux densities at VLA scales can be recovered at VLBI scales \citep[see][]{Orienti2010, Kharb2021}. Since our estimates are based on VLA core flux densities, the jet powers estimated here could be an overestimate.} Following the equations \ref{eq:PJrad_Foschini} and \ref{eq:PJkin_Foschini}, the derived values of ${\rm{log}} P_{\rm{jet,rad}}$, ${\rm{log}} P_{\rm{jet,kin}}$ have been tabulated in Table \ref{tab:tab5}. 

Since none of these relations were estimated for low-luminosity radio-quiet AGN like Seyferts and LINERs, we calculated the jet powers using other empirical relations available in the literature to check for consistency within the error bars. \citet{Merloni2007} provided a relation between the jet kinetic power, estimated from the work done to inflate cavities in the surrounding X-ray emitting gas \citep[see][]{Allen2006, Rafferty2006, Cavagnolo2010}, and the radio core luminosity at 5 GHz for their sample of {low-luminosity radiatively inefficiently accreting sources hosting relativistic jets (e.g., FRI radio galaxies)} as
\begin{equation}
    \log P_{\rm{jet, kin}}=(0.81\pm0.11)\log L_{\rm{core,5GHz}}+11.9^{+4.1}_{4.4}
\end{equation}
Using this equation and the radio core luminosity at 5~GHz, derived from the 10~GHz peak flux using the corresponding $\alpha$, the jet kinetic powers have been tabulated in Table~\ref{tab:tab5}. Both \citet{Foschini2014} and \citet{Merloni2007} provide similar estimates of jet powers from the core luminosities within errors.

Additionally, we have estimated the jet bulk kinetic power using empirical relation given by \citet{Willott1999} \citep[also see][]{Rawlings1991,Falcke1995} where they found a tight correlation among the bulk kinetic power and narrow line region luminosity for radio-loud FRII sources. The relation is given as:
\begin{equation}
    P_{\rm{jet,kin}}\approx 3\times 10^{38}L_{151}^{6/7}\, {\rm{W}},
    \label{eq:willottjetpower}
\end{equation}
where $L_{151}$ is the lobe radio luminosity at 151~MHz in the units of 10$^{28}$~W~Hz$^{-1}$~Sr$^{-1}$. {We note that jet powers derived using 151 MHz luminosities from the TIFR GMRT Sky Survey (TGSS) and equation \ref{eq:willottjetpower} above yielded similar numbers as from the other two methods.}


{Our selected sources being low luminosity and radiatively inefficiently accreting sources, though the presence of relativistic jets is not yet confirmed for all of them, \citet{Merloni2007} relations remain best-suited for determining the jet powers as used for other RQ sources or Seyferts \citep[e.g.,][]{Doi2012, Zakamska2014,Silpa2021}.} 
The total jet power ($P_{\rm{jet}}$) is computed as the sum of $P_{\rm{jet,kin}}$ and $P_{\rm{jet,rad}}$ \citep[see][]{Angelakis2015}. The estimated jet powers for our sample of Seyferts and LINERs are about $2-4$ orders lower than the median of RL jets' powers \citep[$>10^{45}$~erg~s$^{-1}$;][]{Liu2006,Croston2018, Fan2019}. 
This suggests that apart from the very dense environment hindering the jet propagation for these sources, intrinsically these jets are low-powered consistent with inefficient accretion in these systems \citep{Heckman2014,Berton2020}.

\subsection{Energetics}\label{energetics}
Assuming equipartition between relativistic particles and magnetic fields, we could estimate the total radio luminosity ($L_\mathrm{rad}$), magnetic field strength ($B_\mathrm{min}$), total energy ($E_\mathrm{Total}$) of the particle and fields, and electron/particle lifetimes including synchrotron and IC CMB losses ($\tau$), following the relations in \citet{OdeaOwen1987,vanderlaan1969}. In these calculations, the ratio of the ion energy to the electron energy ($k$) and the volume filling factor ($\phi$) are both assumed to be unity. The upper and lower cutoff frequencies ($\nu_u$ and $\nu_l$) are assumed to be 15~GHz and 100~MHz, respectively. The value of the function $c_{12}$ in the equipartition calculation is obtained from \citet{Pacholczyk1970}. A cylindrical or spherical volume (for sources with circular morphology) is assumed for the sources. The equipartition estimates are presented in Table~\ref{tab:tab3}. {These estimates are similar to those presented for other Seyfert galaxies and RQ AGN in the literature \citep[e.g.,][]{Kharb2016,Silpa2021MNRAS,rao2023}.} {The errors in $B_\mathrm{min}$ and $E_\mathrm{Total}$ have been estimated by propagating errors from $\alpha$ and $S_C$.} The B-field strengths in these sources range from $1-3~\mu$G, except for NGC 1068, which exhibits a high B-field of approximately $100~\mu$G, likely influenced by galactic B-field contributions. The total energy of these sources spans from $10^{52}$ to $10^{55}$~ergs, and the lifetimes of the relativistic particles range from $10^5$ to $10^7$ yr.

\subsection{Correlations}\label{sec:correlations}
{We present here the results of correlations between several global properties of our Seyfert and LINER sample. Due to the small sample size and to avoid biases due to outliers, we have primarily relied on Kendall's Tau (KT) correlation test. We have also performed partial correlation tests to look for the dominance of a given parameter by accounting for the effects of others using the \textsc{python} package \textsc{pingouin} for Spearman's rank (SR) partial correlation test. To account for the upper limits, \textsc{fortran} routines in the \textsc{asurv} package \citep{asurv1992} have been used. Survival analysis tests like the generalized KT test \citep{generalizedKT1973} was used for all the censored datasets. We note that discrepancies can arise in comparisons due to the varying methodologies used in deriving the literature values for these AGN. For instance, M$_{\rm{BH}}$ for most of type 1 AGN in this study has been estimated using the most reliable reverberation mapping technique. In contrast, type 2 AGN rely on more indirect methods, such as the black hole mass–stellar velocity dispersion relation or the BLR size–optical luminosity relation \citep[see][]{Woo2002}. This bias in the estimate may give rise to discrepancies in the correlations. In the cases of obvious outliers, we have assessed whether the correlations persist or change after removing outliers. For correlations involving polarization estimates, we have only considered the 10 sources with 10~GHz data for uniformity.}

Our sample of Seyferts and LINERs are RQ sources and fall on the lower BH mass end compared to RL AGN \citep{Laor2000,Sikora2007}. {We find a strong correlation between ${R}$ and the SMBH masses (KT test probability, p = 0.006, r = 0.63). The trend shown in Figure~\ref{fig:RvsMBH} follows a much tighter correlation than was obtained for a sample of Seyferts, LINERs, RL quasars and broad-line radio galaxies by \citet{Sikora2007}, where ${R}$ was as defined in \citet{Kellermann1989}. }

\begin{figure*}
    \centering
    \includegraphics[width=10cm]{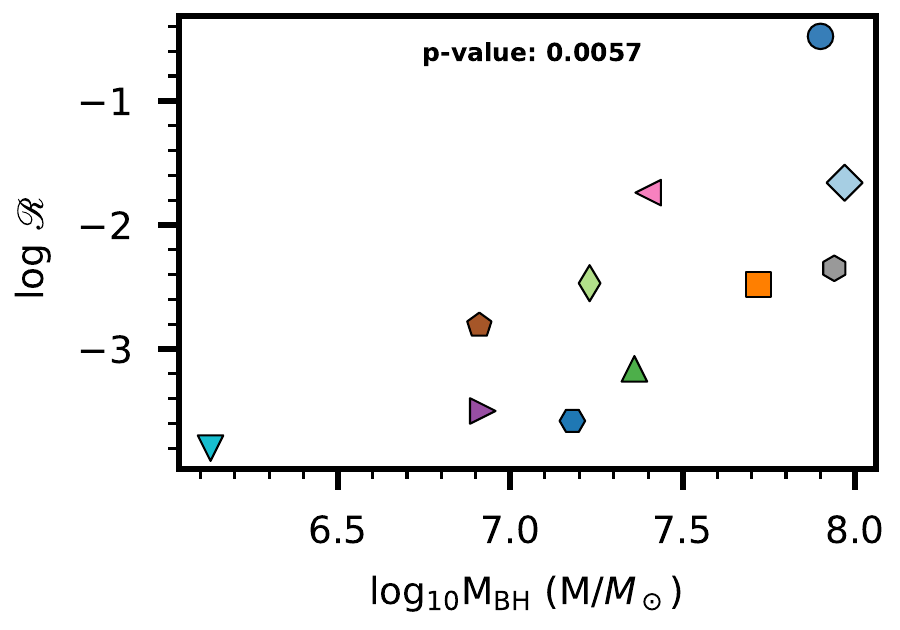}
\caption{\small Radio loudness parameter versus black hole masses for 11 out of 12 sources in the selected sample. We did not have a proper estimate of radio loudness for NGC 4594  due to large uncertainties in the  [OIV]$\lambda$25.89 $\mu$m line luminosity and has thus been excluded. A strong correlation is found between ${R}$ and M$_{BH}$ for these sources (KT test with p = 0.0057). }
    \label{fig:RvsMBH}
\end{figure*}
Figure~\ref{fig:totlumvscorelum} (left panel) shows that most sources in our sub-sample are core-dominated. However, a few sources exhibit {luminous radio} jets or lobes, viz., NGC 1068, NGC 4388, NGC 2992, NGC 3079, and NGC 3516 (listed in decreasing order of their lobe luminosities relative to their core luminosities). The total luminosity of NGC 1068 at 10 GHz is notably high, possibly due to the extrapolation from the 1.4~GHz value using the in-band $\alpha$ value, the detection of more resolved components at 1.4 GHz with VLA BnA$\rightarrow$A array observations, or contamination from galactic emission (though galactic contributions have been subtracted, some residuals may remain). {The core luminosity at 10~GHz for the sample is found to be weakly correlated (KT test p = 0.03) with the black hole masses as shown in the right panel of Figure~\ref{fig:totlumvscorelum}.} 

\begin{figure*}
\centering
\includegraphics[width=8cm]{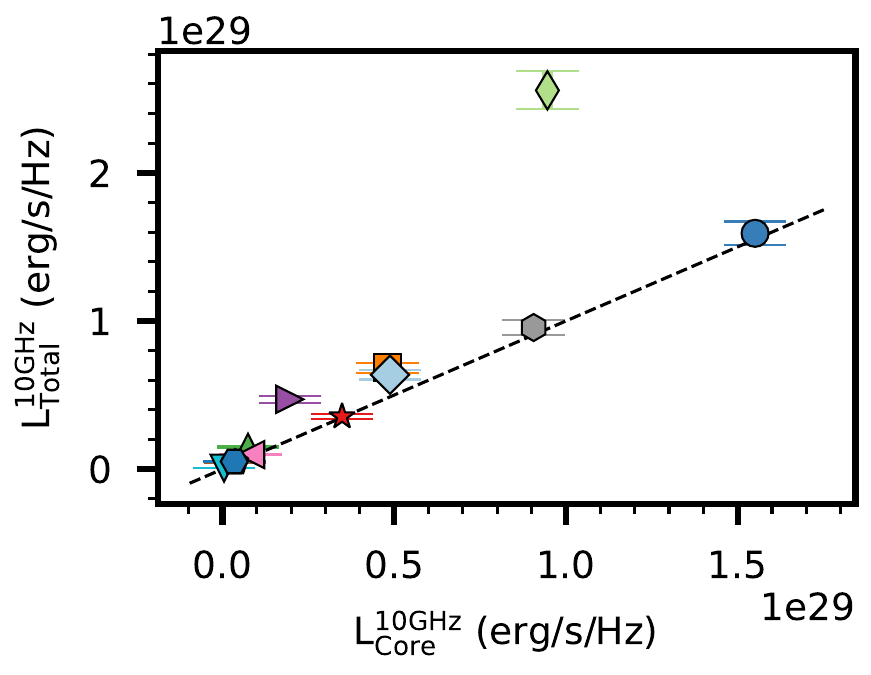}
\includegraphics[width=8.6cm]{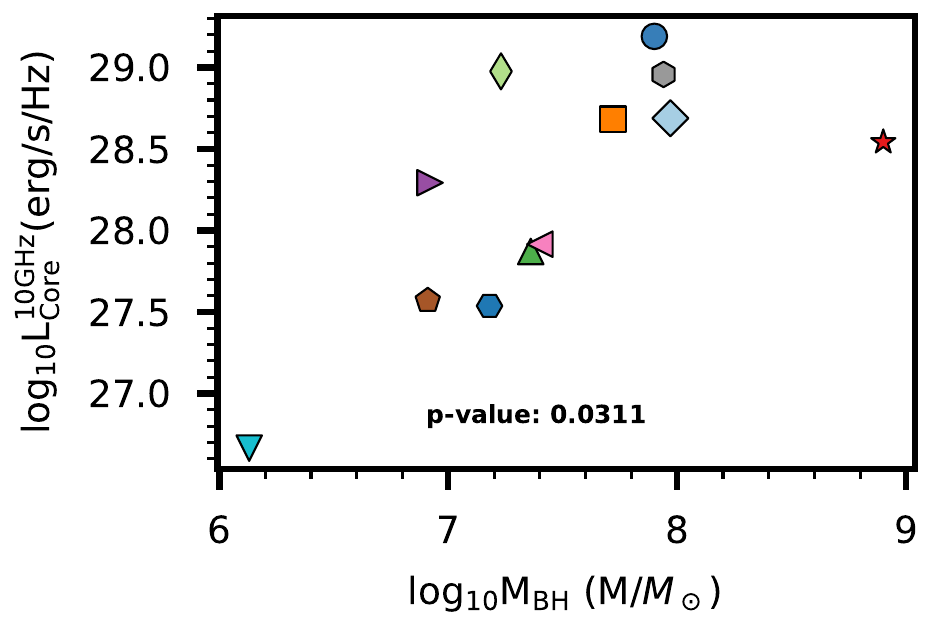}
\caption{\small (Left) Total luminosity versus core luminosity at 10~GHz. The black dashed line indicates the slope equal to unity. 
(Right) Core luminosity at 10 GHz versus black hole mass. A marginal correlation is observed with KT test p = 0.0311. {Error bars are noted along with the symbols}.}
\label{fig:totlumvscorelum}
\end{figure*}

{Figure~\ref{fig:extfluxvscoreflux} (left panel) demonstrates a significant correlation between the polarized flux density and the core flux density at 10 GHz (KT test, p = 0.0041). For comparing the polarised core and total flux densities, we have used average values within regions that are approximately the size of the synthesized beam at the centre.}
Higher polarized flux density typically arises from organized B-fields in the synchrotron plasma. Intrinsically randomized B-fields can reduce overall polarized intensity. Additionally, external Faraday rotating media can decrease polarized flux density by randomizing the polarization angles. The observed correlation between total intensity and polarized intensity at the core suggests that sources with more organized B-fields tend to be more luminous in radio.

{
Although polarized emission is typically not detected in the core regions, it is detected with a high SNR in the extended regions ($\sim$kpc scales) for all sources. To calculate the polarized flux densities in the extended region, we subtracted the core's contribution from the total polarized emission. The KT test shows a moderate correlation with p = 0.019 with the core flux density (see Table~\ref{tab:tab6}). The f$_p$ indicator (see Figure~\ref{fig:extfluxvscoreflux}, right panel) shows that the extended structures are highly polarized, suggesting a dependence of the kpc-scale B-field ordering in the jets or lobes of these AGN.}

\begin{figure*}
\centering
\includegraphics[width=8cm]{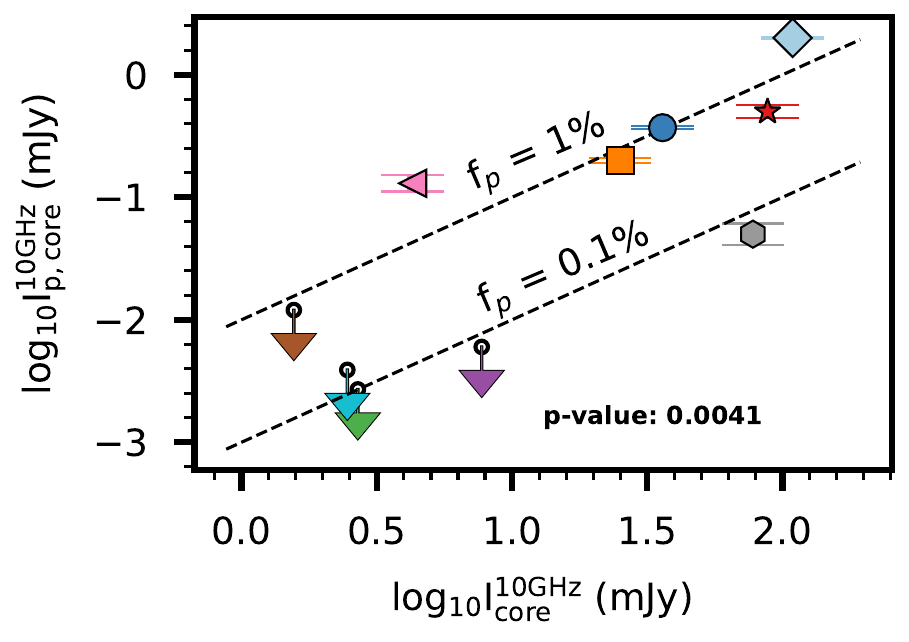}
\includegraphics[width=8cm]{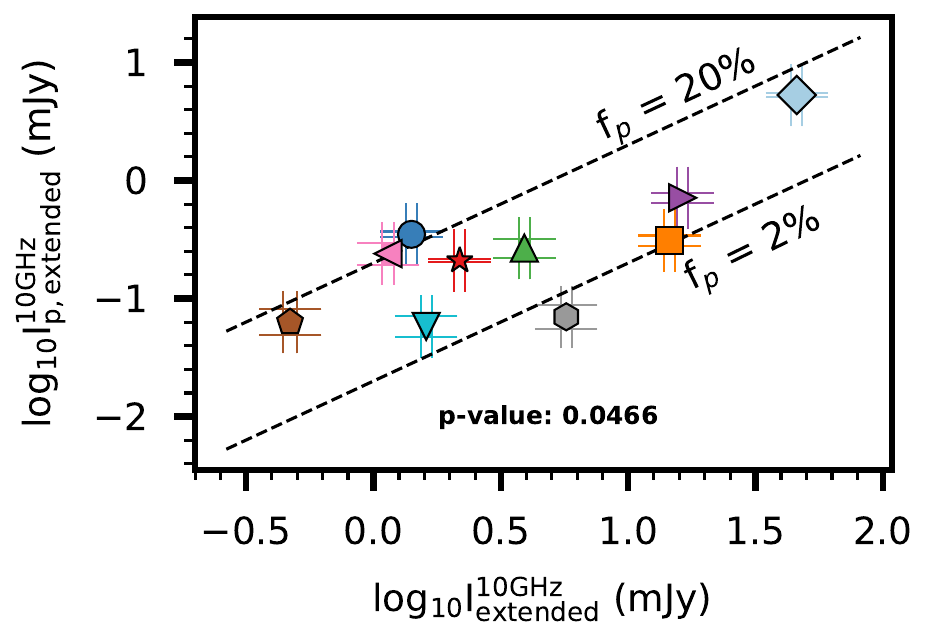}
{\caption{\small (Left) Logarithm of polarized flux density versus logarithm of total flux density in the core at 10~GHz. The errors on log$_{10}$y-value have been calculated as (errors on y-value)/(y-value$\times$ log$_e$10) using error propagation methods. The generalized KT test indicates a significant correlation with p = 0.0041. The dashed lines show the various fractional polarization values. (Right) Logarithm of polarized flux density from extended emission (excluding the core) versus the logarithm of total flux density at 10~GHz. The KT test shows a mild correlation with p = 0.0466. Error bars are noted along with the symbols, and upper limits are noted as downward-facing arrows.}}
\label{fig:extfluxvscoreflux}
\end{figure*}
The Eddington luminosity for the sources has been obtained using the relation \citep{Rybicki1979}:
\begin{equation}
     L_{\rm{Edd}}(\rm{in\,erg/s})=1.26\times10^{38}M_{\rm{BH}}/M_\odot
\end{equation}

\begin{figure*}
\centering
    \includegraphics[width=10cm]{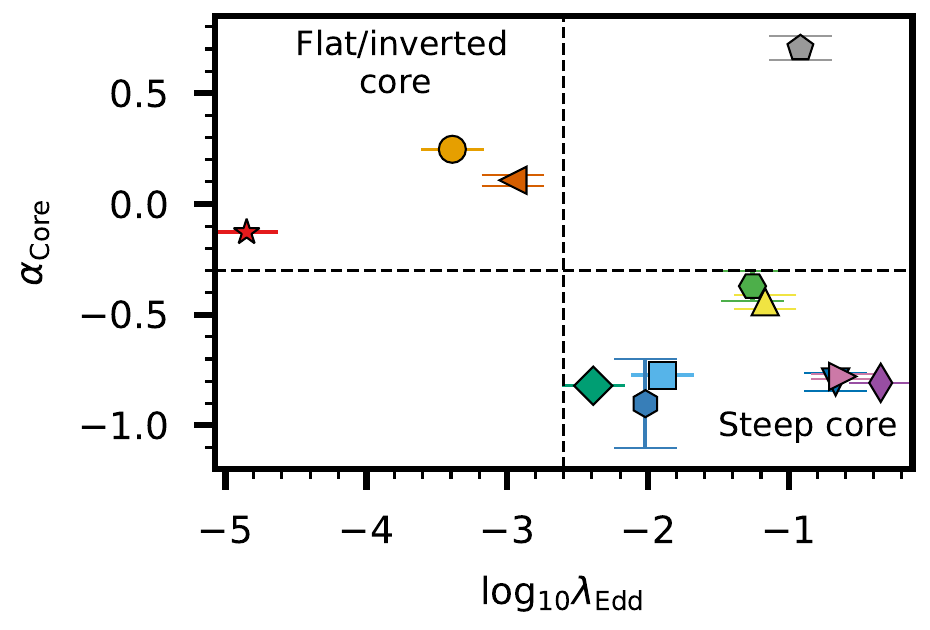}
\caption{\small In-band core spectral indices versus the Eddington ratios. The horizontal black dashed line indicates a spectral index of $-0.3$. The vertical dashed line indicates the central value ([minimum+maximum]/2)of $\lambda_{Edd}$. Values greater than $\alpha=-0.3$ are considered flat while lower values are considered steep. {Error bars are noted along with the symbols.}}
\label{fig:SpixvsLambdaedd}
\end{figure*}

We further obtained the Eddington ratio as $\lambda_{\rm{Edd}}=L_{\rm{bol}}/L_{\rm{Edd}}$, where $L_{\rm{bol}}$ is obtained as in Table \ref{tab:tab1}. Since we do not have direct estimates of the accretion rates for these sources, we assume $\lambda_{\rm{Edd}}$ to represent the accretion scenario. We compared the values obtained from our radio-polarimetric observations with the intrinsic properties of the central engine to identify the primary factors responsible for producing KSRs from RQ nuclei. 

\begin{figure*}
\centering{
\includegraphics[width=8cm]{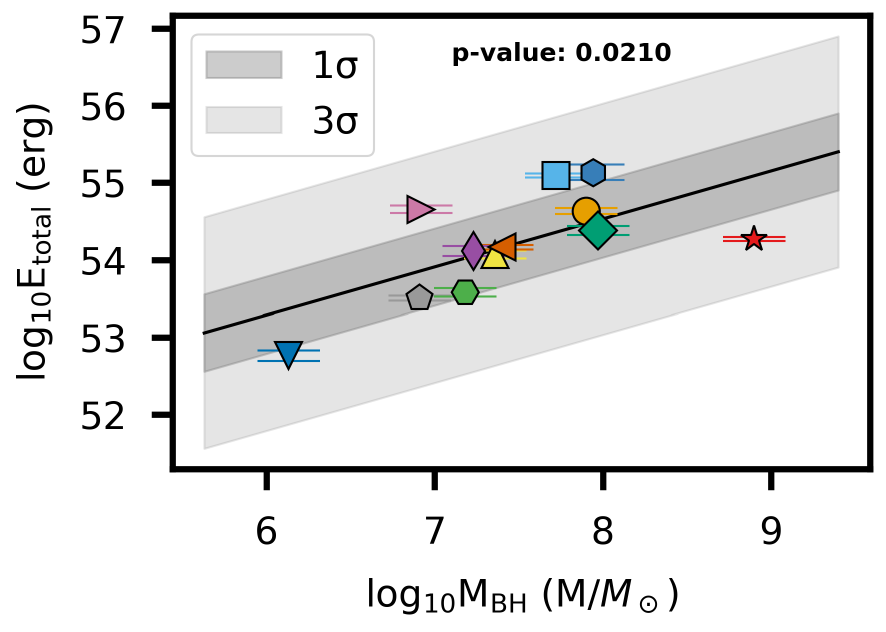}
\includegraphics[width=8cm]{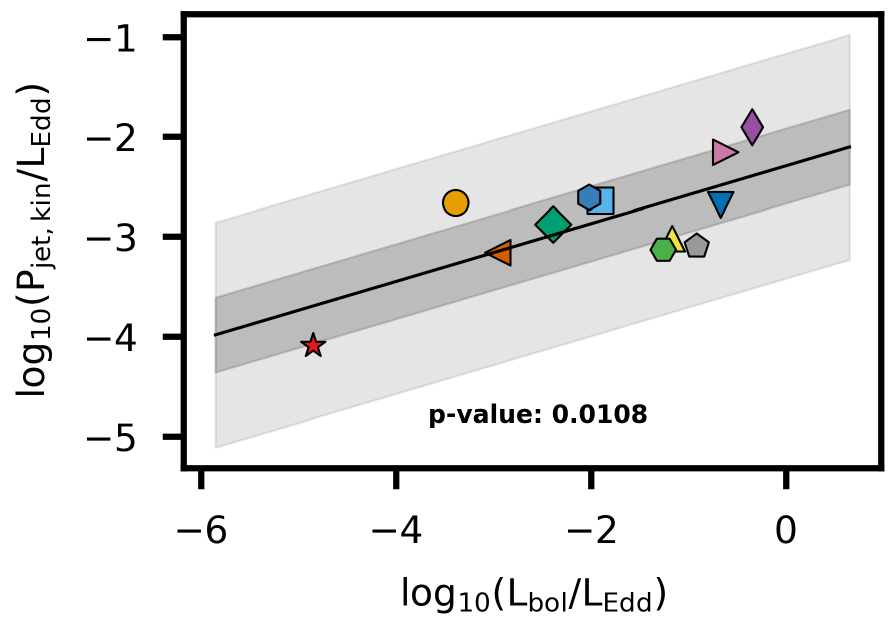}}
\caption{\small (Left) Total energy (particle + fields) versus the black hole mass in log-log scale. The KT test indicates a marginal correlation (p = 0.0210). A linear regression model has been fit here to obtain the best-fit line shown in black. The $1\sigma$, $3\sigma$ and $5\sigma$ standard deviations are shown with filled regions of different transparencies. (Right) The ratio of kinetic jet power to Eddington luminosity versus the ratio of bolometric luminosity to Eddington luminosity in log-log scale. The KT test suggests a weak correlation (p = 0.0628). A linear regression model with the best-fit line shown in black suggests a slightly stronger correlation (p = 0.0108). The 1$\sigma$ and 3$\sigma$ standard deviations are shown with filled regions of different transparencies. {Error bars are noted along with the symbols.}}
\label{fig:totalenergyvsmbh}
\end{figure*}

Figure~\ref{fig:SpixvsLambdaedd} shows the distribution of core spectral indices versus the Eddington ratios for the sources. The {in-band $\alpha$ image of NGC 4593 (Figure~\ref{fig:NGC4593fpol&spix}; right panel) is complex; apart from the likely presence of an inverted spectrum core slightly offset from the peak radio emission, a spectral steepening is observed in east-west direction north of the peak emission region, consistent with the small double-jetted source observed in the higher resolution image of \citet{Gallimore2004}. We observe a bimodality in the spectral index distribution with respect to Eddington ratios after excluding the source NGC~4593. Although we note that after combining the 3~GHz data from \citet{rao2023} and our work, the integrated $\alpha$ value obtained for NGC~4593 is $-0.48\pm0.04$, which would also make NGC~4593 fall in the bottom right quadrant along with the majority of the sources.} We have defined $\alpha>-0.3$ to be flat and $\alpha<-0.3$ to be a steep spectral index. We find that sources with higher $\lambda_{Edd} (>2.5\times10^{-3})$ have a steep spectrum core, whereas lower Eddington ratio sources appear to have a flat core. A similar trend has been observed for other RQ sources \citep[e.g.,][]{Laor2019}  suggesting a common physical driver linking these two quantities. Notably, most of the compact sources in our sample exhibit steep spectra. This could be due to presence of jets in these sources which are not resolved in our images \citep[e.g.,][]{ODea1998, Moran2000, ODea2021}.

We observe a marginal correlation between the total energy (particle + B-fields) and the black hole masses (see Figure \ref{fig:totalenergyvsmbh}, left panel). The best-fit line has a slope of 0.62 and an intercept of 49.27 with a standard deviation of 0.5. No correlation was found between other equipartition estimates and the intrinsic properties of the central engine. Since total energy is related to (radiative) jet power, this correlation suggests a dependence of jet power on black hole mass as has also been found by \citet{McLure2004, Chen2015} in RL sources. Additionally, we found a strong correlation (p = 0.0008 with partial SR test, see Table~\ref{tab:tab6}) between the jet kinetic power, derived using the method of \citet{Merloni2007}, and black hole mass \citep[see also][]{Liu2006}. This indicates that black hole mass is a major driver for jets in RQ AGN with KSRs.

The ratio of jet kinetic power to Eddington luminosity shows a weak correlation with the Eddington ratio (see Figure~\ref{fig:totalenergyvsmbh}, right panel). This relationship can be expressed as $P_{\rm{jet,kin}}/L_{\rm{Edd}} = (0.29 \pm 0.09) \lambda_{\rm{Edd}} - (2.29 \pm 0.21)$. This observed dependence has a shallower slope than that reported by \citet{Merloni2007}, who suggested a radiatively inefficient accretion flow as proposed by \citet{Fender2003}. {This dissimilarity may indicate that the Seyfert and LINER galaxies in our sample are not as dominated by relativistic jets as the low luminosity RL sources investigated by \citet{Merloni2007}.} Rather, only a fraction of the bolometric luminosity is converted into jet kinetic and radiative power in the RQ sources {consistent with \citet{Padovani2017}}.

\begin{figure*}
\centering
\includegraphics[width=8cm]{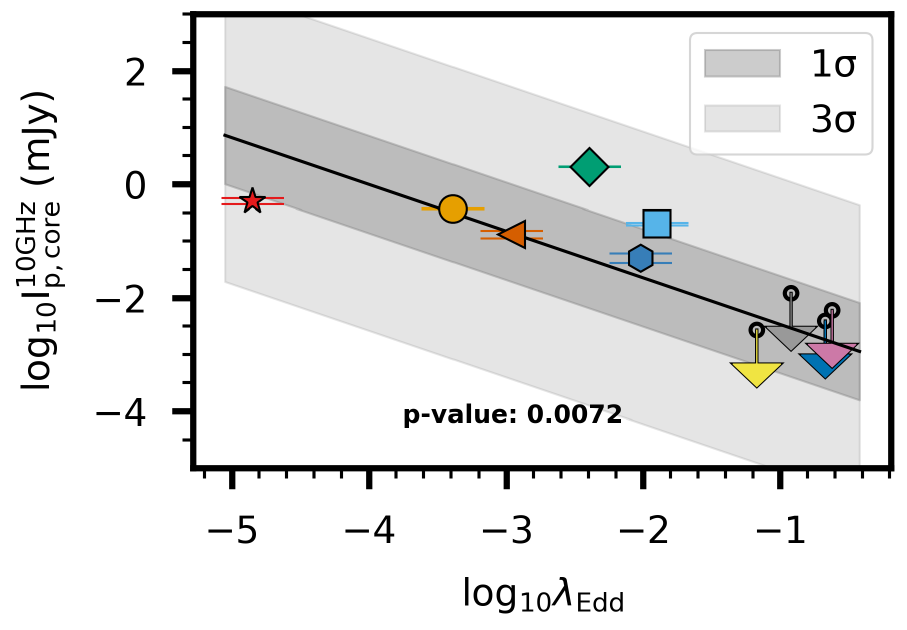}
\includegraphics[width=8cm]{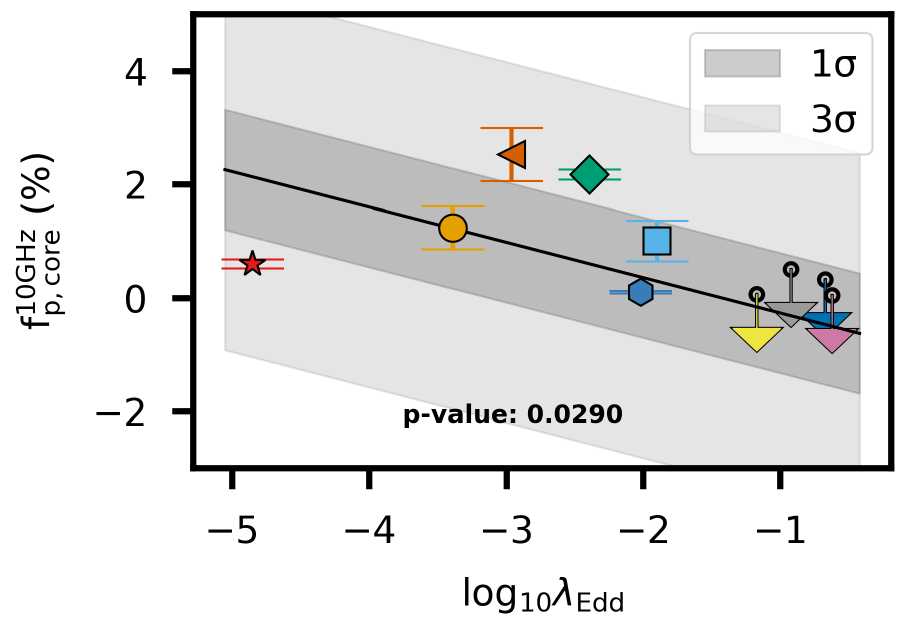}
\includegraphics[width=8cm]
{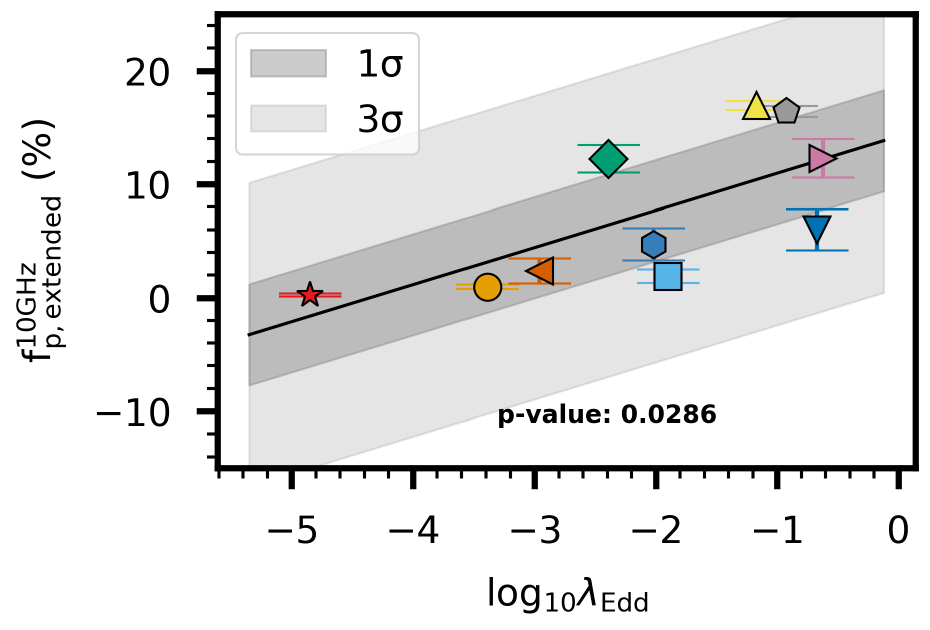}
{\caption{\small (Top left) Log-log plot of the core polarized flux density versus the Eddington ratios. The generalized KT test indicates a strong correlation (p = 0.0072). A linear regression model has been fit here to obtain the best-fit line shown in black. The 1$\sigma$ and 3$\sigma$ deviations are shown with filled regions of different transparencies. (Top right) Plot of core fractional polarization versus logarithm of the Eddington ratios. The generalized KT test indicates a correlation (p=0.0290). The linear regression model best-fit line is shown in black with parameters: slope $= -0.57$, intercept $= 0.08$, standard error = 0.19. The 3$\sigma$ and 5$\sigma$ standard deviations are shown with filled regions of different transparencies. (Bottom) Polarization fraction for the extended emission (excluding the core) versus logarithm of the Eddington ratios. The KT test indicates a weak positive correlation with p=0.0286. Error bars are noted along with the symbols, and the upper limits are noted as downward facing arrows.}}
\label{fig:corepolfluxvseddratio}
\end{figure*}

We observe a strong correlation between the core polarized flux density and Eddington ratios (Figure~\ref{fig:corepolfluxvseddratio}; {top} left panel; generalized KT test p = 0.007). The best-fit line has a slope of $-0.82$, intercept of $-3.29$ and a standard deviation of $0.85$. Notably, the core polarization intensity for NGC~3079 is significantly high, exceeding the $1\sigma$ deviation, which is consistent with its high total intensity at the core. To account for the bias from the total radio emission, we have also plotted the core polarization fraction against Eddington ratios (Figure~\ref{fig:corepolfluxvseddratio}; {top} right panel). This analysis reveals a moderate correlation (generalized KT test p = 0.029; partial SR test p = 0.025, keeping $\mathrm{M_{BH}}$ fixed). The best-fit line has a slope of $= -0.62$, intercept $= -0.89$ and standard deviation $=1.1$. {These correlations suggest that B-field ordering in the core regions is higher in the lower accretion systems. This can be possible because higher accretion may accumulate more Faraday rotating and hence depolarising material in the core regions.} Overall, these results may indicate that in lower accretion systems, wind-like outflows are also present along with the jets. This results in decreasing the B-field ordering as well as polarized intensity \citep[e.g.,][]{Silpa2021, silpa2022}. {On the other hand, Figure~\ref{fig:corepolfluxvseddratio} (bottom panel) shows a positive correlation between polarization fraction and Eddington ratios in the extended radio lobes (KT test p=0.029). This may suggest that higher accreting sources produce radio emission with more ordered B-field structures.} 

{The current work presents a small sample size, hence the above correlation results represent small-number statistics, potentially applicable to the RQ population of low-luminosity AGN showing larger extents of radio emission. In the future, we intend to include the remaining sources from the sample, adding more KSRs and point sources at $\sim15\arcsec$ resolution, to provide a more comprehensive analysis of the entire population of Seyferts and LINERs with significant AGN contributions.}
\section{Discussion}\label{discussions}
The origin of radio emission in RQ AGN remains a topic of significant investigation and debate after years of study \citep{Ishibashi2011, Zakamska2014, Panessa2019, Radcliffe2021}. 
Unlike RL AGN exhibiting powerful radio jets extending far beyond the host galaxy into the IGM, RQ AGN typically show much more compact radio emissions, which makes them more difficult to study and understand. Understanding the origin of the radio emission in these RQ sources is crucial not only for interpreting these systems but also for comprehending AGN feedback and galaxy evolution, as these constitute the dominant ($\sim$80\%) population of the AGN. AGN feedback, which includes both the release of energy into the surrounding medium and the regulation of star formation and black hole growth \citep{King2015}, is less obvious and harder to detect in RQ AGN \citep{Santoro2020, Harrison2024}. {The following sub-sections discuss the properties of the selected sample and their relation with the host galaxies.}

\subsection{Radio-FIR correlation} Most of the sources in the studied sample lie above the tight radio-FIR correlation derived by \citet{Yun2001} for a large sample of AGN and star-forming galaxies, indicating that starburst activity alone cannot account for the observed radio excess in these galaxies \citep{Gallimore2006, Sebastian2020}. Typically, radio emission from stellar activity and winds exhibits a morphology similar to that of the host galaxy \citep{Panessa2019}. While this is observed in some of the sample sources, the Seyfert outflow is often oriented at a large angle to the galactic radio continuum. 

\subsection{Star formation rates}
We have estimated the star formation rates in these galaxies using the relation SFR $=4.5\times 10^{-44}$ $L_{\rm{FIR}}$ \citep{Kennicutt1998}, $L_{\rm{FIR}}=(1+S_{100\mu m}/2.58S_{60\mu m})L_{60\mu m}$ \citep{Helou1988,Yun2001}, and FIR luminosities from \citet{Fullmer1989, Moshir1990,Moshir1993}. The SFR in these Seyfert and LINER galaxies are approximately $1-2$ M$_\sun$/yr except in NGC~1068 ($\sim$14 M$_\sun$/yr), NGC~3079 ($\sim$4 M$_\sun$/yr) and NGC~4388 ($\sim$4 M$_\sun$/yr), where radio emission from the galactic disk is also detected. Therefore, the observed radio synchrotron emission in these galaxies with KSRs can be attributed to AGN-driven jets and winds, rather than starburst superwinds.

\subsection{Polarization and B-field properties}
We have detected polarized emission in all the sources. The sources with detected galactic emission clearly show a difference in the fractional polarization compared to the Seyfert or LINER radio outflow. The highly polarized extended radio emission, displaying a well-ordered B-field structure anchored to the central engine, supports the suggestion that these KSRs are AGN-driven \citep[e.g.,][]{Sebastian2020}. 

Jets can be launched by extracting energy from the rotating SMBH, where B-fields are twisted and amplified by the black hole, generating a strong Poynting flux \citep[][BZ mechanism]{Blandford1977, Qian2018}. Alternatively, jets can derive energy from the accretion disk, generating strong toroidal B-fields that drive the jet through magnetocentrifugal forces \citep[][BP mechanism]{Blandford1982}, or through a combination of both mechanisms \citep{Wang2008, Garofalo2010, Talbot2021}. The BZ mechanism is thought to be more effective for high black hole spin parameters and is often associated with launching axial jets \citep{Wang2008, Qian2018}, whereas the BP mechanism is more effective for lower black hole spin parameters and is often associated with launching disk winds \citep{Li2008, Qian2018}. AGN-driven winds refer to magnetically-driven outflows that can be the broadened base of the jet \citep{Mehdipour2019, Silpa2021}, the broken/decollimated part of the jet \citep{Veilleux2005, Mukherjee2018}, or from the inner accretion disk \citep{Fukumura2010}, rather than radiatively/thermally-driven winds from the accretion disk/torus/BLR \citep[e.g.,][]{Begelman1983, Proga2004, Mizumoto2019}. 
{Our study supports the suggestion of jet launching accompanied either by an accretion disk wind or broadened jets with transverse velocity structures. Based on the jet simulation studies of \citet{Tchekhovskoy2016, Barniol2017}, the magnetic field structures observed in our sources may support the presence of the BZ mechanism.}

\begin{table*}
\small{
\caption{Correlations and probability (p) values}
\tabcolsep=0.1cm
\begin{center}
\begin{tabular}{cccccc}
\hline \hline

Quantity 1&Quantity 2&{p-value}&{Correlation coefficient}&{Test type}\\\hline

M$_{\rm{BH}}$&$\lambda_{\rm{Edd}}$&0.0001&-0.91&Partial SR keeping all fixed\\
P$_{\rm{jet,kin}}$&M$_{\rm{BH}}$&0.0008&0.85&Partial SR keeping $\lambda_{\rm{Edd}}$ fixed\\
P$_{\rm{jet,kin}}$&$\lambda_{\rm{Edd}}$&0.0123&0.72&Partial SR keeping M$_{\rm{BH}}$ fixed\\
f$_{\rm{p,core}}$&M$_{\rm{BH}}$&0.2145&0.43&Partial SR keeping $\lambda_{\rm{Edd}}$ fixed\\
f$_{\rm{p,core}}$&$\lambda_{\rm{Edd}}$&0.0251&-0.70&Partial SR keeping M$_{\rm{BH}}$ fixed\\
f$_{\rm{p,extended}}$&M$_{\rm{BH}}$&0.067&-0.6&Partial SR keeping $\lambda_{\rm{Edd}}$ fixed\\
f$_{\rm{p,extended}}$&$\lambda_{\rm{Edd}}$&0.0133&0.75&Partial SR keeping M$_{\rm{BH}}$ fixed\\
P$_{\rm{jet,kin}}$&M$_{\rm{BH}}$&0.0209&0.52&KT\\
P$_{\rm{jet,kin}}$&$\lambda_{\rm{Edd}}$&0.3807&-0.21&KT\\
P$_{\rm{jet,kin}}$/L$_{\rm{Edd}}$&$\lambda_{Edd}$&0.0108&0.70&Linear regression$^\ast$\\
${R}$&M$_{\rm{BH}}$&0.0057&0.63&KT\\
L$_{\rm{core,10GHz}}$&M$_{\rm{BH}}$&0.0311&0.48&KT\\
L$_{\rm{bol}}$&M$_{\rm{BH}}$&0.0287&-0.51&KT\\
B$_{\rm{min}}$&M$_{\rm{BH}}$&0.6384&0.12&KT\\
E$_{\rm{total}}$&M$_{\rm{BH}}$&0.0210&0.51&KT\\
$\tau$&M$_{\rm{BH}}$&0.7373&-0.09&KT\\
I$_{\rm{p,core}}$&I$_{\rm{core}}$&0.0041&2.874$^\dagger$&Generalized KT\\
I$_{\rm{p,extended}}$&I$_{\rm{extended}}$&0.0466&0.51&KT\\
I$_{\rm{p,extended}}$&I$_{\rm{core}}$&0.0196&0.58&KT\\
I$_{\rm{p,core}}$&$\lambda_{\rm{Edd}}$&0.0072&2.69$^\dagger$&Generalized KT\\
f$_{\rm{p,core}}$&$\lambda_{\rm{Edd}}$&0.0290&2.183$^\dagger$&Generalized KT\\
f$_{\rm{p,extended}}$&$\lambda_{\rm{Edd}}$&0.0286&0.553&KT\\

 \hline
\end{tabular}
\end{center}

{\small \textit{Note:} SR = Spearman's rank test, KT = Kendall's Tau, $^\ast$linear regression by EM algorithm from ASURV package, $^\dagger$z-value from Generalized KT test \citep{asurvcoef}. M$_{\rm{BH}}$ is the mass of the black hole, $\lambda_{\rm{Edd}}$ is the Eddington ratio, P$_{\rm{jet,kin}}$ is the jet kinetic power, f$_{\rm{p,core}}$ is the core fractional polarization, L$_{\rm{Edd}}$ is the Eddington luminosity, ${R}$ is the radio luminosity, L$_{\rm{core,10GHz}}$ is the core luminosity at 10 GHz, L$_{\rm{bol}}$ is the bolometric luminosity, B$_{\rm{min}}$ is the magnetic field under equipartition condition, E$_{\rm{total}}$ is the total energy under equipartition condition, $\tau$ is the electron lifetime with synchrotron and IC CMB losses, I$_{\rm{p,core}}$ is polarized intensity at the core, I$_{\rm{p,extended}}$ is the polarized intensity in the extended region including the jets and the lobes, and f$_{\rm{p,extended}}$ is the fractional polarization in the same region.}
}
\label{tab:tab6}
\end{table*}

The 10~GHz $\sim$7$\arcsec$ cores of most of the Seyfert galaxies (e.g., NGC 3516, NGC 4593, NGC 5506, NGC 1320) are unpolarized in our observations to within $\sim$10$\mu$Jy~beam$^{-1}$. This may be due to the presence of denser material (cold gas) in the cores causing Faraday depolarization \citep{Ioannis2015}, inhomogeneous polarised emission, or Faraday rotation within the source depth \citep{Cioffi1980}. Those with detected polarization (e.g., NGC 4235, NGC 4388, NGC 4594) primarily show an inferred toroidal field component in the cores. This may suggest the presence of AGN winds or jet sheaths along with the jets in these sources \citep[e.g.,][]{Mehdipour2019, Silpa2021}. The presence of winds or sheaths may in turn be the cause of the stunted growth of the jets, as they act like a cocoon around the jet and hinder the jet's motion.

{In several sources (e.g., NGC~1320, NGC~3516, NGC~3079, NGC~4388), we observe jet bending with the B-field lines following the bend suggesting that poloidal magnetic fields are  `frozen' in the synchrotron plasma of the jet.}
Moreover, jet bending may suggest that the denser environment in the spiral host galaxies of Seyferts or LINERs can be an important factor in their stunted growth \citep{Gallimore2006b,Nolting2019, Silpa2023}. In some sources, the dense medium causes the jet to break up and channel through the medium into the IGM along the least hindered paths \citep[e.g.,][]{Saxton2005,Mukherjee2018}.

The inferred B-field orientations in the radio outflows of all the Seyfert and LINER galaxies presented here support the presence of ordered B-fields even on kpc-scales. The B-field structures observed in some sources (e.g., NGC 3516) are similar to those observed in RL FRII radio galaxies where the hotspot shows B-field compression perpendicular to the jet direction \citep[e.g.,][]{Conway1993,Kharb2008,Baghel2023}. The hotspot in NGC 3516 additionally shows a spectral index flattening signifying particle re-acceleration at the terminal shock region of the radio outflow. This suggests that some RQ sources are not intrinsically different from RL sources in terms of their jet-launching mechanisms. 

\subsection{Host-AGN association}
The Seyfert-starburst composite galaxy NGC 3079 shows a lack of atomic neutral hydrogen gas along the Seyfert outflow direction indicating that AGN activity is either ionizing or pushing away gas, depleting the neutral hydrogen gas reservoir. This in turn quenches star formation along the radio outflow. On the other hand, in NGC 4388, we observe an extension of the H$\alpha$ gas along the jet, which may indicate the presence of star-forming regions along the jet. Additionally, the second fainter component of H$\alpha$ in NGC 2992 coincides with the region of possible `relic' emission. Similarly, we see [O III]~5007\AA~emission \citep{Schmitt2003} following the S-shape of the Seyfert radio jet in NGC~3516 but on smaller scales, which may indicate cold gas entrainment by the jet. Therefore, overall, we observe a close connection between the emission-line gas and radio outflows, and an indication of both positive and negative AGN feedback, similar to those seen in numerical simulations of jets by \citet{Mukherjee2018, Meenakshi2023}.

\subsection{Drivers of the radio outflow}
The radio outflow properties in our small sample show a strong connection with the black hole masses. Both the jet power and the total energy (particle + B-field energy) correlate strongly with M$_{BH}$, but not significantly with $\lambda_{Edd}$ {when checked using partial SR test}. This indicates that the driving factor is the mass of the black holes rather than the accretion rate. The correlation suggests that the B-field strength at the jet launching sites is stronger for sources with higher black hole masses, resulting in stronger jets. This finding potentially supports the BZ mechanism of jet launching \citep[see][]{Tchekhovskoy2016}.

We observe a bimodality in terms of Eddington ratios. Radio cores with flatter spectra are associated with sources with lower Eddington ratios, whereas the steeper cores are associated with higher Eddington ratios \citep[also observed in ][]{Laor2008, Chen2023}. Similarly, an anti-correlation is observed between the polarization fraction and the Eddington ratios (partial correlation test p = 0.025, see Table~\ref{tab:tab6}) in the cores. {This could be consistent with greater Faraday-rotating medium and greater net depolarization closer to the jet launching sites in highly accreting sources, or more radiation-pressure-driven winds in them. Interestingly, we observe a weak positive trend in the fractional polarization versus Eddington ratios (partial SR correlation test p = 0.013) for lobe emission. This may suggest that higher accreting sources are more effective at sustaining ordered magnetic fields on larger spatial scales.} 

Low-luminosity AGN are hypothesized to possess inner radiatively inefficient accreting flow (RIAF) disk and outer thin disk systems \citep{Lasota1996, Ho2000, Ho2008}. Moreover, the study by \citet{Ho2009} suggests that RIAF systems power Seyferts/LINERs with radiative efficiency much less than {$0.1$}. This implies that higher Eddington ratio sources, although accreting at a higher rate, cannot efficiently convert gravitational potential energy to radiative energy. Rather a significant part of the accretion power gets advected onto the black hole \citep[advection-dominated accretion flow, ADAF;][]{Narayan1998} or gets transferred to jet power \citep[e.g.,][]{Falcke2001}. {The spectral steepening with higher $\lambda_{\rm{Edd}}$ supports the idea that sources with flatter core spectra are currently in an active jet phase, accreting via ADAF/RIAF modes. In contrast, sources with steeper cores likely experienced jet activity in the past and are now accreting efficiently in a radiatively dominant mode.}

\section{Summary \& Conclusions}\label{summary}
In this paper, we have presented radio polarimetric images of 12 Seyfert galaxies with KSRs belonging to the CfA+12 micron sample at 
1.4$-$6~GHz and 10 GHz
using the VLA BnA$\rightarrow$A or D array configurations, respectively. We have attempted to understand the origin and impact of the radio outflows from these RQ AGN. The outcomes of this study are summarised below.

\begin{enumerate}
\item Bubble-like or lobe-like radio structures with dominant cores are detected in the 12 sources. The in-band spectral indices range from inverted to steep ($\alpha\sim+0.5$ to $-1.2$) at the core. We find signatures of organized B-field structures in the kpc-scale cores, jets, and lobes of these galaxies with the fractional polarization varying between a few per cent in the radio cores to up to $47\pm18\%$ in the lobes. These results are consistent with the AGN being the primary driver of radio outflows in these galaxies. 
\item In the four galaxies showing host galaxy emission along with the KSRs, viz., NGC 4388, NGC 3079, NGC 4594, and NGC1068, the apparent B-fields are aligned along the spiral arms of the galactic disks with the degrees of polarization typically being lower than the KSR values in that source. 
\item Based on global correlations we find a strong correlation between the mass of the supermassive black holes and the radio outflow properties like jet kinetic power, with accretion rate playing a secondary role in these RQ sources. This further attests to the dominance of AGN jets in producing KSRs in Seyfert and LINER galaxies. 
\item We find a significant correlation between total and polarized core intensities in these sources. This implies that a more organized B-field which results in greater polarized emission also produces a more luminous radio source. This attests to the existence of dominant MHD outflows in RQ AGN. The B-fields are inferred to be toroidal in the cores of several sources which could suggest the presence of either a jet sheath surrounding the jet spine or a wind component surrounding the jet. The spectral index versus Eddington ratio relation is consistent with this suggestion. 
\item An anti-correlation is observed between the core fractional polarization and Eddington ratios. {This could suggest the presence of greater depolarizing media in sources with greater accretion rates, or more radiation-pressure-driven winds in them. A weak positive trend in the fractional polarization versus Eddington ratios for lobe emission, on the other hand, may suggest that higher accreting sources are more effective at sustaining ordered magnetic fields on kpc-scales.}

\item An examination of the radio properties along with emission line gas reveals the signatures of both positive and negative jet-related AGN feedback in these sources (e.g., NGC~3079, NGC~2992). This supports the idea that radio-quiet AGN can significantly impact their host galaxy environments in spite of producing smaller outflows. 
\item Several sources show the signatures of jet bending, flaring or disruption indicating strong jet-medium interaction. The spectra and/or polarization structures indicate the presence of relic lobes and thereby episodic AGN activity in some of the sources (e.g., NGC~1068, NGC~2992, NGC 4593). 
\end{enumerate}    

Overall, we find that AGN-related radio outflows are prominent in radio-quiet AGN and could be driven by strong magnetic fields anchored to the black holes themselves. These jets interact strongly with their surrounding media which in turn affects their total radio extents making them radio-quiet.

\section{Acknowledgements}
{We thank the referee for their suggestions which have improved this manuscript significantly.}
SG and PK acknowledge the support of the Department of Atomic Energy, Government of India, under the project 12-R\&D-TFR-5.02-0700. {SB and CO acknowledge support from the Natural Sciences and Engineering Research Council (NSERC) of Canada.} This research has used data from the National Radio Astronomy Observatory (NRAO) facility. NRAO is the facility of the National Science Foundation operated under a cooperative agreement by Associated Universities, Inc. We thank the staff of the GMRT that made these observations possible. GMRT is run by the National Centre for Radio Astrophysics of the Tata Institute of Fundamental Research. This research has made use of the NASA/IPAC Extragalactic Database (NED), which is operated by the Jet Propulsion Laboratory, California Institute of Technology, under contract with the National Aeronautics and Space Administration.

\bibliography{ms}{}
\bibliographystyle{aasjournal}

\section{APPENDIX}
We present here the radio-optical overlays for the four galaxies showing galactic emission along with AGN outflows. 

\begin{figure*}[b]
\centering
\includegraphics[width=8cm]{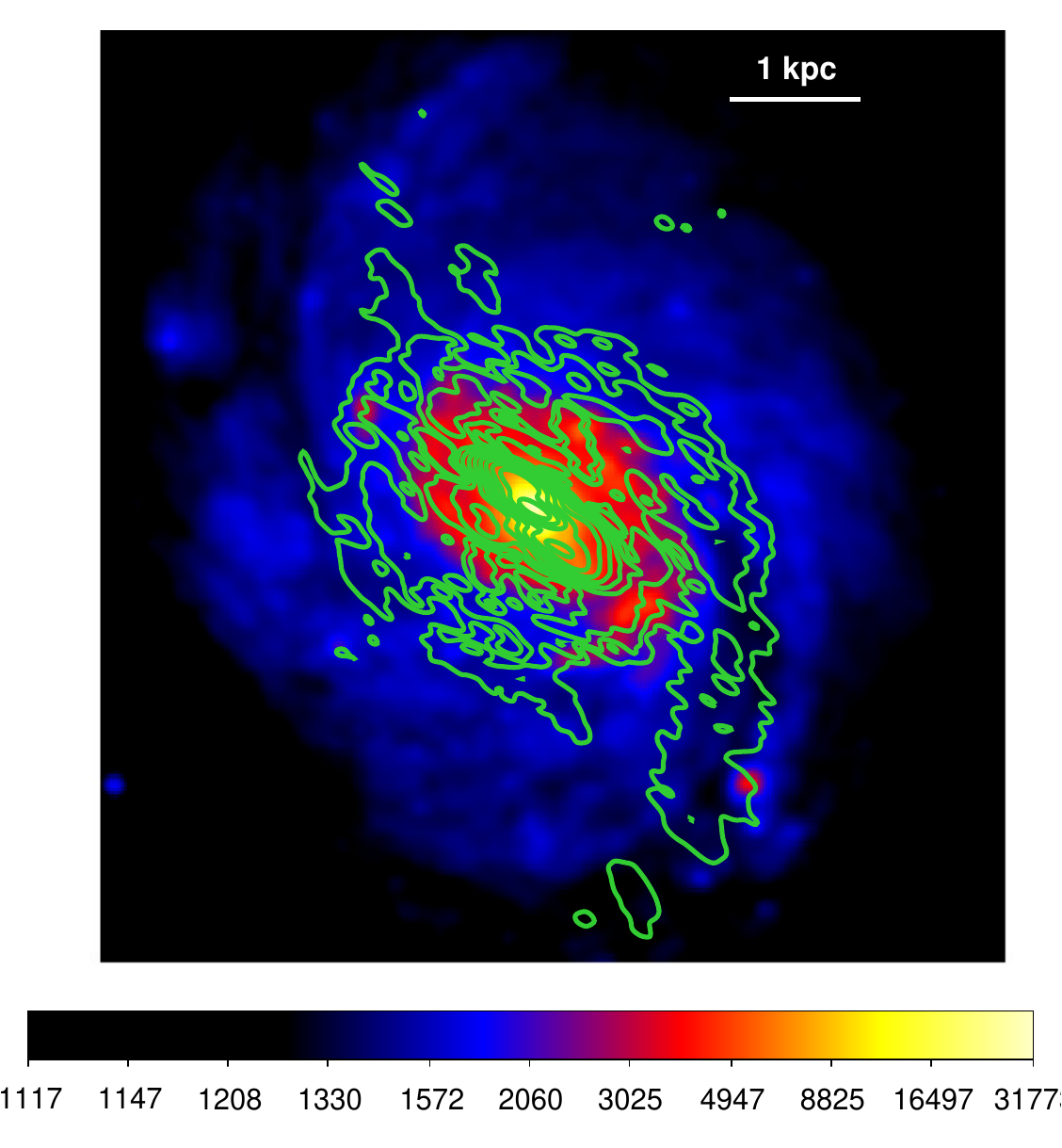}
\includegraphics[width=8cm]{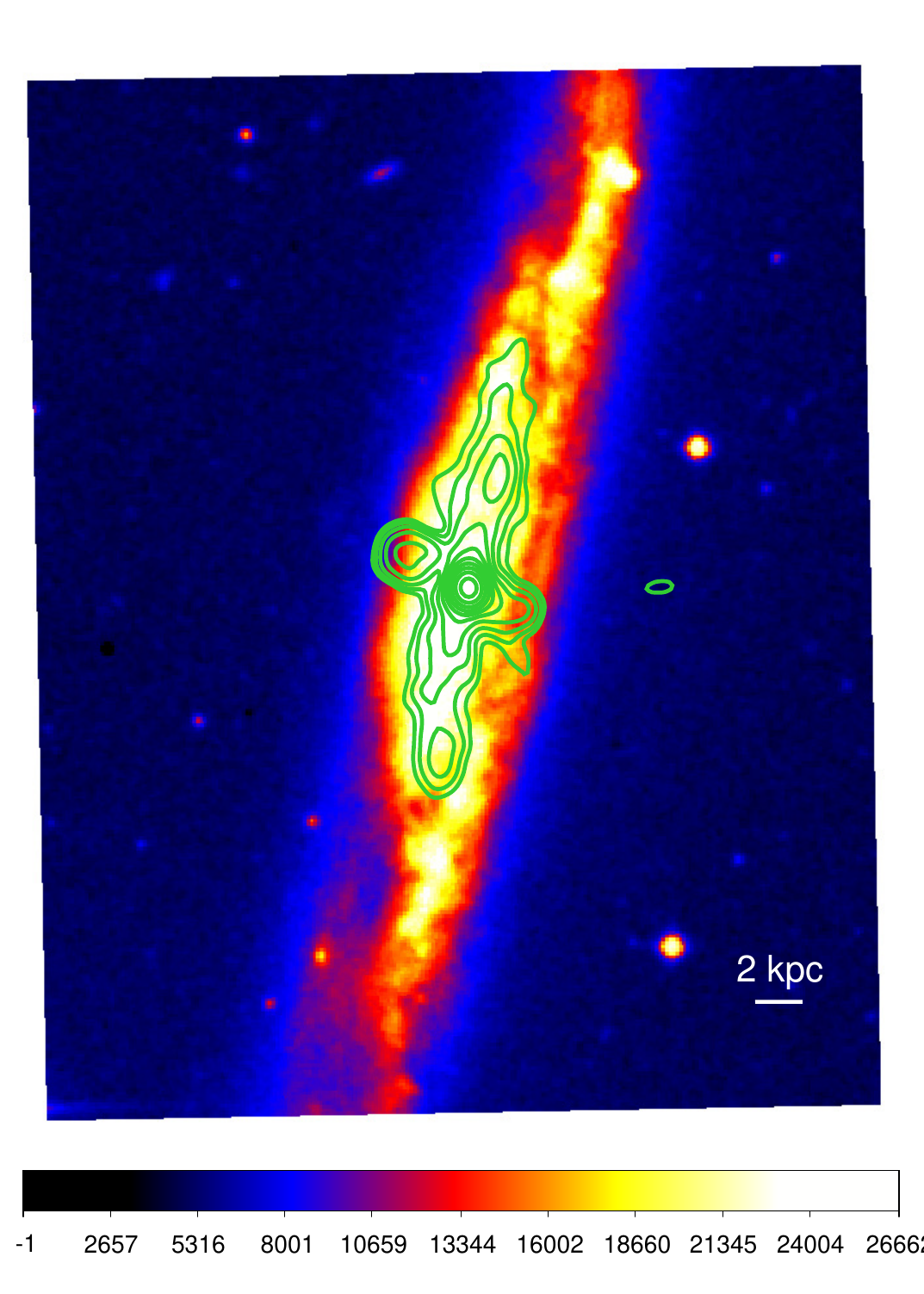}
\includegraphics[width=8cm]{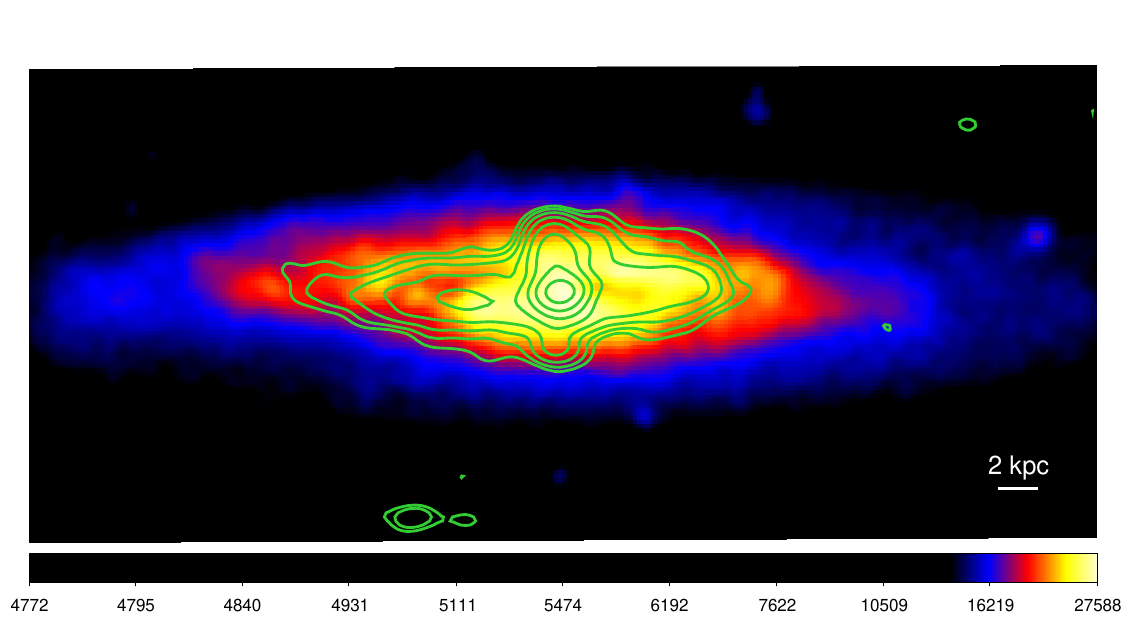}
\includegraphics[width=8cm]{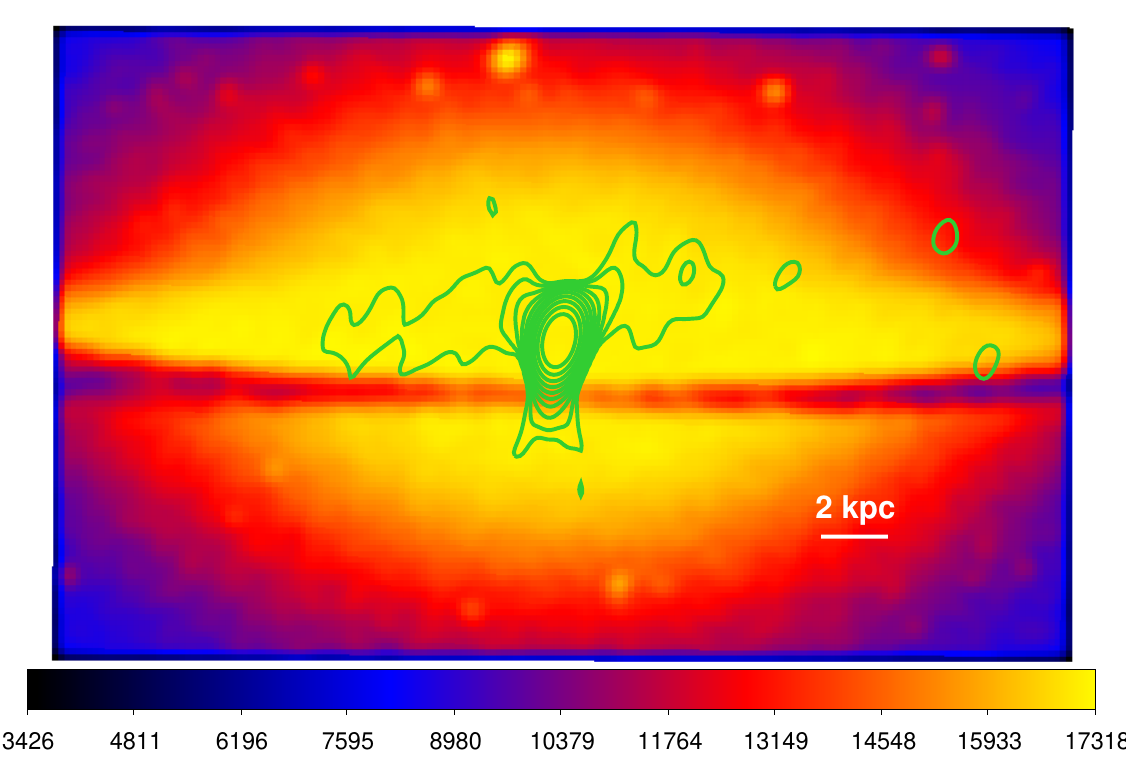}
\caption{\small Radio-optical overlays for galaxies with observed galactic radio emission along with the Seyfert outflows. Galaxies shown clockwise from the left top are NGC 1068, NGC 3079, NGC 4594, and NGC 4388. 1.4~GHz (NGC 1068) or 10~GHz (NGC 3079, NGC 4388, NGC4594) radio contours in green superimposed on SDSS g-filter images of the host galaxies. The radio contour levels are 3$\sigma \times(-1, 1, 2, 4, 8, 16, 32, 64, 128, 256, 512)$ with $\sigma = 250, 50, 20, 22~\mu$Jy beam$^{-1}$ for NGC 1068, NGC 3079, NGC 4594, and NGC 4388, respectively. }
\label{fig:opticalradiooverlay}
\end{figure*}
\end{document}